\documentclass[final,5p,times,twocolumn]{elsarticle}

\usepackage{amsmath}
\usepackage{amssymb}
\biboptions{numbers,sort&compress} 
\usepackage{graphicx}
\usepackage[svgnames,dvipsnames]{xcolor} 
\usepackage{etoolbox} 
\usepackage{bm} 
\usepackage{bbm} 
\usepackage{soul} 
\usepackage{mathtools}
\usepackage[makeroom]{cancel}
\usepackage{xspace}
\usepackage{multirow}
\usepackage{ulem}
\usepackage{booktabs}
\usepackage{tabularx}
\usepackage{makecell}
\usepackage{longtable}
\usepackage{chngcntr}
\newcommand{\hpkot}[1]{#1}
\definecolor{scolor}{rgb}{0,0,0} 
\makeatletter
\def\LT@makecaption#1#2#3{%
  \LT@mcol\LT@cols c{%
    \parbox{\LTcapwidth}{%
      \rightskip=0pt\sffamily\small
      \textbf{\color{scolor}#2}\par #3\par\vskip4pt}}}
\makeatother

\usepackage{hyperref} 
\hypersetup{
    colorlinks=true,
    linkcolor=blue,
    citecolor=blue,
    filecolor=magenta,
    urlcolor=blue
}

\newcommand{\BFO}{BiFeO$_3$}

\usepackage{placeins}
\usepackage{float}  
\usepackage{needspace}  

\begin{document}
\def\floatpagepagefraction{1} \def\textpagefraction{.001}

\begin{frontmatter}

\title{AlterSeeK-Path: Systematic construction of generalized band-structure paths for displaying altermagnetic spin splitting}

\author[1]{Yujia Teng}

\author[1,2]{Mesfin Eshete}

\author[1,3]{Andrea Urru}

\author[1,4]{Daniel Seleznev}

\author[5]{Se Young Park}

\author[6]{Sebastian E. Reyes-Lillo}

\author[1]{Karin M. Rabe\corref{cor1}}
\ead{kmrabe@physics.rutgers.edu}

\cortext[cor1]{Corresponding author}

\affiliation[1]{organization={Department of Physics and Astronomy, Center for Materials Theory}, addressline={Rutgers University}, city={Piscataway}, state={New Jersey}, postcode={08854}, country={USA}}

\affiliation[2]{organization={Department of Industrial Chemistry, Addis Ababa Science and Technology University}, addressline={P.O. Box 16417}, city={Addis Ababa}, country={Ethiopia}}

\affiliation[3]{organization={Dipartimento di Fisica}, addressline={Università di Cagliari, Cittadella Universitaria}, city={Monserrato, CA}, postcode={09042}, country={Italy}}

\affiliation[4]{organization={Department of Physics}, addressline={University of Texas at Austin}, city={Austin}, state={Texas}, postcode={78712}, country={USA}}

\affiliation[5]{organization={Department of Physics and Origin of Matter and Evolution of Galaxies (OMEG) Institute}, addressline={Soongsil University}, city={Seoul}, postcode={06978}, country={Korea}}

\affiliation[6]{organization={Departamento de F\'isica y Astronom\'ia}, addressline={Universidad Andres Bello}, city={Santiago}, postcode={837-0136}, country={Chile}}

\begin{abstract}
Altermagnetic materials exhibit spin splitting in their electronic band structures while maintaining zero net magnetization. However, conventional high-symmetry k-paths generally hide this splitting because they follow symmetry lines that in most cases enforce spin degeneracy. We present AlterSeeK-Path, an open-source Python tool that systematically constructs generalized band-structure paths for collinear altermagnets. The method selects the centroid of the conventional irreducible wedge used in routine band-structure calculations as the representative general k-point, maps it to a spin-flip-related partner, and inserts paired segments through these points into the standard high-symmetry path; these segments systematically sample the interior of the irreducible wedge. We demonstrate the construction for all 54 three-dimensional combinations of extended Bravais lattice type and spin Laue group across the six crystal systems that support collinear altermagnetism, and for the 12 two-dimensional cases spanning the four compatible two-dimensional Bravais lattices. Representative band structures are shown for the distinct lattice/path cases. With AlterSeeK-Path, these band structures can be constructed with essentially the same effort as conventional band structures. 

\vskip 1em
\noindent \textbf{Program Summary} \\
\textit{Program title:} AlterSeeK-Path \\
\textit{Developer's repository link:} \url{https://github.com/yujia-teng/AlterSeeK-Path} \\
\textit{Licensing provisions:} MIT \\
\textit{Programming language:} Python \\
\textit{External routines/libraries used:} FINDSPINGROUP, SeeK-path, spglib, pymatgen, ASE, NumPy, SciPy, SymPy, Matplotlib \\
\textit{Nature of problem:} To systematically display the spin splitting in the band structure of altermagnetic crystals in an automated way. \\
\textit{Solution method:} To extend the conventional band structure path by sampling lines from the center of the irreducible Brillouin zone to high-symmetry points on its boundary, with construction of the extended path requiring only the magnetic crystal structure information as input.\\ 
\textit{Additional comments:} VASP, Quantum ESPRESSO, and ABINIT are fully supported and guidance for other packages is provided.\\
Restrictions: Currently limited to collinear magnetic order.
\end{abstract}


\end{frontmatter}

\section{Introduction}
Altermagnetism is a new class of collinear antiferromagnetic order, distinct from conventional antiferromagnetism, characterized by spin splitting (SS) in the nonrelativistic band structure~\cite{smejkal-prx22,smejkal-prx22a}. The spin splitting at general k-points is due to the breaking of $PT$ and $U\mathbf{t}$ symmetries, where $P$ is space inversion, $T$ is time reversal, $U$ is spin reversal, a $180^\circ$ spin rotation about an axis perpendicular to the collinear spin axis, and $\mathbf{t}$ is a fractional translation that connects opposite spin sublattices. 
While in conventional antiferromagnets, zero magnetization is enforced by $PT$ and $U\mathbf{t}$ or both, in altermagnets, zero magnetization is enforced by a symmetry operation $T\{R|\mathbf{t}\}$ or $U\{R|\mathbf{t}\}$ where $R$ is neither the identity operation $E$ nor $P$. 

While SS in the electronic band structure is a direct computational signature of altermagnetism, SS is not necessarily visible in conventional band structure plots. Such plots are composed of high-symmetry lines, and the symmetries of these lines in many cases enforce spin degeneracy. This is illustrated by the well-known compound \BFO. Although \BFO\ is an altermagnet \cite{smejkal-24,Farooq-prb23,bernardini-jap25,urru-prb25,dong-prb25,dong-prb25b,husain-prl26}, no SS appears along its conventional high-symmetry path \cite{neaton-prb05,urru-prb25,dong-prb25}. 

Current approaches to showing SS in the altermagnetic band structure are of two types~\cite{yuan-prb20, ding-prlett.24,yu-npjquantummater.25,reimers-natcommun24,liu-prlett.24,jiang-nat.phys.25,lee-prlett.24,xu-prb25,adamantopoulos-npjspintronics24,yang-natcommun25,rooj-25,ma-natcommun21,han-sciadv24}. The first is to choose an arbitrary general k-point and make an auxiliary plot showing the path from $\Gamma$ to this point together with the path from $\Gamma$ to the spin-flip image of this point. The second is to add an extra manually selected low-symmetry line (or lines) to the conventional band structure path. In general, the nonzero splitting displayed on the added path demonstrates altermagnetism but depends strongly on the ad hoc choices made and is not representative of the spin splitting across the Brillouin zone (BZ) in a way that allows prediction of physical properties or comparison of different materials.
What is needed is a sampling of the interior of the BZ centered on a general k-point defined by the same prescription in every BZ, on a path constructed following a universal convention.

Here, we introduce AlterSeeK-Path, a tool to systematically construct band paths that display the characteristic SS of collinear altermagnets.
The key idea is to choose the  representative general k-point as the centroid of the nonmagnetic irreducible BZ and sample the interior of the BZ by lines connecting the representative general k-point to points on the BZ boundary.
It incorporates the capabilities of SeeK-path~\cite{seek-path}, which constructs the conventional path of high-symmetry lines, and FINDSPINGROUP~\cite{findspingroup,chen-prx24,liu-nature26,liu-prx22} which determines spin space group (SSG) operations for a given magnetically ordered crystal. 
In Section 2, we describe the method in detail.
In Section 3, we present illustrations of the application of the method for various magnetically ordered crystal structures: The complete case library is presented in \ref{sec:case-library}.
In Section 4, we discuss the scope and limitations of the method. 
Section 5 concludes the paper.

\section{Method}

\begin{figure}[t]
    \centering
    \includegraphics[width=8cm]{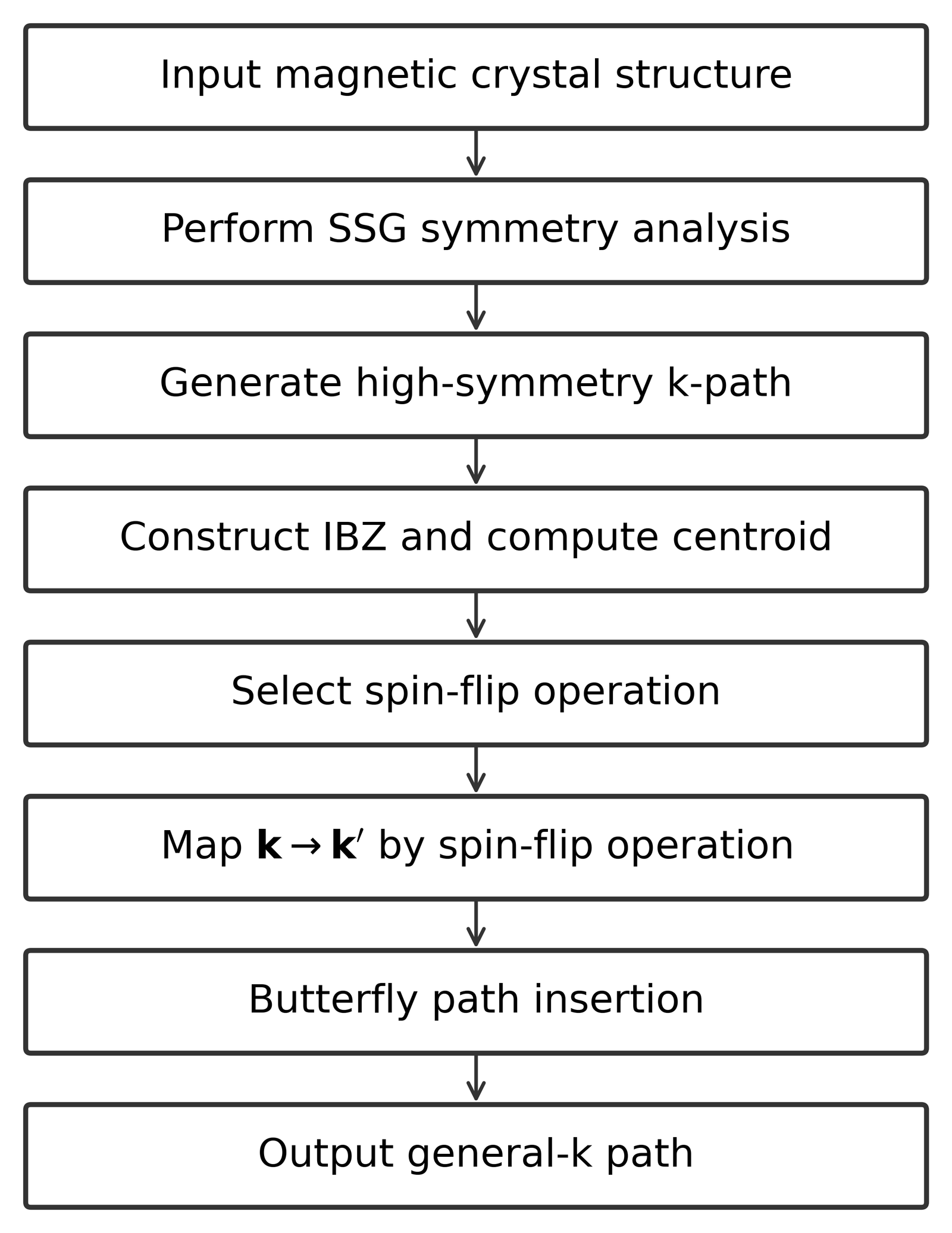}
    \caption{Workflow of AlterSeeK-Path for collinear altermagnets. The input magnetic crystal structure is analyzed with FINDSPINGROUP to identify the spin-flip operations and the space group consistent with the magnetic order. The conventional high-symmetry path is generated with SeeK-path. The general point $\mathbf{k}$ is taken as the centroid of the IBZ. One spin-flip operation is then selected, $\mathbf{k}$ is mapped to its spin-flipped partner $\mathbf{k}'$, and butterfly blocks are inserted into the high-symmetry path to produce the final general-$k$ path.}
    \label{fig:workflow}
\end{figure}

The construction begins with an input crystal structure with collinear magnetic moments assigned to magnetic atoms. 
From this information, AlterSeeK-Path constructs the irreducible Brillouin zone (IBZ). 
The volume centroid point of the IBZ is chosen as the representative general k-point $\mathbf{k}$. 
To display the alternating character of the altermagnetic SS, we identify the SSG and the ``MSG without SOC'' (magnetic space group without spin-orbit coupling) and select a particular symmetry operation that interchanges the up and down sublattices. We then construct the 
spin-flip IBZ (sfIBZ), which is the image of the IBZ under the point part of this operation, with volume centroid point $\mathbf{k}'$, which is the spin-flip image of $\mathbf{k}$.
AlterSeeK-Path then inserts $\mathbf{k}$ and $\mathbf{k}'$ into the standard high-symmetry path in a structured ``butterfly'' pattern. The resulting path samples general k-points and paired segments in both the IBZ and sfIBZ, revealing SS without manual path design. The overall workflow is summarized in Fig.~\ref{fig:workflow}; the method is described in detail in the following subsections.

\subsection{General k-point: IBZ centroid}
\label{sec:centroid}
First, we discuss the construction of the IBZ and the identification of the general k-point $\mathbf{k}$ as the volume centroid of the IBZ.
In a nonrelativistic collinear magnet, the spin-only symmetry $UT$ ensures that the energy in each spin channel is even in momentum,
\begin{equation}
E_\sigma(\mathbf{k})=E_\sigma(-\mathbf{k}).
\label{eq:spin_even_k}
\end{equation}
Thus, the band structure is always inversion symmetric, whether or not inversion is a symmetry of the magnetic crystal. For constructing the IBZ, we therefore include inversion in the crystallographic point-group operations, and use the resulting Laue group to identify symmetry-equivalent k-points. 

The choice of IBZ is not unique. Here, we obtain the high-symmetry points and lines from SeeK-path and construct the IBZ as the volume defined by high-symmetry points as vertices. 

For certain monoclinic cases, the tabulated high-symmetry vertices do not enclose the correct IBZ region. In these cases, AlterSeeK-Path constructs the IBZ using a Voronoi partition of symmetry-related images.
A further special case is handled automatically: for the Laue groups $6/m$, $\bar{3}m$, and $4/m$, the tabulated vertices enclose only half of the true IBZ. We double this to the full IBZ as described in more detail in Sec.~\ref{sec:enlarged} and \ref{sec:doubling-choices}.

When the assignment of collinear magnetic moments to the magnetic atoms lowers the symmetry of the crystal structure, but the atomic positions are not relaxed or adjusted accordingly, the input to SeeK-path needs to be modified for construction of the correct IBZ.
We add a marker atom at a general point and all points related by the operations in the MSG without SOC so that SeeK-path recognizes the symmetries of the magnetically ordered structure. More details are in Sec.~\ref{sec:magnetic-symmetry-reduction}.

The representative general k-point $\mathbf{k}$ is chosen as the volume centroid of the IBZ. The volume centroid $\mathbf{k}_c$ of the IBZ is computed numerically using a tetrahedral decomposition, as follows. The IBZ vertices are first identified from the standard high-symmetry k-point tables~\cite{seek-path}. The resulting IBZ is represented as a convex hull, which is then decomposed into tetrahedra. The centroid of each tetrahedron is the simple average of its four vertices, and the overall IBZ centroid is obtained as a volume-weighted sum of the tetrahedral centroids:
\begin{equation}
    \mathbf{k}_c = \frac{\sum_i V_i \, \mathbf{k}_i^{\rm tet}}{\sum_i V_i},
\end{equation}
where $V_i$ and $\mathbf{k}_i^{\rm tet}$ are the volume and centroid of the $i$-th tetrahedron, respectively.

\subsection{Spin-flip operation \texorpdfstring{$R$}{R}}
\label{sec:R}

In an altermagnet, the up- and down-spin sublattices are exchanged by SSG elements of the form $U\{R|\mathbf{t}\}$ and $U\{R|\mathbf{t}\}^{-1}$ in which the spin reversal $U$ is accompanied by the spatial part $\{R|\mathbf{t}\}$, consisting of a point operation $R$ and a translation $\mathbf{t}$. 
Because these sublattice-exchange symmetries ensure that 
\begin{equation}
E_{\sigma}(R\mathbf{k})=E_{-\sigma}(\mathbf{k}).
\label{eq:spin-flip_k}
\end{equation}
we refer to $R$ as a ``spin-flip operation''. 
For a selected spin-flip operation $R$, the general k-point $\mathbf{k}$ in the IBZ is mapped to its spin-flipped partner $\mathbf{k}'=R\mathbf{k}$, and the sfIBZ is the image of the IBZ under $R$.

The spatial operations $R$ associated with spin-flip symmetries are identified automatically from the magnetic structure using FINDSPINGROUP. There can be more than one spin-flip operation; choosing a different spin-flip operation generates a different, but symmetry-equivalent sfIBZ, and the resulting band-structure plot is the same (Sec.~\ref{sec:discussion}).

\subsection{Butterfly path construction}
\label{sec:butterfly}
After choosing the general point $\mathbf{k}$ and a spin-flip operation $R$, AlterSeeK-Path constructs a band path 
by inserting segments into the conventional high-symmetry path that connect the high-symmetry points to $\mathbf{k}$ and $\mathbf{k}'$, sampling the interior of the BZ and showing the altermagnetic SS.

For a conventional high-symmetry line $A$--$B$, the elementary butterfly block is
\begin{equation}
    A \text{--} k
    \,|\,
    k' \text{--} A' \text{--} B' \text{--} k'
    \,|\,
    k \text{--} B ,
    \label{eq:butterfly_block}
\end{equation}
where $A'=RA$ and $B'=RB$.  
We use $k$ and $k'$ as band path labels for $\mathbf{k}$ and $\mathbf{k}'=R\mathbf{k}$.
For a high-symmetry line on which spin splitting is symmetry forbidden, the bands along $A^\prime$--$B^\prime$  are identical to those along $A$--$B$. 
The $\Gamma$ point is self-conjugate, $\Gamma^\prime=\Gamma$, and therefore keeps its unprimed label.

Applying Eq.~\eqref{eq:butterfly_block} to every segment in the conventional path would lead to unnecessary duplication of segments connecting $\mathbf{k}$ and $\mathbf{k}'$ to high-symmetry points.  AlterSeeK-Path therefore uses the following rules:
\begin{enumerate}
    \item \textit{Chain rule}: the conventional path is first separated into connected chains at each path break.
    \item \textit{Alternation rule}: within each connected chain, conventional high-symmetry segments and butterfly segments are alternated. The first chain starts with a conventional segment, while later disconnected chains start with a butterfly segment.
    \item \textit{Endpoint rule}: once a high-symmetry endpoint has been sampled by a butterfly block, it is not expanded again. If the last point of a chain has not yet been sampled, a compact endpoint path $B\text{--}k\,|\,k'\text{--}B'$ is appended. 
    \item \textit{Isolated-vertex rule}: any high-symmetry vertex not part of a connected chain is appended as a standalone path $C\text{--}k\,|\,k'\text{--}C'$.
\end{enumerate}

The following short path illustrates how these rules act when a connected chain is followed by a disconnected segment and one endpoint appears more than once.
For the conventional path, adapted from the hexagonal path used in Sec.~\ref{sec:gag}, $\Gamma\text{--}M\text{--}K\text{--}\Gamma\,|\,M\text{--}L$, the generated path is
\begin{equation}
\begin{split}
\Gamma
&\text{--}
\underbrace{M\text{--}k\,|\,k'\text{--}M'}_{}
\text{--}
\underbrace{K'\text{--}k'\,|\,k\text{--}K}_{}
\text{--}
\underbrace{\Gamma\text{--}k\,|\,k'\text{--}\Gamma}_{}
\\
&\,|\,
M\text{--}
\underbrace{L\text{--}k\,|\,k'\text{--}L'}_{}.
\end{split}
\end{equation}
Here the first segment $\Gamma$--$M$ is kept as the conventional reference segment, the segment $M$--$K$ is expanded into a full butterfly block, and the terminal $\Gamma$ point receives the compact endpoint path. After the path break, the segment $M$--$L$ starts a new chain; since $M$ has already been sampled, only the new endpoint $L$ is expanded. We thus see that this construction keeps a direct reference to the conventional high-symmetry path and minimizes the length of the generated band-path file while sampling the corresponding paired unprimed and primed path segments needed for systematic display of altermagnetic SS.

\subsection{Two-dimensional altermagnets}
For two-dimensional altermagnets represented as periodic slabs, in addition to $PT$ and $U\mathbf{t}$ symmetry, the symmetries $C_{2z}T$ and $Um_z$ must also be broken~\cite{zeng-prb24-2d,zeng-prb24-bilayer,pan-prl24}. AlterSeeK-Path constructs the two-dimensional band path in the physical reciprocal plane perpendicular to the selected vacuum axis, conventionally $k_z=0$ when the slab normal is $z$. The representative general point is the area centroid of the in-plane IBZ, which is determined by the two-dimensional Laue group. Although the slab is represented by a three-dimensionally periodic cell and its symmetry is analyzed in three dimensions, its in-plane reciprocal-space geometry and band path are determined by the corresponding two-dimensional symmetry. The oblique Bravais lattice cannot support altermagnetic spin splitting because its only nontrivial point operation is $C_{2z}$, whose spin-flipping form enforces spin degeneracy. The 12 allowed spin-layer-group cases therefore span the remaining four two-dimensional Bravais lattices. The distinct in-plane spin-layer-group patterns and paths are summarized in Table~\ref{tab:spin-layer-2d}.

\subsection{Implementation}
The method is implemented in the open-source Python package AlterSeeK-Path, available at \url{https://github.com/yujia-teng/AlterSeeK-Path}. The package requires Python $\geq$ 3.11. It uses \texttt{spglib}~\cite{spglib}, FINDSPINGROUP~\cite{findspingroup,chen-prx24,liu-nature26,liu-prx22}, and SeeK-path~\cite{seek-path} for symmetry analysis and standard k-path conventions; pymatgen~\cite{pymatgen} and ASE~\cite{ase-paper} for structure handling; and NumPy, SciPy, and Matplotlib for numerical computation and visualization.

AlterSeeK-Path is run with the command \texttt{alterseek-path}, which reads a POSCAR, CIF, or MCIF structure together with a collinear magnetic-moment specification, extracted automatically from MCIF files or supplied manually for POSCAR/CIF inputs. FINDSPINGROUP is called to identify the SSG, from which we identify the MSG without SOC. 

The reciprocal-space geometry is obtained from SeeK-path. Using the standardized reciprocal cell and high-symmetry points from SeeK-path, AlterSeeK-Path constructs the IBZ, computes its volume centroid $\mathbf{k}$, and generates the IBZ, sfIBZ, and spin pattern. Internally, these quantities are represented in the standardized primitive reciprocal basis returned by SeeK-path. Combining this geometry with the selected spin-flip operation, the workflow constructs the butterfly path described above. The final band-path coordinates are converted back to the input cell basis. Representative examples are discussed in Sec.~\ref{sec:magnetic-symmetry-reduction}.

\begin{figure*}[t]
\centering
\includegraphics[width=0.9\textwidth,keepaspectratio]{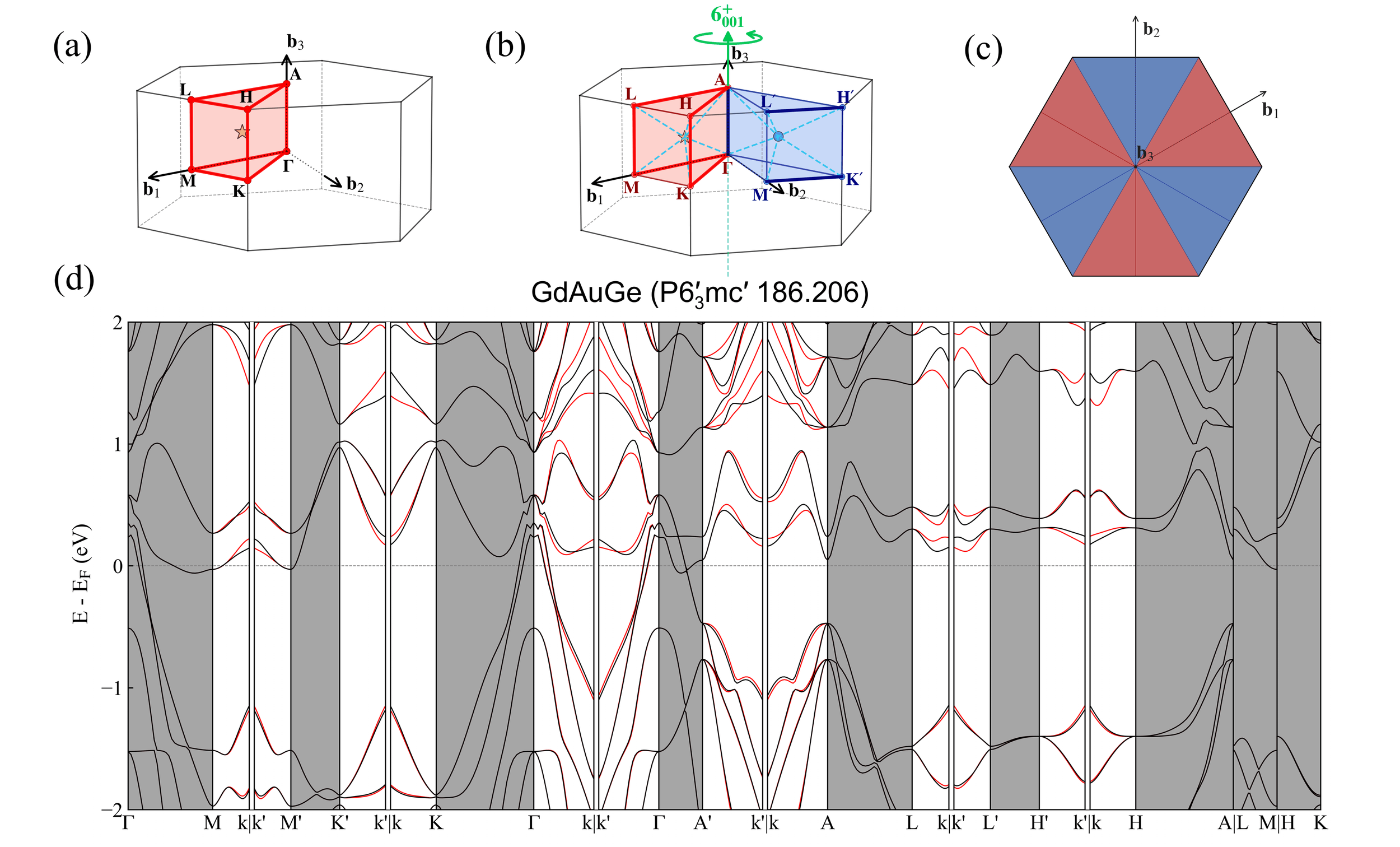}
\caption{Hexagonal hP2 BZ and band structure for the spin-Laue-group $^{2}6/^{2}m^{1}m^{2}m$ (bulk $g$-wave). (a) IBZ. (b) IBZ (red) and sfIBZ (blue), related by the spin-flip operation $6^{+}_{001}$. (c) Top view of the BZ, showing the resulting spin pattern. (d) Band structure along the generalized path generated by AlterSeeK-Path. Black and red curves denote spin-up and spin-down bands, respectively; grey and white background panels mark conventional high-symmetry and inserted general-$k$ paths, respectively. In this example, the two spin channels are degenerate in the grey panels but split in the white panels.}
\label{fig:HEX}
\end{figure*}

\section{Illustrative examples}

\subsection{Hexagonal GdAuGe}
\label{sec:gag}

We first consider the hexagonal compound GdAuGe, previously studied in Refs.~\cite{ram-prb23a,LaDuca-nanoletter24,samanta-arxiv26}, which crystallizes in space group $P6_3mc$ (No.~186) with lattice type hP2 (defined in SeeK-path convention~\cite{seek-path}), point group $6mm$ and Laue group $6/mmm$. Using the notation of Ref.~\cite{ram-prb23a}, its calculated magnetic ground state is the stripe-like AFM5 order. In this subsection, we instead consider AFM1, a simple A-type antiferromagnetic order consisting of ferromagnetic Gd layers stacked antiferromagnetically along the $c$ direction. The MSG without SOC is $P6_3'mc'$ (BNS 186.206, Type III). This configuration provides a simple example of an altermagnet whose SS is hidden along conventional high-symmetry lines.

Symmetry analysis identifies the AFM1 configuration as an altermagnet and therefore allows spin-split bands in the BZ. Nevertheless, this altermagnetic character is not visible along the conventional high-symmetry lines, $\Gamma\text{--}M\text{--}K\text{--}\Gamma\text{--}A\text{--}L\text{--}H\text{--}A \,|\, L\text{--}M \,|\, H\text{--}K$. Along these high-symmetry lines, the spin-up and spin-down bands remain degenerate, as shown in grey panels in Fig.~\ref{fig:HEX}. Thus, the conventional band structure alone does not reveal the SS expected from the symmetry classification.

AlterSeeK-Path supplements the conventional path with paired segments passing through the IBZ centroid $\mathbf{k}$ and its spin-flipped partner $\mathbf{k}'$. The generated path is
\begin{equation}
\begin{split}
\Gamma
&\text{--}
\underbrace{M\text{--}k\,|\,k'\text{--}M'}_{}
\text{--}
\underbrace{K'\text{--}k'\,|\,k\text{--}K}_{}
\text{--}
\underbrace{\Gamma\text{--}k\,|\,k'\text{--}\Gamma}_{}
\\
&\text{--}
\underbrace{A'\text{--}k'\,|\,k\text{--}A}_{}
\text{--}
\underbrace{L\text{--}k\,|\,k'\text{--}L'}_{}
\text{--}
\underbrace{H'\text{--}k'\,|\,k\text{--}H}_{}
\text{--}A
\\
&\,|\,L\text{--}M
\,|\,H\text{--}K.
\end{split}
\end{equation}
The conventional high-symmetry segments (grey panels) provide the conventional reference, while the inserted general-$k$ paths (white panels) reveal the SS away from symmetry-protected lines. Figure~\ref{fig:HEX} therefore directly contrasts the absence of splitting along the conventional path with the finite splitting exposed by the AlterSeeK-Path construction.

\subsection{Magnetic-order-induced symmetry reduction}
\label{sec:magnetic-symmetry-reduction}
\begin{figure*}[t]
    \centering
    \includegraphics[width=0.9\textwidth,keepaspectratio]{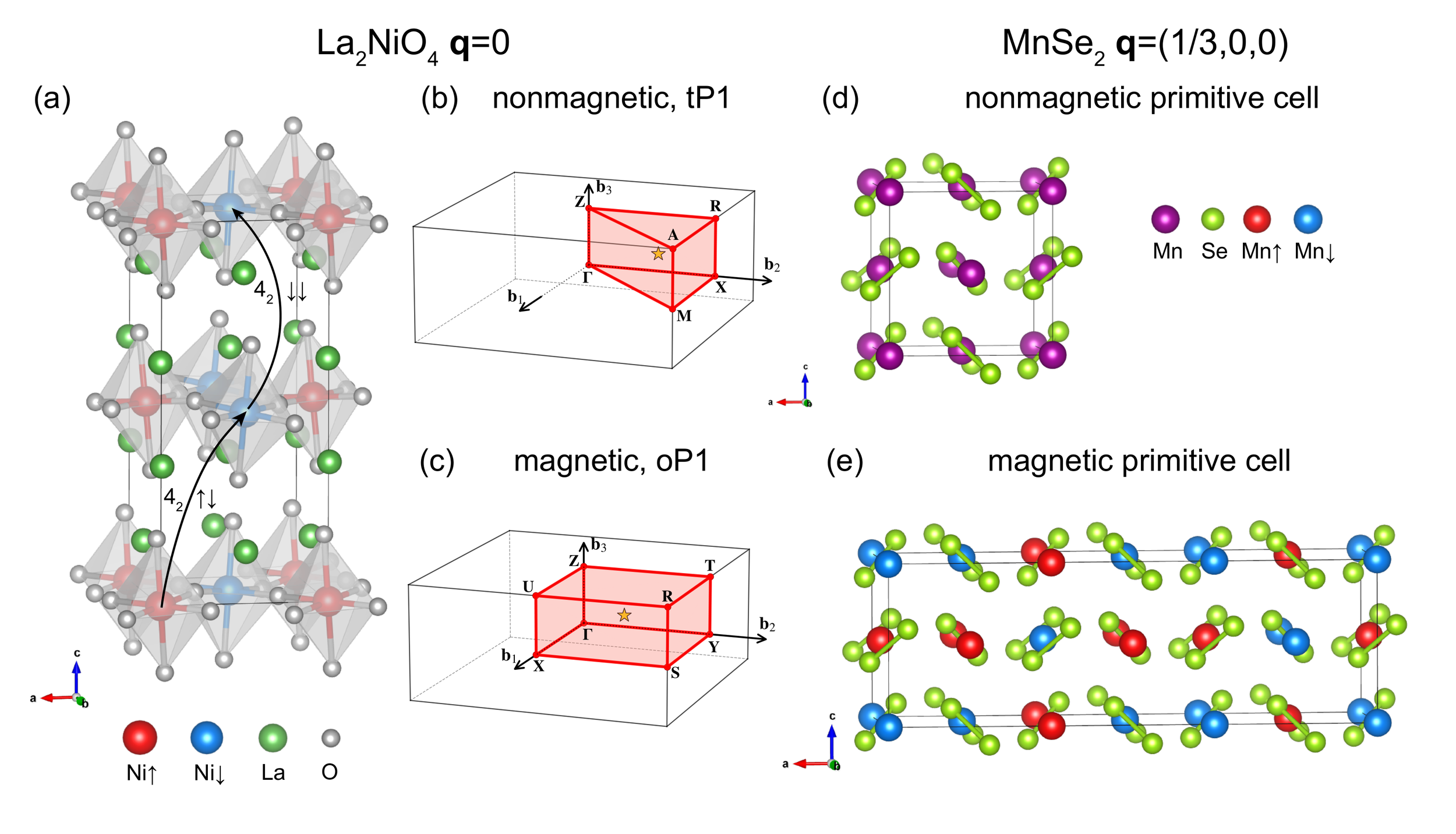}
    \caption{Magnetic-order-induced symmetry reduction at $\mathbf{q}=0$ and $\mathbf{q}\neq0$. (a)--(c) La$_2$NiO$_4$ (MAGNDATA 0.45) with $\mathbf{q}=0$. (a) Crystal structure with the collinear Ni moments. The nonmagnetic and magnetic primitive cells coincide. The magnetic order lowers the crystal symmetry from tetragonal $P4_2/ncm$ (No.~138), to orthorhombic $Pccn$ (No.~56). Black arrows: fourfold screw axis $4_2$ connecting both same-spin ($\uparrow \uparrow$) and opposite-spin ($\uparrow \downarrow$) Ni sites. Because each symmetry operation of a collinear magnet must either preserve or reverse the magnetic sublattices, the screw axis is not a symmetry of the magnetic state, and the fourfold axis is broken. Similarly, the diagonal mirror plane is also broken. (b)--(c) BZ and IBZ constructed using (b) the nonmagnetic symmetry, lattice type tP1, and (c) magnetic symmetry, lattice type oP1; the star marks the volume centroid of each IBZ (red shading). Because the two cells are the same, both panels share the same BZ, and only the IBZ differs. (d)--(e) MnSe$_2$ (MAGNDATA 1.0.47) with $\mathbf{q}=(1/3,0,0)$: (d) Cubic $Pa\bar{3}$ (No.~205) nonmagnetic primitive cell; (e) Orthorhombic $Pbca$ (No.~61) magnetic primitive cell, which contains three nonmagnetic primitive cells. The larger magnetic cell changes the BZ from cubic to orthorhombic, so both the BZ shape and the IBZ differ from the nonmagnetic case.}
    \label{fig:magnetic-symmetry-reduction}
\end{figure*}

Magnetic order can change the symmetry-adapted IBZ and band path in two distinct ways. The first occurs when a $\mathbf{q}=0$ order changes the Laue group while preserving the translational periodicity of the nonmagnetic parent; in this case, the BZ remains unchanged, but the IBZ and symmetry-adapted path can change. The second occurs when a nonzero propagation vector $\mathbf{q}$ changes the translational periodicity and enlarges the magnetic primitive cell, thereby changing the BZ and reciprocal basis as well. The effects of both mechanisms on the IBZ and band-path construction are illustrated in Fig.~\ref{fig:magnetic-symmetry-reduction}.

La$_2$NiO$_4$ (MAGNDATA 0.45) provides the $\mathbf{q}=0$ example. The lattice is unchanged while alternating bc planes of up and down Ni atoms lower the point symmetry from tetragonal to orthorhombic [Fig.~\ref{fig:magnetic-symmetry-reduction}(a)]. Its nonmagnetic structure has tetragonal space group $P4_2/ncm$ (No.~138,  tP1), Laue group $4/mmm$, whereas the MSG without SOC is orthorhombic $Pc'cn'$ (BNS 56.370, Type III) with magnetic Laue group $m'mm'$. The breaking of the fourfold screw axis symmetry $4_2$ by magnetic order is shown in detail in Fig.~\ref{fig:magnetic-symmetry-reduction}(a), where it can be seen that the screw axis neither interchanges the up and down spin magnetic sublattices nor takes each magnetic sublattice to itself. The diagonal mirror plane symmetries can similarly be seen to be broken. 

SeeK-path, given the La$_2$NiO$_4$ structure directly, would identify the nonmagnetic space group $P4_2/ncm$ and construct the tetragonal tP1 IBZ [Fig.~\ref{fig:magnetic-symmetry-reduction}(b)], which does not reflect the magnetic symmetry lowering. AlterSeeK-Path resolves this by first performing the SSG analysis to determine the space group $Pccn$ consistent with the magnetic order, then decorating the structure with marker atoms that lower the space group to $Pccn$ before passing it to SeeK-path. The result is the correct orthorhombic oP1 IBZ [Fig.~\ref{fig:magnetic-symmetry-reduction}(c)]. Because the lattice is unchanged, both panels share the same BZ; only the IBZ differs, with the orthorhombic wedge having twice the volume of the tetragonal one due to the lower symmetry.

MnSe$_2$, one of the supercell-altermagnet examples~\cite{jaeschke-ubiergo-prb24}, illustrates the complementary nonzero-$\mathbf{q}$ mechanism. Here, we use the magnetic structure reported as MAGNDATA 1.0.47 as the example, which has $\mathbf{q}=(1/3,0,0)$ relative to the cubic $Pa\bar{3}$ (No.~205) nonmagnetic primitive cell. Its magnetic primitive cell contains three copies of the nonmagnetic primitive cell [Fig.~\ref{fig:magnetic-symmetry-reduction}(d)--(e)] with MSG without SOC $Pb'c'a$ (BNS 61.436, Type III). Unlike the $\mathbf{q}=0$ case, the nonzero propagation vector changes the translational periodicity, so the BZ itself is different: the nonmagnetic cell has a cubic BZ, while the magnetic primitive cell has an orthorhombic BZ. SeeK-path, given only the nonmagnetic primitive cell, would construct the cubic $Pa\bar{3}$ path, which does not describe the magnetic state. When the magnetic primitive cell is submitted to AlterSeeK-Path, the SSG analysis determines the space group $Pbca$ consistent with the magnetic order, and the correct orthorhombic IBZ and band path are constructed. The alternative MnSe$_2$ magnetic structure, MAGNDATA 1.0.48, is also classified as an altermagnet and has the same propagation vector and threefold enlargement of the magnetic primitive cell, so the same path construction applies.

\begin{figure*}[t]
\centering
\includegraphics[width=0.9\textwidth,keepaspectratio]{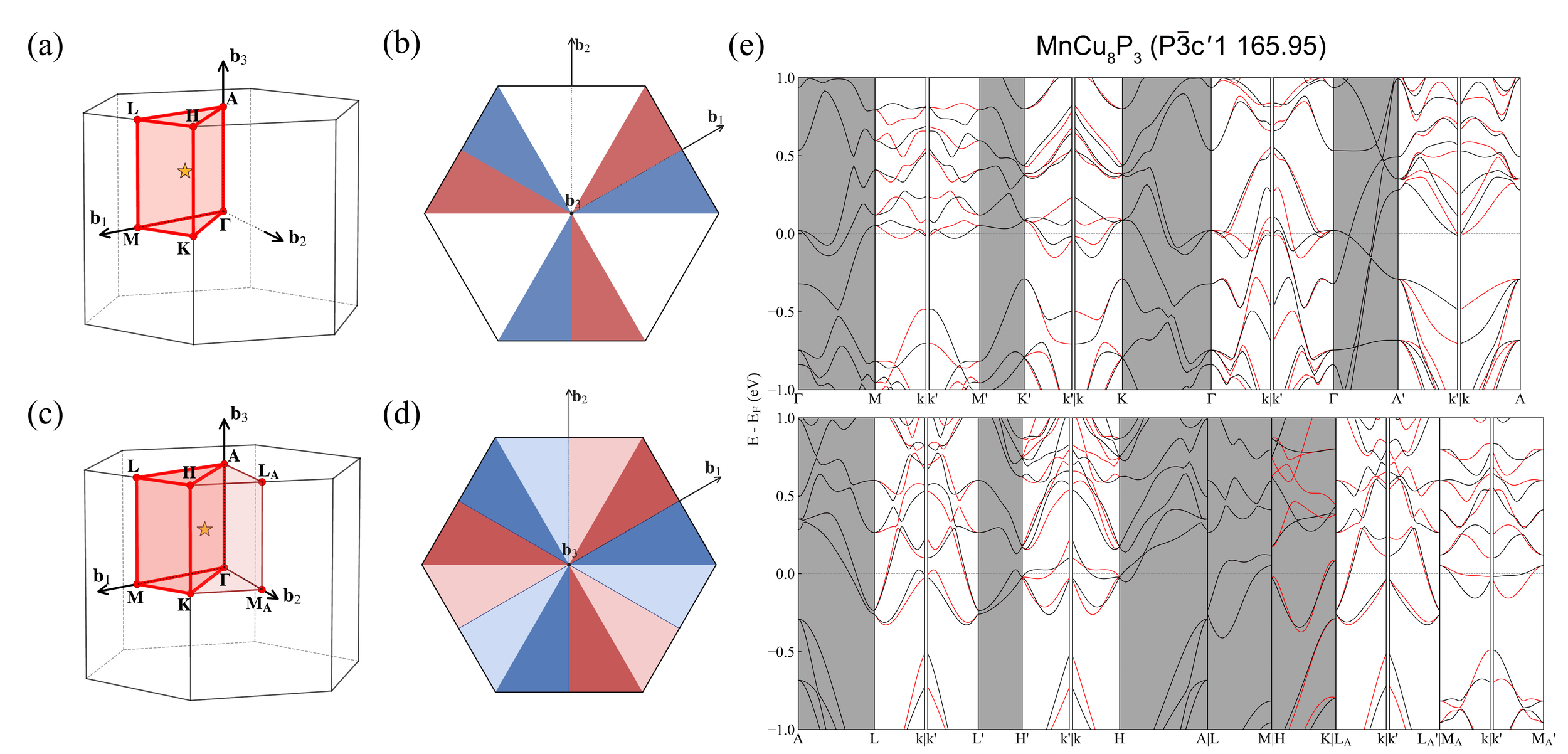}
\caption{hP2 BZ and band structure for the trigonal spin-Laue-group $^{1}\bar{3}^{2}m$ (bulk $g$-wave). (a) Conventional hexagonal-lattice IBZ. (b) Top-view spin pattern as in Fig.~\ref{fig:HEX}(c); applying the Laue-group operations to this IBZ leaves half the BZ uncovered (white). (c) True IBZ for this case: the shape from (a) plus an additional region (light red) not related to it by any Laue group operation. (d) Top-view spin pattern based on the true IBZ, filling the gaps left uncovered in (b); light and dark shades of blue show spin-flip images of the light and dark red regions, respectively. (e) Band structure along the generalized path generated by AlterSeeK-Path (grey: conventional path; white: inserted general-$k$ path).}
\label{fig:doubled-IBZ}
\end{figure*}

\subsection{Special cases with enlarged IBZ}
\label{sec:enlarged}

For most cases, the compact region bounded by the conventional high-symmetry vertices can be chosen as the IBZ.  The hexagonal crystal structures with Laue groups $6/m$ and $\bar{3}m$ and the tetragonal crystal structures with Laue group $4/m$ require special consideration. 
The IBZs in these cases have double the volume of the lattice IBZ, and can be constructed from the lattice IBZ in different ways depending on which additional operation of the lattice Laue group is used for the doubling. In each case, we aim to place the conventional high-symmetry points at corners of the enlarged IBZ rather than inside its edges, so that the lengths of the inserted general path do not change too much, and choose the doubling accordingly. The doublings available for each extended Bravais lattice type, and the one used, are given in \ref{sec:doubling-choices}.
The same geometric enlarging occurs for the cubic crystal structures with Laue group $m\bar{3}$ and trigonal crystal structures with Laue group $\bar{3}$; however, these groups do not support collinear altermagnetism and are therefore not included in the altermagnetic case library. 

Here we take the trigonal $\bar{3}m$ case as a representative example. We consider MnCu$_8$P$_3$, derived from the Cu$_3$P structure reported in the Materials Project as \texttt{mp-359}~\cite{materials-project1-nm25,materials-project2-apl13}. Guided by the Wyckoff-position construction discussed in Ref.~\cite{schiff-prresearch25}, the two Cu atoms on the $2b$ Wyckoff orbit were replaced by Mn, while the Cu atoms on the $12g$ and $4d$ orbits and the P atoms on the $6f$ orbit were retained. The resulting structure preserves space group $P\bar{3}c1$ (No.~165) with MSG without SOC $P\bar{3}c'1$ (BNS 165.95, Type III). The lattice type is hP2, with Laue group $\bar{3}m$. Although its BZ has the familiar hexagonal-prism shape, the trigonal Laue group $\bar{3}m$ contains only 12 operations, half the 24 operations of the hexagonal Laue group $6/mmm$. Its IBZ must therefore have twice the volume of the conventional hexagonal IBZ.

Fig.~\ref{fig:doubled-IBZ}(a)--(b) show what happens if we still use the usual IBZ for the hexagonal lattice with Laue group $6/mmm$. Applying all 12 operations of the trigonal Laue group to this volume covers only half of the BZ, leaving the other half uncovered. The failure of these symmetry-related copies to tile the complete BZ demonstrates that the volume surrounded by conventional high-symmetry lines of the hexagonal lattice is not the true IBZ in the trigonal system. Fig.~\ref{fig:doubled-IBZ}(c)--(d) show the correct, doubled construction: a second part, which is not related to the original part by any operation of the trigonal Laue group, is added to form the complete IBZ. Applying the same 12 operations to this doubled IBZ fills the entire BZ.

In Figs.~\ref{fig:doubled-IBZ}(c)--(d), the dark and light red colors distinguish the original and additional parts of the doubled IBZ. They do not represent different spin: dark and light red both denote spin-up regions, while dark and light blue both denote spin-down regions. If the original and additional parts were shown using the same color, their distinction would be obscured and the top-view projection in (d) could misleadingly resemble a planar $i$-wave pattern. Retaining this color distinction makes the three mirror planes of the trigonal system visible, consistent with its bulk $g$-wave classification~\cite{urru-prb25,schiff-prresearch25}.

The conventional high-symmetry path for this type of lattice is $\Gamma\text{--}M\text{--}K\text{--}\Gamma\text{--}A\text{--}L\text{--}H\text{--}A \,|\, L\text{--}M \,|\, H\text{--}K$. The doubled IBZ also changes the AlterSeeK-Path construction: the general point $\mathbf{k}$ is computed as the centroid of the complete true IBZ. The additional part of the IBZ contributes the vertices $M_A$ and $L_A$, the images of $M$ and $L$ under the mirror that completes the doubling; they lie in the same $k_z$ planes as $M$ and $L$, on the far side of the $\Gamma\text{--}K$ plane. Neither vertex is connected to the conventional path by a plotted edge, so each is appended as an independent excursion to the general point $\mathbf{k}$, following the isolated-vertex rule (Sec.~\ref{sec:butterfly}). The generated path is
\begin{equation}
\begin{split}
\Gamma
&\text{--}
\underbrace{M\text{--}k\,|\,k'\text{--}M'}_{}
\text{--}
\underbrace{K'\text{--}k'\,|\,k\text{--}K}_{}
\text{--}
\underbrace{\Gamma\text{--}k\,|\,k'\text{--}\Gamma}_{}
\\
&\text{--}
\underbrace{A'\text{--}k'\,|\,k\text{--}A}_{}
\text{--}
\underbrace{L\text{--}k\,|\,k'\text{--}L'}_{}
\text{--}
\underbrace{H'\text{--}k'\,|\,k\text{--}H}_{}
\text{--}A
\\
&\,|\,L\text{--}M\,|\,H\text{--}K\,|\,\underbrace{L_A\text{--}k\,|\,k'\text{--}L_A'}_{}
\,|\,\underbrace{M_A\text{--}k\,|\,k'\text{--}M_A'}_{}
.
\end{split}
\end{equation}

\begin{figure*}[t]
\centering
\includegraphics[width=0.9\textwidth,keepaspectratio]{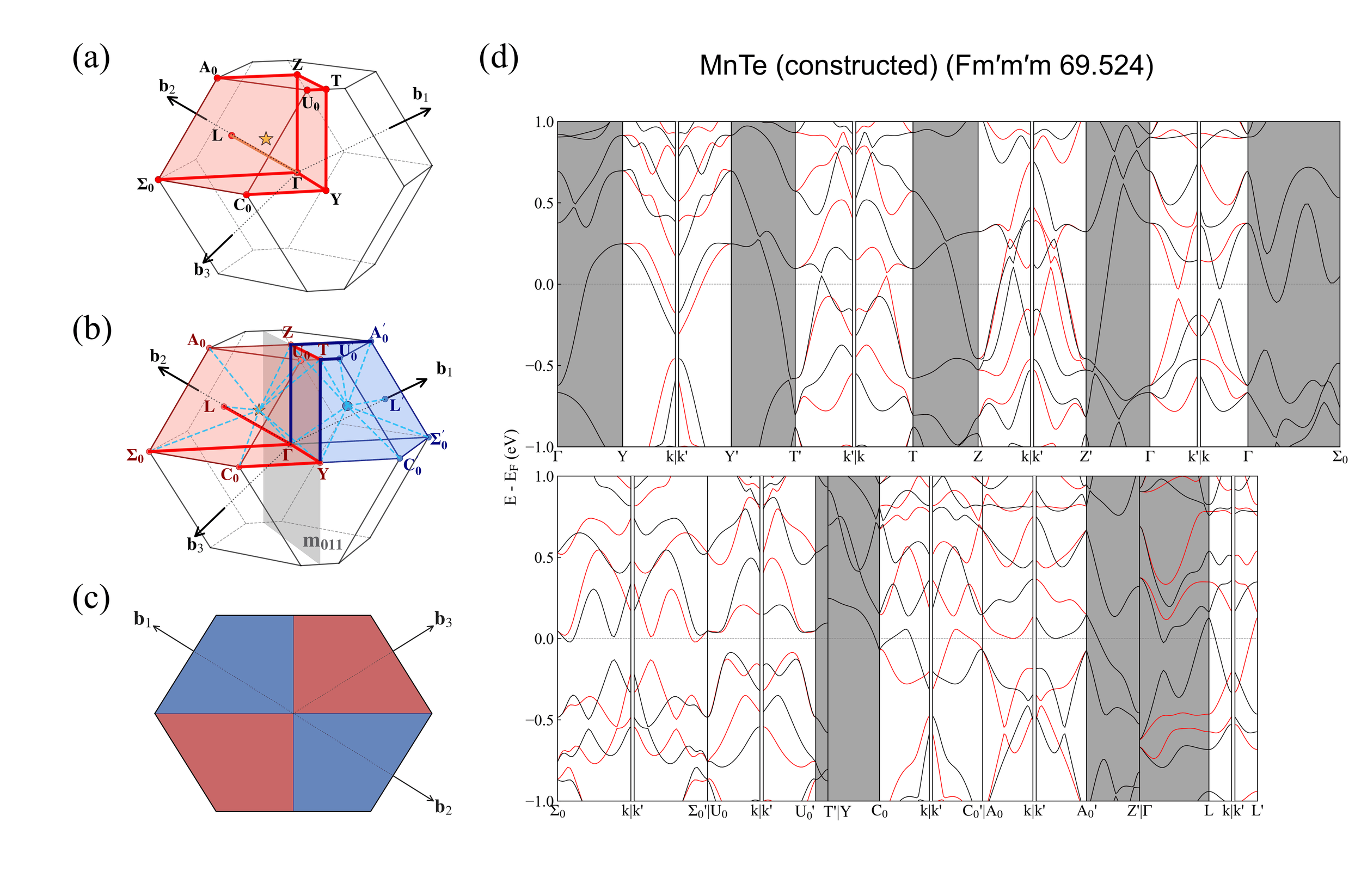}
\caption{oF1 BZ and band structure for the spin-Laue-group $^{2}m^{2}m^{1}m$ (planar $d$-wave). (a) IBZ, an irregular eight-vertex polyhedron whose centroid $\mathbf{k}$ is the true volume-weighted centroid rather than the vertex average. (b) sfIBZ, obtained from the IBZ by the spin-flip operation $m_{011}$. (c) Top view of the BZ, showing the resulting spin pattern. (d) Band structure along the generalized path generated by AlterSeeK-Path (grey: conventional path; white: inserted general-$k$ path), split into two stacked panels for clarity because the path is long.}
\label{fig:oF1-221}
\end{figure*}

\subsection{Irregular IBZ and long band path}
So far, the examples only show regular IBZ shapes, where the centroid coincides with the simple average of all vertices. We next consider space group $Fmmm$ (No.~69) with lattice type oF1 to illustrate an irregular IBZ. The structure is a synthetic Mn--Te compound constructed using the Wyckoff-position method of Ref.~\cite{schiff-prresearch25}, with Laue group $mmm$ and MSG without SOC $Fm'm'm$ (BNS 69.524, Type III). Note that this is distinct from the well-known hexagonal MnTe with space group $P6_3/mmc$ (No.~194); the present $Fmmm$ structure is a model compound built specifically to represent the oF1 altermagnetic case. As shown in Fig.~\ref{fig:oF1-221}(a), the oF1 IBZ has an irregular elongated shape with eight distinct vertices, and its centroid must be computed as the true volume-weighted centroid rather than the simple vertex average.

The conventional high-symmetry path from SeeK-path for oF1 is $\Gamma\text{--}Y\text{--}T\text{--}Z\text{--}\Gamma\text{--}\Sigma_0 \,|\, U_0\text{--}T \,|\, Y\text{--}C_0 \,|\, A_0\text{--}Z \,|\, \Gamma\text{--}L$. Since $T$, $Y$, $Z$, and $\Gamma$ are already sampled in the first chain when the later segments $U_0\text{--}T$, $Y\text{--}C_0$, $A_0\text{--}Z$, and $\Gamma\text{--}L$ are reached, the endpoint rule (Sec.~\ref{sec:butterfly}) is applied.
The generated path is
\begin{equation}
\begin{split}
\Gamma
&\text{--}
\underbrace{Y\text{--}k\,|\,k'\text{--}Y'}_{}
\text{--}
\underbrace{T'\text{--}k'\,|\,k\text{--}T}_{}
\text{--}
\underbrace{Z\text{--}k\,|\,k'\text{--}Z'}_{}
\\
&\text{--}
\underbrace{\Gamma\text{--}k'\,|\,k\text{--}\Gamma}_{}
\text{--}
\underbrace{\Sigma_0\text{--}k\,|\,k'\text{--}\Sigma_0'}_{}
\\
&\,|\,
\underbrace{U_0\text{--}k\,|\,k'\text{--}U_0'}_{}\text{--}T'
\,|\,
Y\text{--}
\underbrace{C_0\text{--}k\,|\,k'\text{--}C_0'}_{}
\\
&\,|\,
\underbrace{A_0\text{--}k\,|\,k'\text{--}A_0'}_{}\text{--}Z'
\,|\,
\Gamma\text{--}
\underbrace{L\text{--}k\,|\,k'\text{--}L'}_{}.
\end{split}
\end{equation}

\section{Discussion}
\label{sec:discussion}
The band structure constructed using AlterSeeK-Path provides systematic information about the SS in an altermagnetic crystal in a compact and easy-to-understand form. In addition, it can be obtained without requiring any computational tools beyond those included in standard first-principles calculations:
the band structure is already computed in essentially every electronic-structure study, and this altermagnetic k-path can be easily substituted for any conventional path.
Using the volume centroid optimizes the sampling of the interior of the BZ: it lies in the interior of the irreducible wedge and away from its boundary high-symmetry planes. SS is therefore symmetry-allowed at $\mathbf{k}$ and generically along the inserted segments connecting the high-symmetry endpoints to $\mathbf{k}$ and $\mathbf{k}'$.

Thus, AlterSeeK-Path should be found to be a new standard replacement for current spin-splitting display approaches. In particular, we recall that
SS is absent wherever a spin-flip operation leaves the k-point invariant, which is frequently the case along the conventional high-symmetry lines; this is why the conventional path can show no splitting at all, as in \BFO. Adding a low-symmetry line or two to this plot does not give a representative measure of the SS in the system; nor does a general point $\mathbf{k}$ chosen by hand for a plot from $\Gamma$ to $\mathbf{k}$.
Plotting of the Fermi surface is of interest for altermagnetic metals, but in general, plotting a constant energy surface at a single energy gives very limited information about the band structure.

The concept of inserting a general k-point into conventional high-symmetry segments was introduced in our earlier work on BiFeO$_3$~\cite{urru-prb25}, where the general k-point and the spin-flip operation were both selected manually based on the specific crystal symmetry of that system. AlterSeeK-Path automates and standardizes this procedure: the general k-point is computed as the IBZ volume centroid, the spin-flip operation is identified automatically by FINDSPINGROUP, and the reference high-symmetry path follows the SeeK-path convention. As a result, the band plot for BiFeO$_3$ in Ref.~\cite{urru-prb25} looks different in some respects from that obtained with AlterSeeK-Path (Fig.~\ref{fig:hR1-3m}) because we now follow a standardized procedure for generating the band path that is consistent across all altermagnetic system types. 

When FINDSPINGROUP identifies multiple valid spin-flip operations for a given magnetic structure, any of them may be used to construct the butterfly path. The choice does not affect the result: because spin-up and spin-down bands are related by the spin-flip symmetry, the equality $E_\uparrow(\mathbf{k}) = E_\downarrow(\mathbf{k}')$ holds for every valid spin-flip operation that maps $\mathbf{k}$ to $\mathbf{k}'$, since symmetry operations do not change band energies. Different operations yield different $\mathbf{k}'$ partners, but the bands are the same.

For orthorhombic $mmm$, three spin Laue groups support altermagnetism, corresponding to the irreps $\Gamma^+_2$, $\Gamma^+_3$, and $\Gamma^+_4$ of the $mmm$ Laue group: ${}^1m{}^2m{}^2m$, ${}^2m{}^1m{}^2m$, and ${}^2m{}^2m{}^1m$, respectively (Table~\ref{tab:alterseek-case-summary} and Table~I of Ref.~\cite{urru-prb25}). At the level of the symmetry-classified spin pattern, these three cases are interconverted by permuting the orthorhombic axes: each represents the same $d$-wave nodal pattern viewed with a different axis labeling. We retain all three entries because a given conventional cell fixes a particular labeling, but they do not constitute distinct pattern topologies. This equivalence under axis permutation is specific to orthorhombic $mmm$. For tetragonal $4/mmm$ and hexagonal $6/mmm$, the principal $c$ axis is crystallographically distinct from the in-plane axes, so the corresponding spin Laue groups cannot be interconverted by an equivalent change of viewing direction.

The altermagnets reported so far are concentrated in a few lattice types and spin Laue groups. High-throughput screenings of reported magnetic structures find candidates across all the crystal systems that permit altermagnetism, but most of them are simple orthorhombic and hexagonal, and the remainder are spread thinly over the other crystal systems~\cite{song-nat.rev.mater.,bai-adv.funct.mater.,guo-materialstodayphysics23,wan-prlett.25}. Many of the 54 combinations covered here therefore have no suitable candidate: either none has been reported, or the reported ones contain many more atoms per unit cell than are needed to demonstrate the path construction. We use a reported material from the Materials Project~\cite{materials-project1-nm25,materials-project2-apl13}, MAGNDATA~\cite{MAGNDATA1,MAGNDATA2}, or the Computational 2D Materials Database (C2DB)~\cite{c2db-haastrup-2dmater18,c2db-gjerding-2dmater21} wherever a suitable one is available, and otherwise construct a minimal structure following Ref.~\cite{schiff-prresearch25}, so that every case is demonstrated at a comparable unit cell size and number of occupied bands.

AlterSeeK-Path is designed for collinear altermagnets in the nonrelativistic limit, where spin is a good quantum number and each band carries a well-defined global spin-up or spin-down label. The main key concept of sampling the interior of the IBZ by identification of the volume centroid and plotting along the lines that connect the volume centroid to high-symmetry points on the boundary of the zone naturally extends to all crystalline materials. The conventional high-symmetry path is retained and supplemented by connections through the general point. For example, in a nonmagnetic 2D simple hexagonal crystal, the conventional path $\Gamma\text{--}M\text{--}K\text{--}\Gamma$ becomes $\Gamma\text{--}M\text{--}K\text{--}\Gamma\text{--}k\text{--}M\,|\,K\text{--}k$. The second key concept, of mapping to the sfIBZ to systematically display altermagnetic SS, can be tailored in the general case to the physics of interest.
In particular, altermagnetism is only one member of the broader family of unconventional magnets~\cite{luo-26a,luo-26}, which also includes odd-parity, hybrid-parity, and unconstrained-parity magnets with noncollinear or noncoplanar spins. In the odd-parity magnets, the spin-flip construction has a natural implementation with inversion as the spin-flip operation; extending the method to the rest of these cases is a direction for future work.

\section{Conclusion}
We presented AlterSeeK-Path, an automated tool for constructing generalized band paths that reveal altermagnetic SS. Given a magnetic crystal structure, AlterSeeK-Path identifies valid spin-flip operations via FINDSPINGROUP, constructs the conventional IBZ from the SeeK-path high-symmetry points, and computes its centroid $\mathbf{k}$ as the representative general k-point. It then inserts paired general-$k$ path segments, enabling altermagnetic splitting to be displayed without case-specific manual selection of low-symmetry paths.

We applied AlterSeeK-Path to all 54 distinct three-dimensional combinations of extended Bravais lattice type and spin Laue group that support collinear altermagnetism, covering the six compatible crystal systems. Representative band structures are shown for the 28 distinct lattice/path geometries, with spin-Laue variants that share the same path construction grouped together. We also applied AlterSeeK-Path to the 12 two-dimensional cases, spanning the four two-dimensional Bravais lattices that support collinear altermagnetism, with representative band structures shown for the six distinct path geometries. The current implementation is for collinear altermagnets in the nonrelativistic limit. Extending the approach to the broader family of unconventional magnets is a direction for future work.

\section*{CRediT authorship contribution statement}

\textbf{Yujia Teng}: Conceptualization, Data curation, Formal analysis, Investigation, Methodology, Software, Validation, Visualization, Writing -- original draft, Writing -- review \& editing.
\textbf{Mesfin Eshete}: Writing -- review \& editing.
\textbf{Andrea Urru}: Conceptualization, Methodology, Writing -- review \& editing. 
\textbf{Daniel Seleznev}: Conceptualization, Writing -- review \& editing. 
\textbf{Se Young Park}: Writing -- review \& editing.
\textbf{Sebastian E. Reyes-Lillo}: Writing -- review \& editing. 
\textbf{Karin M. Rabe}: Conceptualization, Funding acquisition, Project administration, Resources, Supervision, Writing -- review \& editing.

\section*{Declaration of competing interest}
The authors declare that they have no known competing financial interests or personal relationships that could have appeared to influence the work reported in this paper.

\section*{Data availability}

The AlterSeeK-Path source code is openly available at \url{https://github.com/yujia-teng/AlterSeeK-Path}.
The data supporting this work are available~\cite{teng-alterseek-data}.

\section*{Acknowledgments}

Y.T. acknowledges support from the Wisconsin Materials Research Science and Engineering Center (NSF DMR-2309000). M.E., A.U., D.S., and K.M.R. acknowledge support from the Office of Naval Research Grant No. N00014-21-1-2107. M.E. was also supported by the Gordon and Betty Moore Foundation. A.U. was also supported by the Italian Fund for Science (FIS), grant No. FIS-2024-04793. D.S. was also supported by the W. M. Keck Foundation under grant 996588. S.Y.P. was supported by the National Research Foundation of Korea (NRF) grant funded by the Korea government (MSIT) (RS-2024-00358551). S.E.R.-L. acknowledges ANID Fondecyt regular grant number 1260824. Part of the work by Y.T. and K.M.R. was done at the Aspen Center for Physics, which is supported by the National Science Foundation grant PHY-2210452. Computational facilities were provided by the Beowulf cluster at the Department of Physics and Astronomy of Rutgers University.

\bibliographystyle{elsarticle-num}
\bibliography{cite}

\clearpage
\onecolumn
\appendix
\pretocmd{\subsection}{\Needspace{0.75\textheight}}{}{}
\newcounter{appendixcaseblock}
\pretocmd{\subsubsection}{%
  \stepcounter{appendixcaseblock}%
  \ifnum\value{appendixcaseblock}>2\relax
    \Needspace{0.72\textheight}%
  \fi
}{}{}
\counterwithin*{figure}{section}
\counterwithin*{table}{section}
\counterwithin*{equation}{section}

\section{Computational details}
All Density Functional Theory (DFT) calculations were performed using VASP~\cite{vasp1,vasp2} with the projector augmented wave (PAW) method~\cite{paw} and the PBE exchange-correlation functional~\cite{gga}. A Hubbard $U$ correction in the Dudarev formulation~\cite{dft+u} was applied, with $U = 3$~eV for $d$-electron magnetic atoms and $U = 6$~eV for $f$-electron magnetic atoms. The plane-wave energy cutoff was set to 500~eV and the self-consistent field convergence criterion was $10^{-6}$~eV. The k-point mesh was generated using VASPKIT~\cite{VASPKIT} with a reciprocal-space resolution of 0.03~\AA$^{-1}$. All calculations were performed without SOC. For two-dimensional systems, slab calculations were performed with a vacuum region of at least 15~\AA.

\section{Choice of the IBZ in certain hexagonal and tetragonal crystals}
\label{sec:doubling-choices}
\begin{figure}[H]
\centering
\includegraphics[width=\textwidth,keepaspectratio]{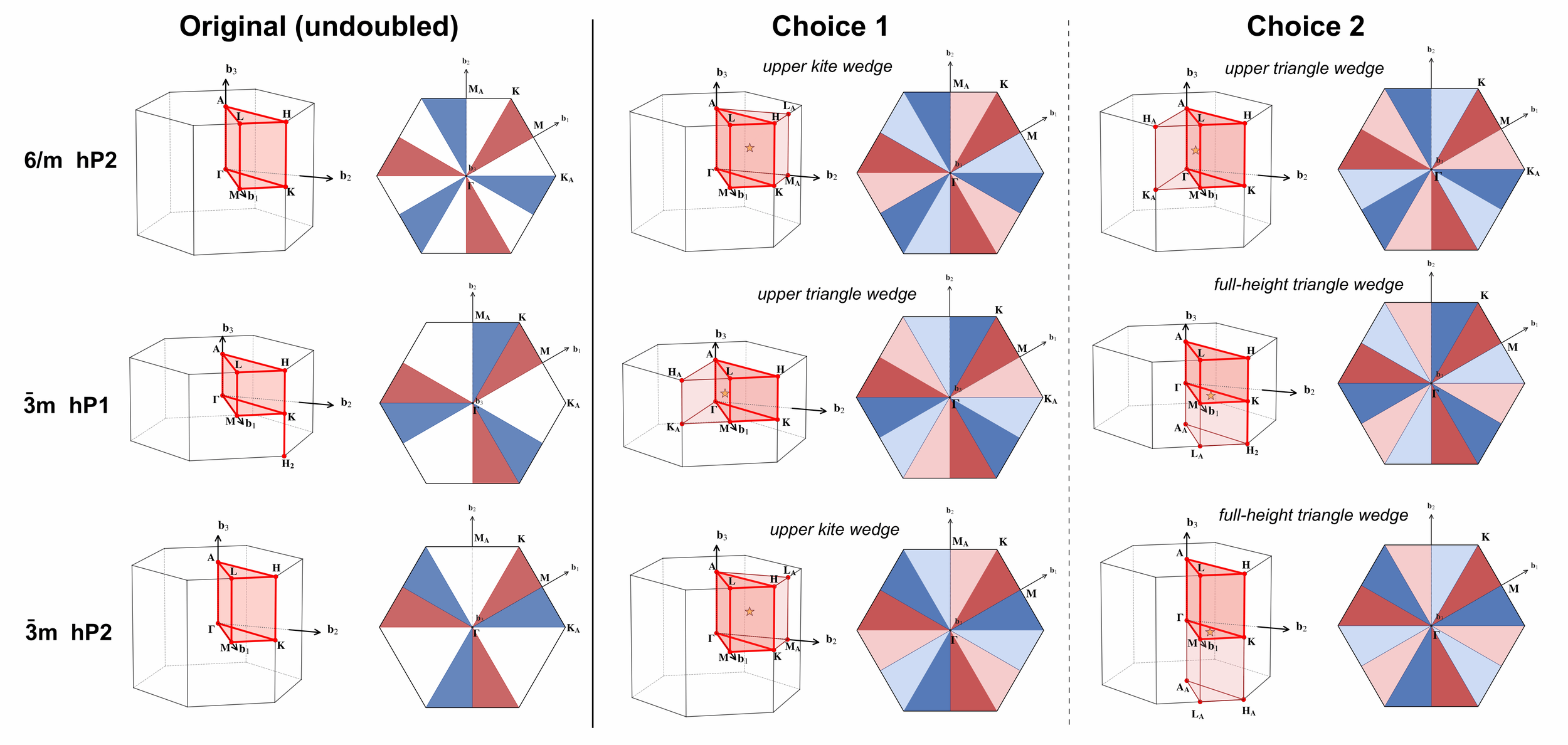}
\caption{Two ways of choosing the IBZ. Each row is one Laue group and extended Bravais lattice type; each entry shows the BZ view, with the star marking the IBZ centroid, and the top-view spin pattern as in Fig.~\ref{fig:HEX}(c). Left: the lattice IBZ, whose symmetry images leave half of the BZ uncovered (white). Choice 1 and choice 2: the two completions, with the added half drawn in a lighter shade; both tile the BZ completely.}
\label{fig:doubling-choices-hex}
\end{figure}

\begin{figure}[H]
\centering
\includegraphics[width=\textwidth,keepaspectratio]{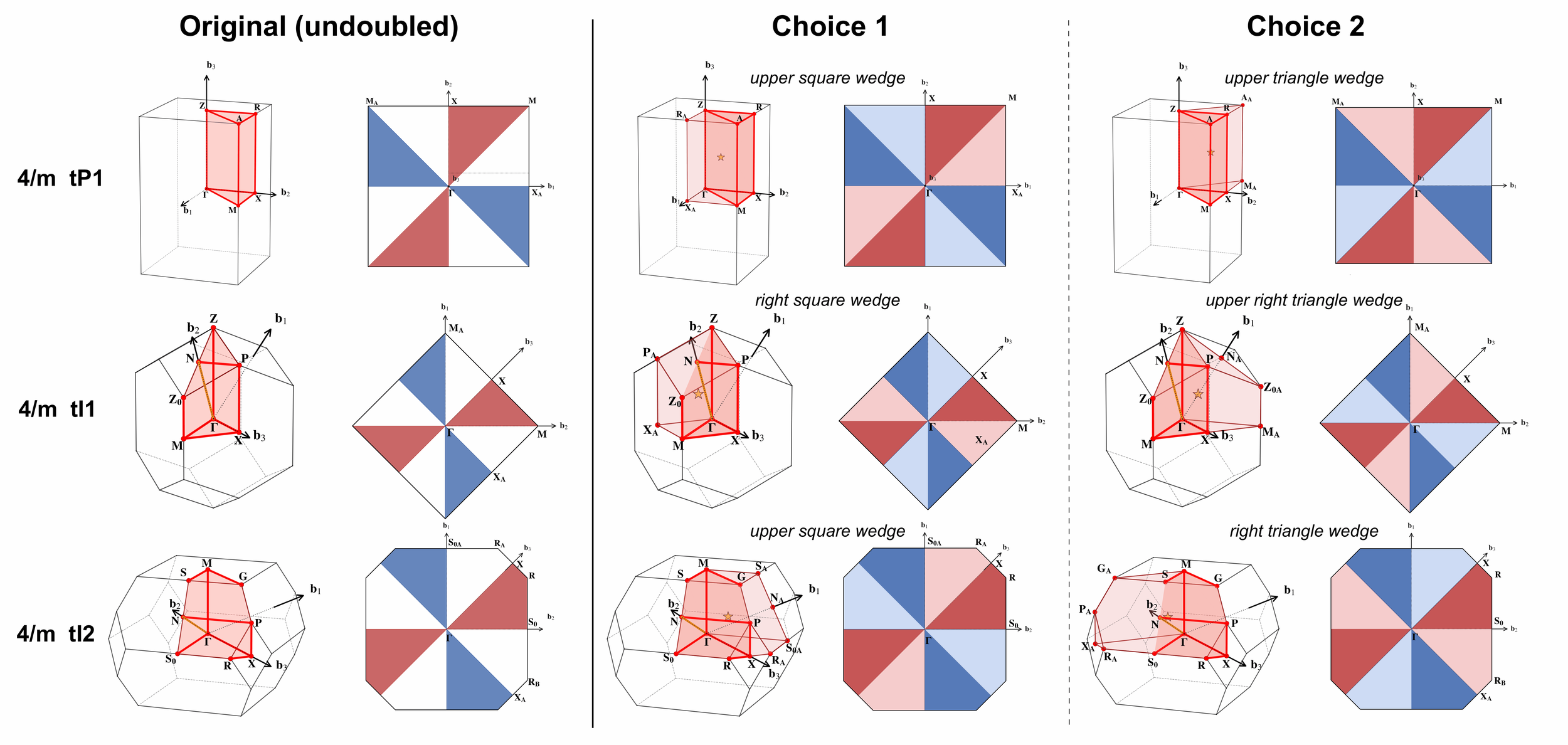}
\caption{As Fig.~\ref{fig:doubling-choices-hex}, for the tetragonal cases with Laue group $4/m$.}
\label{fig:doubling-choices-tet}
\end{figure}

In Figs.~\ref{fig:doubling-choices-hex} and~\ref{fig:doubling-choices-tet} we show the construction of the IBZ  for the hexagonal and tetragonal crystals for which the Laue group is a proper subgroup of the Laue group of the lattice, so that the IBZ has twice the volume of the lattice IBZ.
For Laue group $6/m$, doubling of the lattice IBZ gives upper kite wedge (choice 1) or upper triangle wedge (choice 2). 
For Laue group $\bar{3}m$ hP1, we get upper triangle wedge (choice 1) or full-height triangle wedge (choice 2).
For $\bar{3}m$ hP2, we get upper kite wedge (choice 1) or full-height triangle wedge (choice 2).
For Laue group $4/m$ tP1, tI1 and tI2, we get upper square wedge (choice 1) or upper triangle wedge (choice 2). 
In each case, we prefer choice 1 over choice 2 because in choice 1 the lengths of the generalized path segments are more similar.

\section{Complete case library}
\label{sec:case-library}
\ref{sec:case-library-cubic}--\ref{sec:case-library-monoclinic} cover all 54 distinct combinations of extended Bravais lattice type and spin Laue group that support altermagnetism in three dimensions, organized by crystal system, and \ref{sec:case-library-2d} covers the 12 two-dimensional cases, organized by two-dimensional Bravais lattice. When different spin Laue groups share the same k-path geometry for a given extended Bravais lattice type, they are grouped together and one representative band structure is shown. In the two-dimensional centered-rectangular lattice, the $a<b$ and $a>b$ metric branches are presented separately, because they have different BZ shapes and different high-symmetry point labels. The complete case lists are summarized in Table~\ref{tab:alterseek-case-summary} for three-dimensional crystals and Table~\ref{tab:spin-layer-2d} for two-dimensional crystals.

As discussed in Sec.~\ref{sec:discussion}, we use a reported material from the Materials Project~\cite{materials-project1-nm25,materials-project2-apl13}, MAGNDATA~\cite{MAGNDATA1,MAGNDATA2}, or the Computational 2D Materials Database (C2DB)~\cite{c2db-haastrup-2dmater18,c2db-gjerding-2dmater21} wherever a suitable one is available, and otherwise construct a minimal structure following Ref.~\cite{schiff-prresearch25}, so that every case is demonstrated at a comparable size. In a few cases, a reported structure serves as the starting point but is reduced or has one species substituted following the same Wyckoff-position construction, e.g. Co$_4$O$_4$P$_4$ (based on MAGNDATA 0.262, CoFePO$_5$) and MnCu$_8$P$_3$ (based on Materials Project mp-359, Cu$_3$P). The band structure plots follow the conventions of Fig.~\ref{fig:HEX}(d).

\renewcommand{\arraystretch}{1.3}
\sffamily\small
\begin{longtable}{@{}l@{}}
\caption{AlterSeeK-Path cases organized by crystal system, Laue group, extended Bravais lattice type, irrep, spin Laue group, and altermagnetic spin pattern. The final column gives the case number; each case is listed in the appendix below.}
\label{tab:alterseek-case-summary}\\
\toprule
\begin{tabular}[c]{@{}l@{\hspace{8pt}}l@{\hspace{8pt}}l@{\hspace{8pt}}l@{\hspace{8pt}}l@{\hspace{8pt}}l@{\hspace{28pt}}l@{}}
\makebox[0.120\textwidth][l]{\makecell[l]{Crystal\\system}} & \makebox[0.060\textwidth][l]{\makecell[l]{Laue\\group}} & \makebox[0.055\textwidth][l]{\makecell[l]{Lattice\\type}} & \makebox[0.045\textwidth][c]{Irrep} & \makebox[0.115\textwidth][l]{\makecell[l]{Spin Laue\\group}} & \makebox[\dimexpr 0.225\textwidth+8pt\relax][c]{Altermagnetic spin pattern} & \makebox[0.090\textwidth][l]{\makecell[l]{Case\\number}} \\
\end{tabular} \\
\midrule
\endfirsthead
\toprule
\begin{tabular}[c]{@{}l@{\hspace{8pt}}l@{\hspace{8pt}}l@{\hspace{8pt}}l@{\hspace{8pt}}l@{\hspace{8pt}}l@{\hspace{28pt}}l@{}}
\makebox[0.120\textwidth][l]{\makecell[l]{Crystal\\system}} & \makebox[0.060\textwidth][l]{\makecell[l]{Laue\\group}} & \makebox[0.055\textwidth][l]{\makecell[l]{Lattice\\type}} & \makebox[0.045\textwidth][c]{Irrep} & \makebox[0.115\textwidth][l]{\makecell[l]{Spin Laue\\group}} & \makebox[\dimexpr 0.225\textwidth+8pt\relax][c]{Altermagnetic spin pattern} & \makebox[0.090\textwidth][l]{\makecell[l]{Case\\number}} \\
\end{tabular} \\
\midrule
\endhead
\midrule
\multicolumn{1}{r@{}}{Continued on next page} \\
\endfoot
\bottomrule
\endlastfoot
\begin{tabular}[c]{@{}l@{\hspace{8pt}}l@{}}
\makebox[0.120\textwidth][l]{Cubic} &
\begin{tabular}[c]{@{}l@{\hspace{8pt}}l@{}}
\makebox[0.060\textwidth][l]{$m\bar{3}m$} &
\begin{tabular}[c]{@{}l@{\hspace{8pt}}l@{\hspace{8pt}}l@{\hspace{8pt}}l@{\hspace{8pt}}l@{\hspace{28pt}}l@{}}
\makebox[0.055\textwidth][l]{{\renewcommand{\arraystretch}{1.05}\begin{tabular}[c]{@{}l@{}}\hpkot{cP2}\\ \hpkot{cF2}\\ \hpkot{cI1}\end{tabular}}} & \makebox[0.045\textwidth][c]{$\Gamma^{+}_{2}$} & \makebox[0.115\textwidth][l]{${}^{1}m{}^{1}\bar{3}{}^{2}m$} & \makebox[0.130\textwidth][l]{bulk $i$-wave} & \makebox[0.095\textwidth][c]{\raisebox{-0.5\height}{\includegraphics[width=2.0cm,height=2.0cm,keepaspectratio]{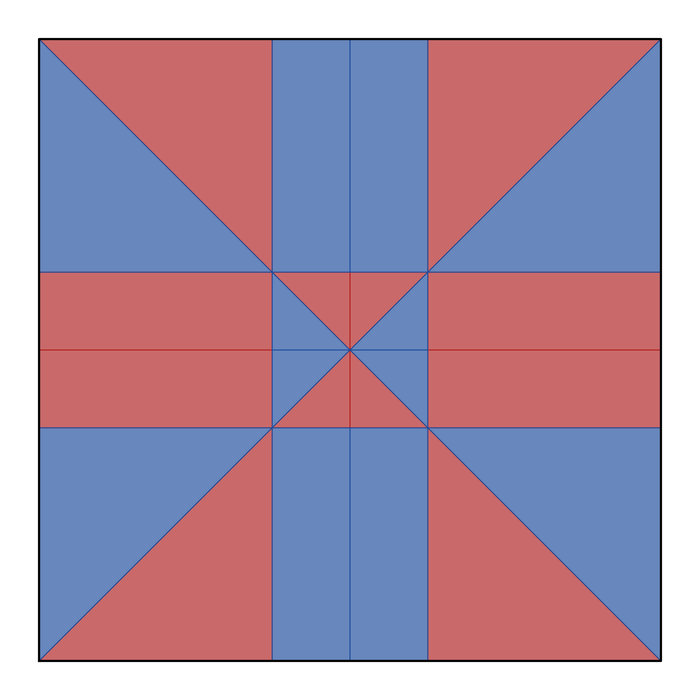}}} & \makebox[0.090\textwidth][l]{{\renewcommand{\arraystretch}{1.05}\begin{tabular}[c]{@{}l@{}}\hyperref[fig:cP2]{Case 01}\\ \hyperref[fig:cF2]{Case 02}\\ \hyperref[fig:cI1]{Case 03}\end{tabular}}} \\
\end{tabular} \\
\end{tabular} \\
\end{tabular} \\
\midrule
\begin{tabular}[c]{@{}l@{\hspace{8pt}}l@{}}
\makebox[0.120\textwidth][l]{Hexagonal} &
\begin{tabular}[c]{@{}l@{\hspace{8pt}}l@{}}
\makebox[0.060\textwidth][l]{$6/mmm$} &
\begin{tabular}[c]{@{}l@{\hspace{8pt}}l@{\hspace{8pt}}l@{\hspace{8pt}}l@{\hspace{8pt}}l@{\hspace{28pt}}l@{}}
\makebox[0.055\textwidth][l]{\hpkot{hP2}} & \makebox[0.045\textwidth][c]{$\Gamma^{+}_{3}$} & \makebox[0.115\textwidth][l]{${}^{2}6/{}^{2}m{}^{2}m{}^{1}m$} & \makebox[0.130\textwidth][l]{bulk $g$-wave} & \makebox[0.095\textwidth][c]{\raisebox{-0.5\height}{\includegraphics[width=2.0cm,height=2.0cm,keepaspectratio]{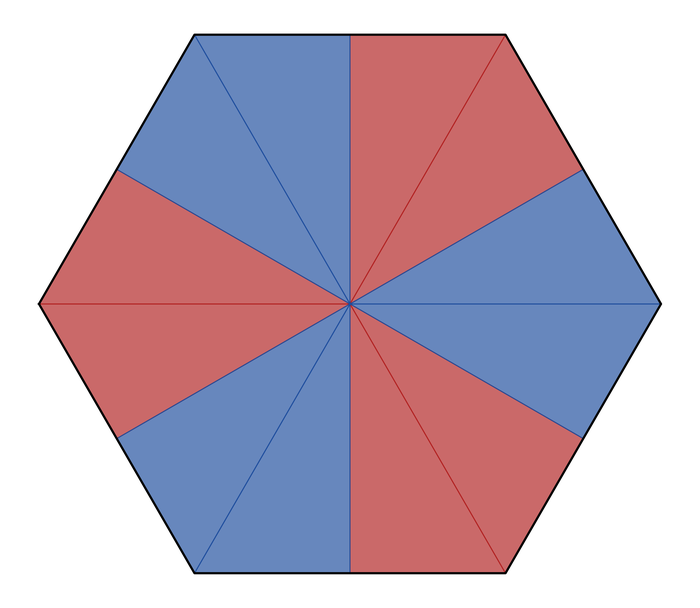}}} & \makebox[0.090\textwidth][l]{\hyperref[fig:hP2-6/mmm]{Case 04}} \\
\cmidrule{1-6}
\makebox[0.055\textwidth][l]{\hpkot{hP2}} & \makebox[0.045\textwidth][c]{$\Gamma^{+}_{4}$} & \makebox[0.115\textwidth][l]{${}^{2}6/{}^{2}m{}^{1}m{}^{2}m$} & \makebox[0.130\textwidth][l]{bulk $g$-wave} & \makebox[0.095\textwidth][c]{\raisebox{-0.5\height}{\includegraphics[width=2.0cm,height=2.0cm,keepaspectratio]{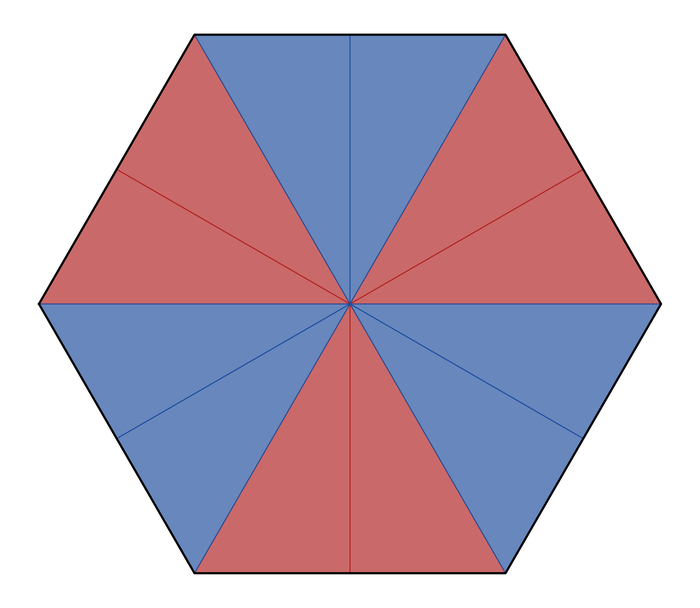}}} & \makebox[0.090\textwidth][l]{\hyperref[fig:hP2-6/mmm]{Case 05}} \\
\cmidrule{1-6}
\makebox[0.055\textwidth][l]{\hpkot{hP2}} & \makebox[0.045\textwidth][c]{$\Gamma^{+}_{2}$} & \makebox[0.115\textwidth][l]{${}^{1}6/{}^{1}m{}^{2}m{}^{2}m$} & \makebox[0.130\textwidth][l]{planar $i$-wave} & \makebox[0.095\textwidth][c]{\raisebox{-0.5\height}{\includegraphics[width=2.0cm,height=2.0cm,keepaspectratio]{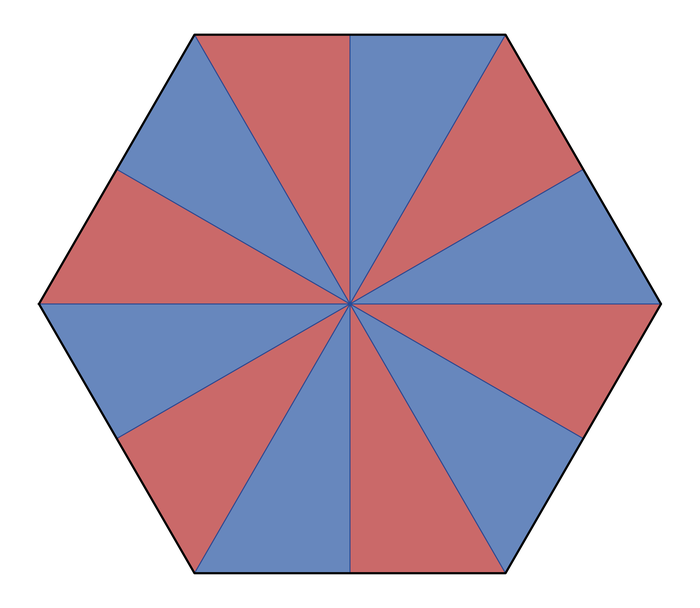}}} & \makebox[0.090\textwidth][l]{\hyperref[fig:hP2-6/mmm]{Case 06}} \\
\end{tabular} \\
\cmidrule{2-2}
\makebox[0.060\textwidth][l]{$6/m$} &
\begin{tabular}[c]{@{}l@{\hspace{8pt}}l@{\hspace{8pt}}l@{\hspace{8pt}}l@{\hspace{8pt}}l@{\hspace{28pt}}l@{}}
\makebox[0.055\textwidth][l]{\hpkot{hP2}} & \makebox[0.045\textwidth][c]{$\Gamma^{+}_{2}$} & \makebox[0.115\textwidth][l]{${}^{2}6/{}^{2}m$} & \makebox[0.130\textwidth][l]{bulk $g$-wave} & \makebox[0.095\textwidth][c]{\raisebox{-0.5\height}{\includegraphics[width=2.0cm,height=2.0cm,keepaspectratio]{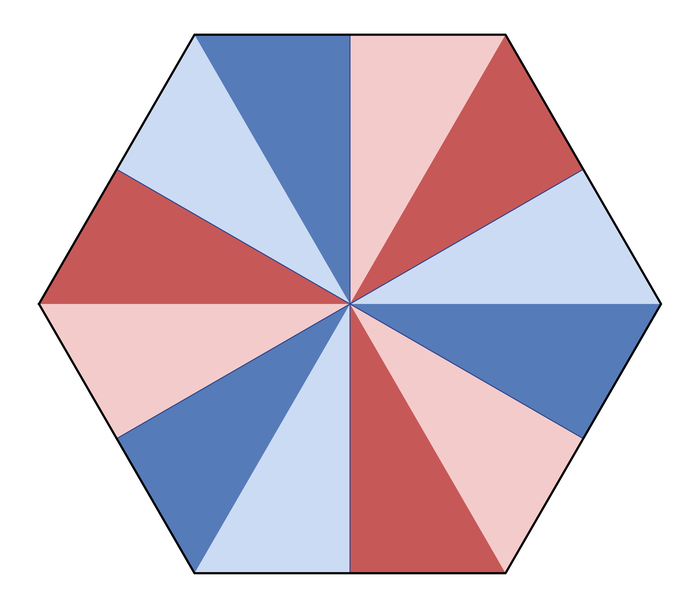}}} & \makebox[0.090\textwidth][l]{\hyperref[fig:hP2-6/m]{Case 07}} \\
\end{tabular} \\
\end{tabular} \\
\end{tabular} \\
\midrule
\begin{tabular}[c]{@{}l@{\hspace{8pt}}l@{}}
\makebox[0.120\textwidth][l]{Trigonal} &
\begin{tabular}[c]{@{}l@{\hspace{8pt}}l@{}}
\makebox[0.060\textwidth][l]{$\bar{3}m$} &
\begin{tabular}[c]{@{}l@{\hspace{8pt}}l@{\hspace{8pt}}l@{\hspace{8pt}}l@{\hspace{8pt}}l@{\hspace{28pt}}l@{}}
\makebox[0.055\textwidth][l]{\hpkot{hP1}} & \makebox[0.045\textwidth][c]{$\Gamma^{+}_{2}$} & \makebox[0.115\textwidth][l]{${}^{1}\bar{3}{}^{2}m$} & \makebox[0.130\textwidth][l]{bulk $g$-wave} & \makebox[0.095\textwidth][c]{\raisebox{-0.5\height}{\includegraphics[width=2.0cm,height=2.0cm,keepaspectratio]{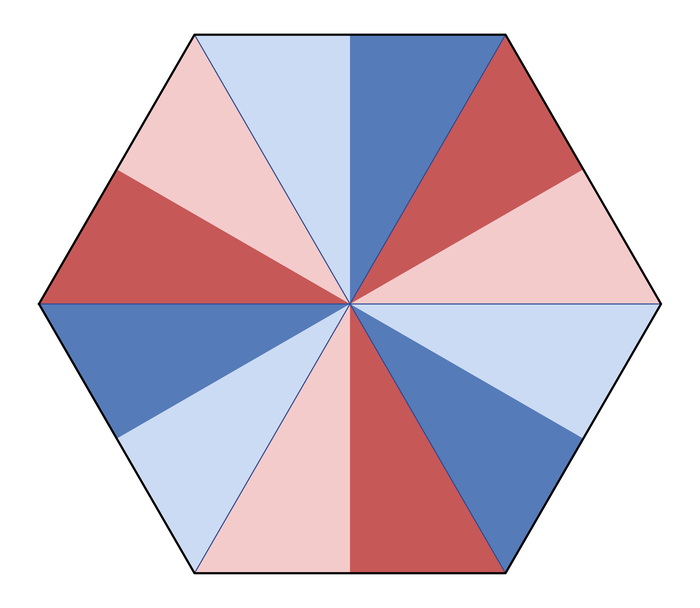}}} & \makebox[0.090\textwidth][l]{\hyperref[fig:hP1-3m]{Case 08}} \\
\cmidrule{1-6}
\makebox[0.055\textwidth][l]{\hpkot{hP2}} & \makebox[0.045\textwidth][c]{$\Gamma^{+}_{2}$} & \makebox[0.115\textwidth][l]{${}^{1}\bar{3}{}^{2}m$} & \makebox[0.130\textwidth][l]{bulk $g$-wave} & \makebox[0.095\textwidth][c]{\raisebox{-0.5\height}{\includegraphics[width=2.0cm,height=2.0cm,keepaspectratio]{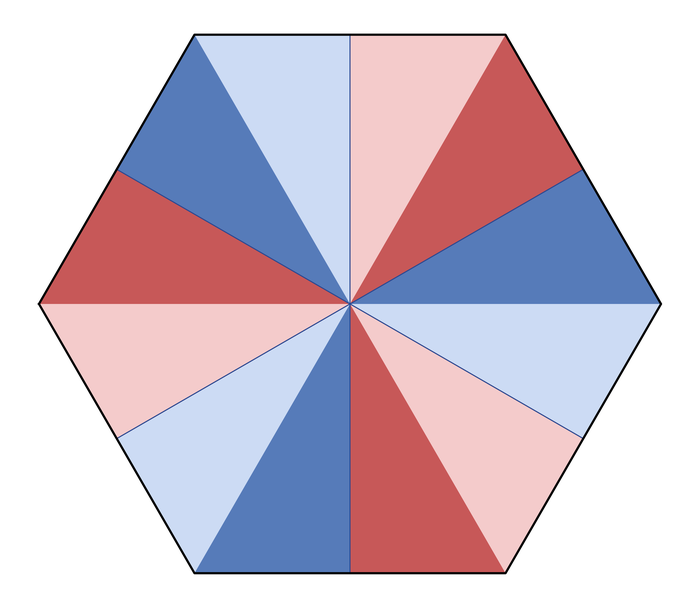}}} & \makebox[0.090\textwidth][l]{\hyperref[fig:hP2-3m]{Case 09}} \\
\cmidrule{1-6}
\makebox[0.055\textwidth][l]{{\renewcommand{\arraystretch}{1.05}\begin{tabular}[c]{@{}l@{}}\hpkot{hR1}\\ \hpkot{hR2}\end{tabular}}} & \makebox[0.045\textwidth][c]{$\Gamma^{+}_{2}$} & \makebox[0.115\textwidth][l]{${}^{1}\bar{3}{}^{2}m$} & \makebox[0.130\textwidth][l]{bulk $g$-wave} & \makebox[0.095\textwidth][c]{\raisebox{-0.5\height}{\includegraphics[width=2.0cm,height=2.0cm,keepaspectratio]{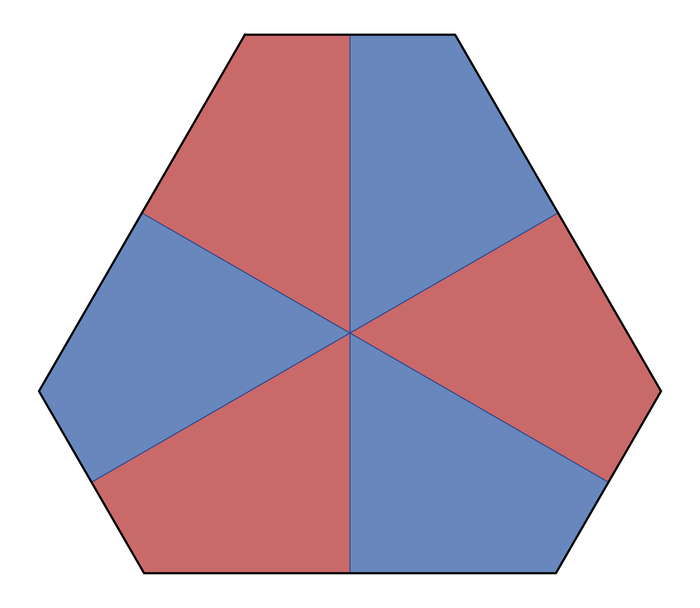}}} & \makebox[0.090\textwidth][l]{{\renewcommand{\arraystretch}{1.05}\begin{tabular}[c]{@{}l@{}}\hyperref[fig:hR1-3m]{Case 10}\\ \hyperref[fig:hR2-3m]{Case 11}\end{tabular}}} \\
\end{tabular} \\
\end{tabular} \\
\end{tabular} \\
\midrule
\begin{tabular}[c]{@{}l@{\hspace{8pt}}l@{}}
\makebox[0.120\textwidth][l]{Tetragonal} &
\begin{tabular}[c]{@{}l@{\hspace{8pt}}l@{}}
\makebox[0.060\textwidth][l]{$4/mmm$} &
\begin{tabular}[c]{@{}l@{\hspace{8pt}}l@{\hspace{8pt}}l@{\hspace{8pt}}l@{\hspace{8pt}}l@{\hspace{28pt}}l@{}}
\makebox[0.055\textwidth][l]{{\renewcommand{\arraystretch}{1.05}\begin{tabular}[c]{@{}l@{}}\hpkot{tP1}\\ \hpkot{tI1}\\ \hpkot{tI2}\end{tabular}}} & \makebox[0.045\textwidth][c]{$\Gamma^{+}_{4}$} & \makebox[0.115\textwidth][l]{${}^{2}4/{}^{1}m{}^{2}m{}^{1}m$} & \makebox[0.130\textwidth][l]{planar $d$-wave} & \makebox[0.095\textwidth][c]{\raisebox{-0.5\height}{\includegraphics[width=2.0cm,height=2.0cm,keepaspectratio]{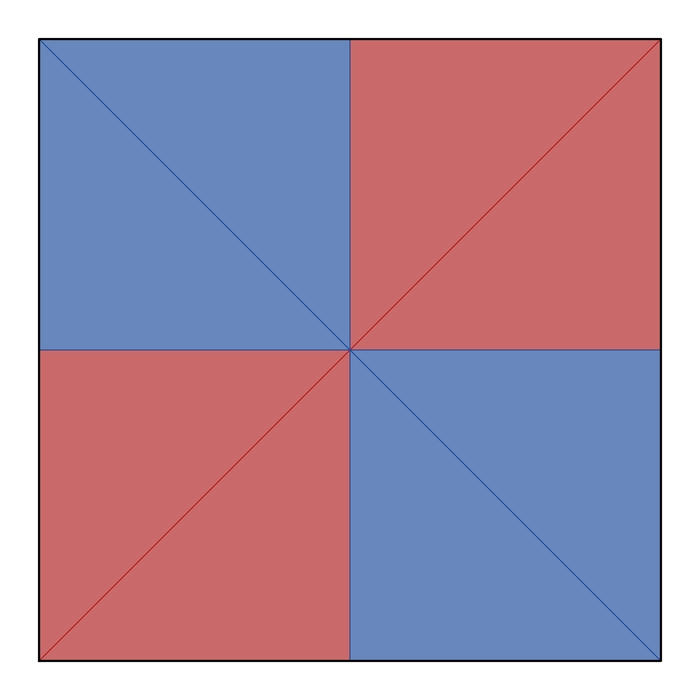}}} & \makebox[0.090\textwidth][l]{{\renewcommand{\arraystretch}{1.05}\begin{tabular}[c]{@{}l@{}}\hyperref[fig:tP1-4/mmm]{Case 12}\\ \hyperref[fig:tI1-4/mmm]{Case 15}\\ \hyperref[fig:tI2-4/mmm]{Case 18}\end{tabular}}} \\
\cmidrule{1-6}
\makebox[0.055\textwidth][l]{{\renewcommand{\arraystretch}{1.05}\begin{tabular}[c]{@{}l@{}}\hpkot{tP1}\\ \hpkot{tI1}\\ \hpkot{tI2}\end{tabular}}} & \makebox[0.045\textwidth][c]{$\Gamma^{+}_{3}$} & \makebox[0.115\textwidth][l]{${}^{2}4/{}^{1}m{}^{1}m{}^{2}m$} & \makebox[0.130\textwidth][l]{planar $d$-wave} & \makebox[0.095\textwidth][c]{\raisebox{-0.5\height}{\includegraphics[width=2.0cm,height=2.0cm,keepaspectratio]{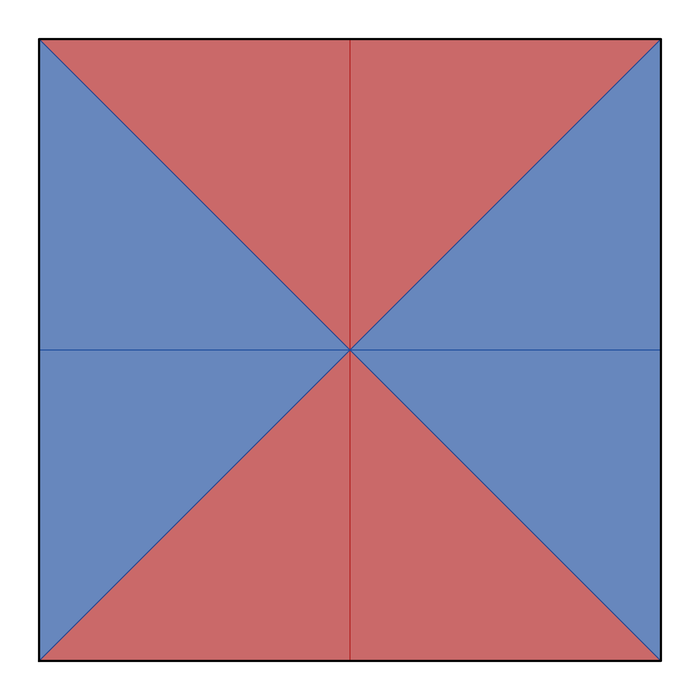}}} & \makebox[0.090\textwidth][l]{{\renewcommand{\arraystretch}{1.05}\begin{tabular}[c]{@{}l@{}}\hyperref[fig:tP1-4/mmm]{Case 13}\\ \hyperref[fig:tI1-4/mmm]{Case 16}\\ \hyperref[fig:tI2-4/mmm]{Case 19}\end{tabular}}} \\
\cmidrule{1-6}
\makebox[0.055\textwidth][l]{{\renewcommand{\arraystretch}{1.05}\begin{tabular}[c]{@{}l@{}}\hpkot{tP1}\\ \hpkot{tI1}\\ \hpkot{tI2}\end{tabular}}} & \makebox[0.045\textwidth][c]{$\Gamma^{+}_{2}$} & \makebox[0.115\textwidth][l]{${}^{1}4/{}^{1}m{}^{2}m{}^{2}m$} & \makebox[0.130\textwidth][l]{planar $g$-wave} & \makebox[0.095\textwidth][c]{\raisebox{-0.5\height}{\includegraphics[width=2.0cm,height=2.0cm,keepaspectratio]{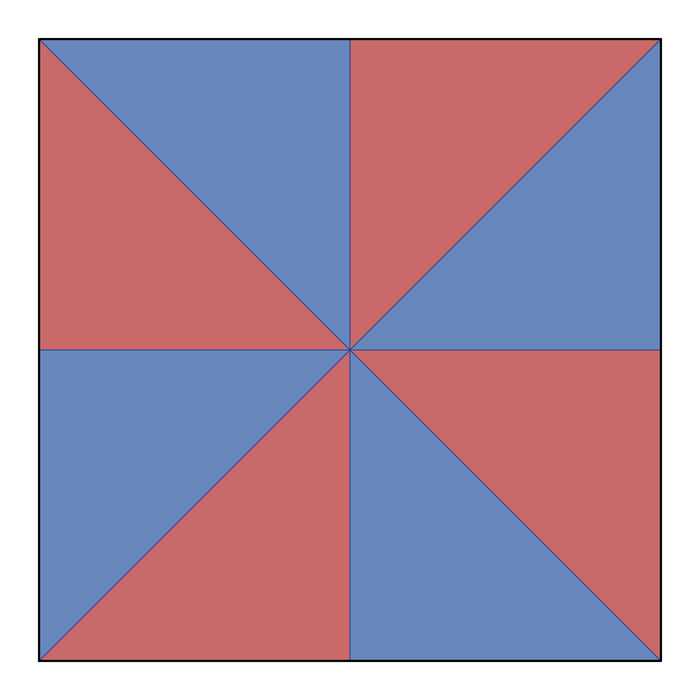}}} & \makebox[0.090\textwidth][l]{{\renewcommand{\arraystretch}{1.05}\begin{tabular}[c]{@{}l@{}}\hyperref[fig:tP1-4/mmm]{Case 14}\\ \hyperref[fig:tI1-4/mmm]{Case 17}\\ \hyperref[fig:tI2-4/mmm]{Case 20}\end{tabular}}} \\
\end{tabular} \\
\cmidrule{2-2}
\makebox[0.060\textwidth][l]{$4/m$} &
\begin{tabular}[c]{@{}l@{\hspace{8pt}}l@{\hspace{8pt}}l@{\hspace{8pt}}l@{\hspace{8pt}}l@{\hspace{28pt}}l@{}}
\makebox[0.055\textwidth][l]{{\renewcommand{\arraystretch}{1.05}\begin{tabular}[c]{@{}l@{}}\hpkot{tP1}\\ \hpkot{tI1}\\ \hpkot{tI2}\end{tabular}}} & \makebox[0.045\textwidth][c]{$\Gamma^{+}_{2}$} & \makebox[0.115\textwidth][l]{${}^{2}4/{}^{1}m$} & \makebox[0.130\textwidth][l]{planar $d$-wave} & \makebox[0.095\textwidth][c]{\raisebox{-0.5\height}{\includegraphics[width=2.0cm,height=2.0cm,keepaspectratio]{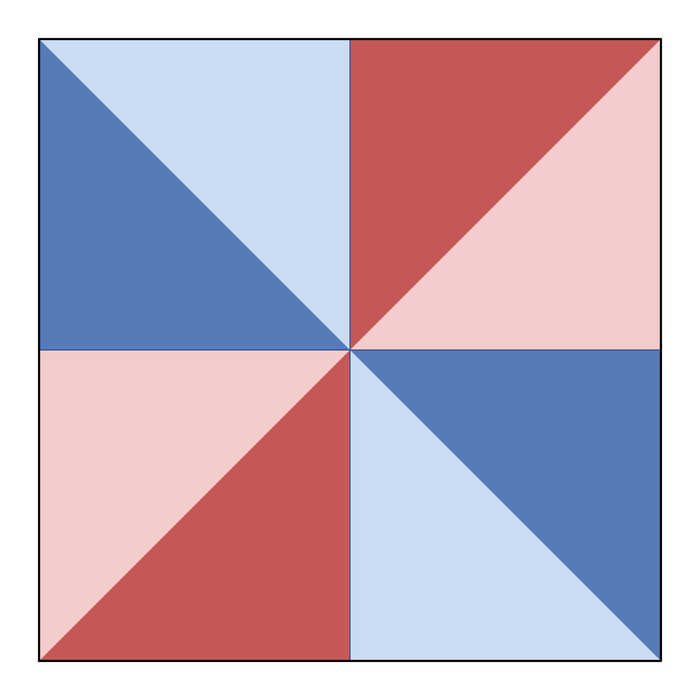}}} & \makebox[0.090\textwidth][l]{{\renewcommand{\arraystretch}{1.05}\begin{tabular}[c]{@{}l@{}}\hyperref[fig:tP1-4/m]{Case 21}\\ \hyperref[fig:tI1-4/m]{Case 22}\\ \hyperref[fig:tI2-4/m]{Case 23}\end{tabular}}} \\
\end{tabular} \\
\end{tabular} \\
\end{tabular} \\
\midrule
\begin{tabular}[c]{@{}l@{\hspace{8pt}}l@{}}
\makebox[0.120\textwidth][l]{Orthorhombic} &
\begin{tabular}[c]{@{}l@{\hspace{8pt}}l@{}}
\makebox[0.060\textwidth][l]{$mmm$} &
\begin{tabular}[c]{@{}l@{\hspace{8pt}}l@{\hspace{8pt}}l@{\hspace{8pt}}l@{\hspace{8pt}}l@{\hspace{28pt}}l@{}}
\makebox[0.055\textwidth][l]{{\renewcommand{\arraystretch}{1.05}\begin{tabular}[c]{@{}l@{}}\hpkot{oP1}\\ \hpkot{oI1}\\ \hpkot{oI2}\\ \hpkot{oI3}\\ \hpkot{oC1$^{*}$}\\ \hpkot{oC2$^{*}$}\\ \hpkot{oF1}\\ \hpkot{oF2}\\ \hpkot{oF3}\end{tabular}}} & \makebox[0.045\textwidth][c]{$\Gamma^{+}_{4}$} & \makebox[0.115\textwidth][l]{${}^{2}m{}^{2}m{}^{1}m$} & \makebox[0.130\textwidth][l]{planar $d$-wave} & \makebox[0.095\textwidth][c]{\raisebox{-0.5\height}{\includegraphics[width=2.0cm,height=2.0cm,keepaspectratio]{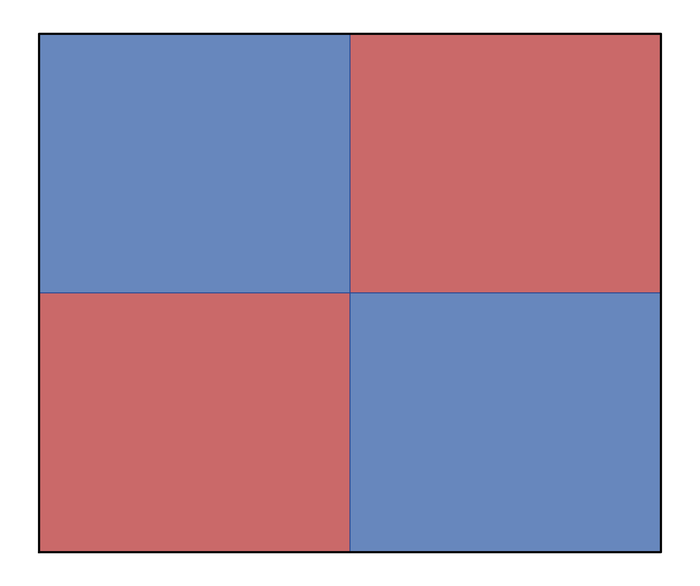}}} & \makebox[0.090\textwidth][l]{{\renewcommand{\arraystretch}{1.05}\begin{tabular}[c]{@{}l@{}}\hyperref[fig:oP1]{Case 24}\\ \hyperref[fig:oI1]{Case 27}\\ \hyperref[fig:oI2]{Case 30}\\ \hyperref[fig:oI3]{Case 33}\\ \hyperref[fig:oC1]{Case 36}\\ \hyperref[fig:oC2]{Case 39}\\ \hyperref[fig:oF1]{Case 42}\\ \hyperref[fig:oF2]{Case 45}\\ \hyperref[fig:oF3]{Case 48}\end{tabular}}} \\
\cmidrule{1-6}
\makebox[0.055\textwidth][l]{{\renewcommand{\arraystretch}{1.05}\begin{tabular}[c]{@{}l@{}}\hpkot{oP1}\\ \hpkot{oI1}\\ \hpkot{oI2}\\ \hpkot{oI3}\\ \hpkot{oC1$^{*}$}\\ \hpkot{oC2$^{*}$}\\ \hpkot{oF1}\\ \hpkot{oF2}\\ \hpkot{oF3}\end{tabular}}} & \makebox[0.045\textwidth][c]{$\Gamma^{+}_{3}$} & \makebox[0.115\textwidth][l]{${}^{2}m{}^{1}m{}^{2}m$} & \makebox[0.130\textwidth][l]{bulk $d$-wave} & \makebox[0.095\textwidth][c]{\raisebox{-0.5\height}{\includegraphics[width=2.0cm,height=2.0cm,keepaspectratio]{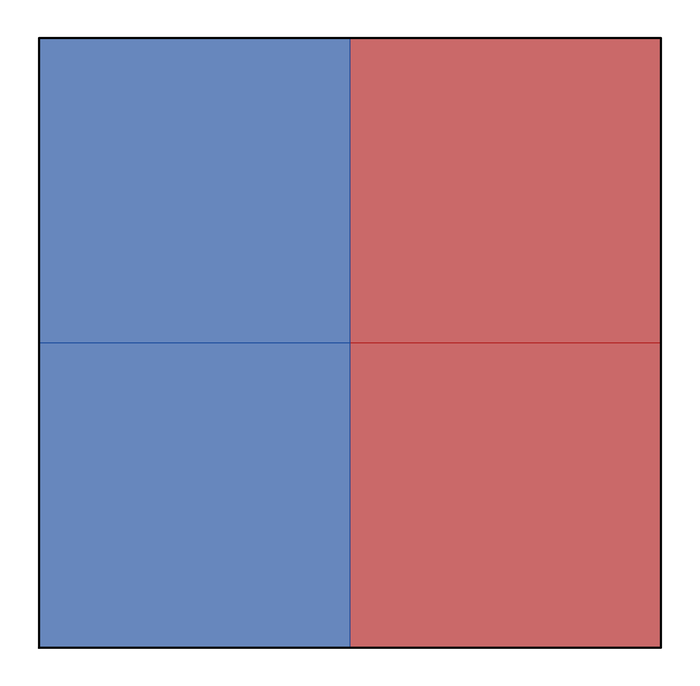}}} & \makebox[0.090\textwidth][l]{{\renewcommand{\arraystretch}{1.05}\begin{tabular}[c]{@{}l@{}}\hyperref[fig:oP1]{Case 25}\\ \hyperref[fig:oI1]{Case 28}\\ \hyperref[fig:oI2]{Case 31}\\ \hyperref[fig:oI3]{Case 34}\\ \hyperref[fig:oC1]{Case 37}\\ \hyperref[fig:oC2]{Case 40}\\ \hyperref[fig:oF1]{Case 43}\\ \hyperref[fig:oF2]{Case 46}\\ \hyperref[fig:oF3]{Case 49}\end{tabular}}} \\
\cmidrule{1-6}
\makebox[0.055\textwidth][l]{{\renewcommand{\arraystretch}{1.05}\begin{tabular}[c]{@{}l@{}}\hpkot{oP1}\\ \hpkot{oI1}\\ \hpkot{oI2}\\ \hpkot{oI3}\\ \hpkot{oC1$^{*}$}\\ \hpkot{oC2$^{*}$}\\ \hpkot{oF1}\\ \hpkot{oF2}\\ \hpkot{oF3}\end{tabular}}} & \makebox[0.045\textwidth][c]{$\Gamma^{+}_{2}$} & \makebox[0.115\textwidth][l]{${}^{1}m{}^{2}m{}^{2}m$} & \makebox[0.130\textwidth][l]{bulk $d$-wave} & \makebox[0.095\textwidth][c]{\raisebox{-0.5\height}{\includegraphics[width=2.0cm,height=2.0cm,keepaspectratio]{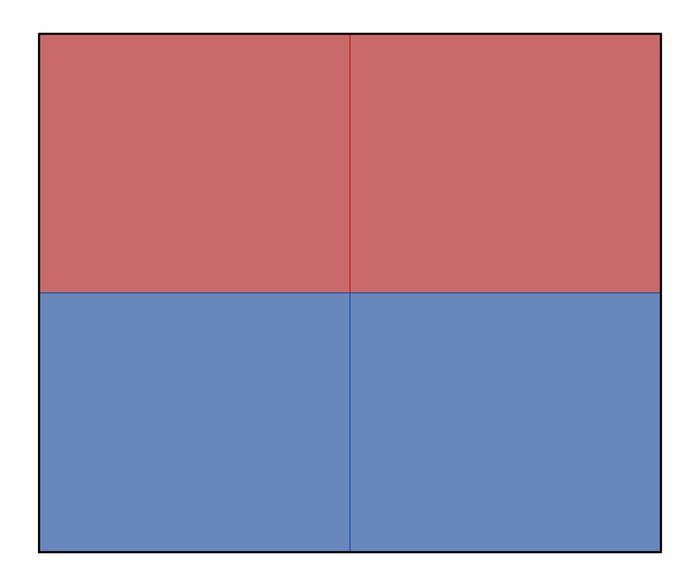}}} & \makebox[0.090\textwidth][l]{{\renewcommand{\arraystretch}{1.05}\begin{tabular}[c]{@{}l@{}}\hyperref[fig:oP1]{Case 26}\\ \hyperref[fig:oI1]{Case 29}\\ \hyperref[fig:oI2]{Case 32}\\ \hyperref[fig:oI3]{Case 35}\\ \hyperref[fig:oC1]{Case 38}\\ \hyperref[fig:oC2]{Case 41}\\ \hyperref[fig:oF1]{Case 44}\\ \hyperref[fig:oF2]{Case 47}\\ \hyperref[fig:oF3]{Case 50}\end{tabular}}} \\
\end{tabular} \\
\end{tabular} \\
\end{tabular} \\
\midrule
\begin{tabular}[c]{@{}l@{\hspace{8pt}}l@{}}
\makebox[0.120\textwidth][l]{Monoclinic} &
\begin{tabular}[c]{@{}l@{\hspace{8pt}}l@{}}
\makebox[0.060\textwidth][l]{$2/m$} &
\begin{tabular}[c]{@{}l@{\hspace{8pt}}l@{\hspace{8pt}}l@{\hspace{8pt}}l@{\hspace{8pt}}l@{\hspace{28pt}}l@{}}
\makebox[0.055\textwidth][l]{{\renewcommand{\arraystretch}{1.05}\begin{tabular}[c]{@{}l@{}}\hpkot{mP1}\\ \hpkot{mC1}\\ \hpkot{mC2}\\ \hpkot{mC3}\end{tabular}}} & \makebox[0.045\textwidth][c]{$\Gamma^{+}_{2}$} & \makebox[0.115\textwidth][l]{${}^{2}2/{}^{2}m$} & \makebox[0.130\textwidth][l]{bulk $d$-wave} & \makebox[0.095\textwidth][c]{\raisebox{-0.5\height}{\includegraphics[width=2.0cm,height=2.0cm,keepaspectratio]{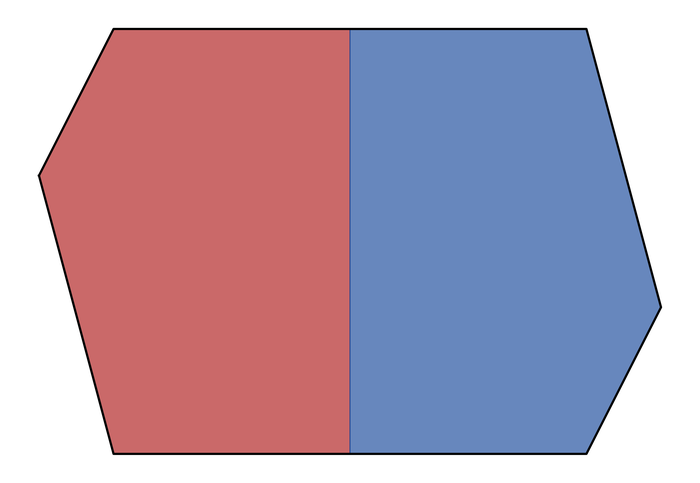}}} & \makebox[0.090\textwidth][l]{{\renewcommand{\arraystretch}{1.05}\begin{tabular}[c]{@{}l@{}}\hyperref[fig:mP1]{Case 51}\\ \hyperref[fig:mC1]{Case 52}\\ \hyperref[fig:mC2]{Case 53}\\ \hyperref[fig:mC3]{Case 54}\end{tabular}}} \\
\end{tabular} \\
\end{tabular} \\
\end{tabular} \\
\end{longtable}
\par\smallskip\noindent\footnotesize\emph{*} In the alternative A-centered conventional setting, oC1 and oC2 are equivalent to oA1 and oA2, respectively.
\normalfont\normalsize

\renewcommand{\arraystretch}{1}

\begin{table}[H]
\centering
\renewcommand{\arraystretch}{1.3}
\caption{AlterSeeK-Path two-dimensional cases for periodic slabs, organized by Bravais lattice, Laue group, irrep, spin layer group, and altermagnetic spin pattern. The final column gives the case number; each case is listed in \ref{sec:case-library-2d}. For the centered-rectangular lattice, the BZ shape depends on whether the conventional lattice vectors satisfy $a<b$ or $a>b$.}
\label{tab:spin-layer-2d}
\begin{tabular}{@{}l@{}}
\toprule
\begin{tabular}[c]{@{}l@{\hspace{8pt}}l@{\hspace{8pt}}l@{\hspace{8pt}}l@{\hspace{8pt}}l@{\hspace{28pt}}l@{}}
\makebox[0.120\textwidth][l]{\makecell[l]{Bravais\\lattice}} & \makebox[0.070\textwidth][l]{\makecell[l]{Laue\\group}} & \makebox[0.050\textwidth][c]{Irrep} & \makebox[0.150\textwidth][l]{\makecell[l]{Spin layer\\group}} & \makebox[\dimexpr 0.100\textwidth+8pt+0.095\textwidth\relax][c]{Altermagnetic spin pattern} & \makebox[0.110\textwidth][l]{\makecell[l]{Case\\number}} \\
\end{tabular} \\
\midrule
\begin{tabular}[c]{@{}l@{\hspace{8pt}}l@{}}
\makebox[0.120\textwidth][l]{Hexagonal} &
\begin{tabular}[c]{@{}l@{\hspace{8pt}}l@{}}
\makebox[0.070\textwidth][l]{{\renewcommand{\arraystretch}{1.05}\begin{tabular}[c]{@{}l@{}}$6/mmm$\\ $\bar{3}m$\end{tabular}}} &
\begin{tabular}[c]{@{}l@{\hspace{8pt}}l@{\hspace{8pt}}l@{\hspace{8pt}}l@{\hspace{28pt}}l@{}}
\makebox[0.050\textwidth][c]{{\renewcommand{\arraystretch}{1.05}\begin{tabular}[c]{@{}c@{}}$\Gamma^{+}_{2}$\\ $\Gamma^{+}_{2}$\end{tabular}}} & \makebox[0.150\textwidth][l]{{\renewcommand{\arraystretch}{1.05}\begin{tabular}[c]{@{}l@{}}${}^{1}6/{}^{1}m{}^{2}m{}^{2}m$\\ ${}^{1}\bar{3}{}^{2}m$\end{tabular}}} & \makebox[0.115\textwidth][l]{$i$-wave} & \makebox[0.095\textwidth][c]{\raisebox{-0.5\height}{\includegraphics[width=2.0cm]{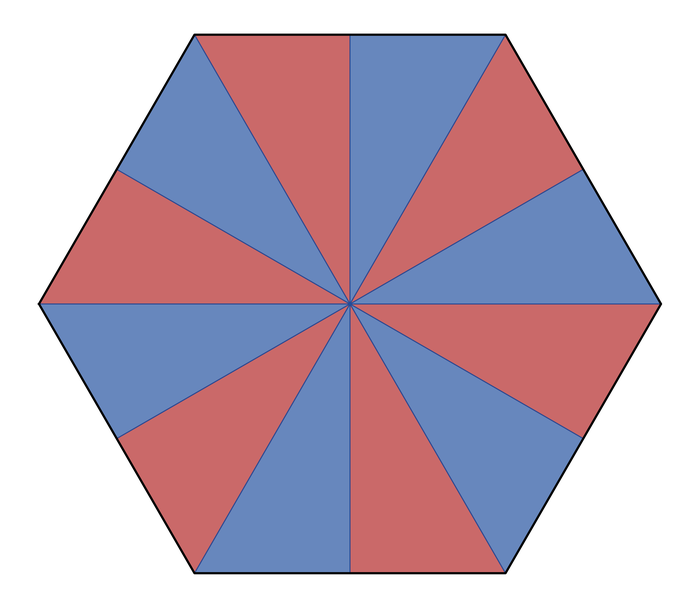}}} & \makebox[0.110\textwidth][l]{{\renewcommand{\arraystretch}{1.05}\begin{tabular}[c]{@{}l@{}}\hyperref[fig:2D-hexagonal]{2D Case 01}\\ \hyperref[fig:2D-hexagonal]{2D Case 02}\end{tabular}}} \\
\end{tabular} \\
\end{tabular} \\
\end{tabular} \\
\midrule
\begin{tabular}[c]{@{}l@{\hspace{8pt}}l@{}}
\makebox[0.120\textwidth][l]{Square} &
\begin{tabular}[c]{@{}l@{\hspace{8pt}}l@{}}
\makebox[0.070\textwidth][l]{$4/mmm$} &
\begin{tabular}[c]{@{}l@{\hspace{8pt}}l@{\hspace{8pt}}l@{\hspace{8pt}}l@{\hspace{28pt}}l@{}}
\makebox[0.050\textwidth][c]{$\Gamma^{+}_{4}$} & \makebox[0.150\textwidth][l]{${}^{2}4/{}^{1}m{}^{2}m{}^{1}m$} & \makebox[0.115\textwidth][l]{$d$-wave} & \makebox[0.095\textwidth][c]{\raisebox{-0.5\height}{\includegraphics[width=2.0cm]{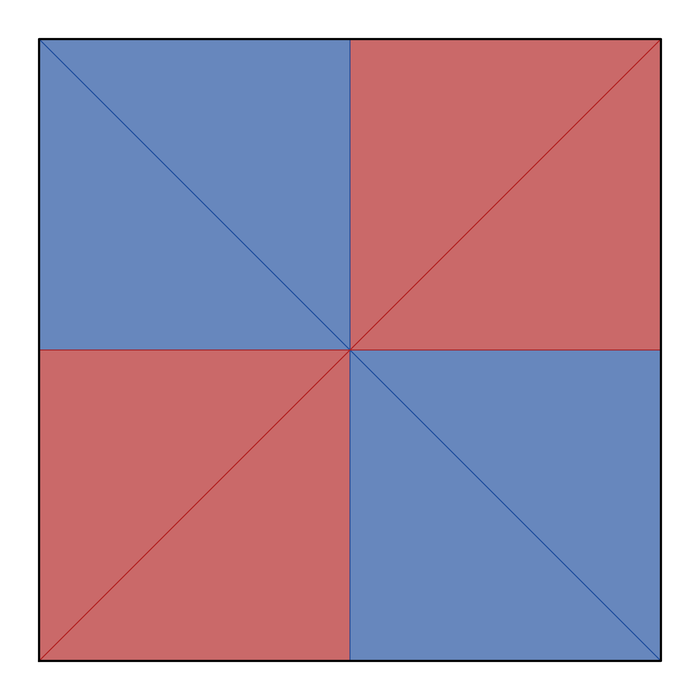}}} & \makebox[0.110\textwidth][l]{\hyperref[fig:2D-square-4mmm]{2D Case 03}} \\
\cmidrule{1-5}
\makebox[0.050\textwidth][c]{$\Gamma^{+}_{3}$} & \makebox[0.150\textwidth][l]{${}^{2}4/{}^{1}m{}^{1}m{}^{2}m$} & \makebox[0.115\textwidth][l]{$d$-wave} & \makebox[0.095\textwidth][c]{\raisebox{-0.5\height}{\includegraphics[width=2.0cm]{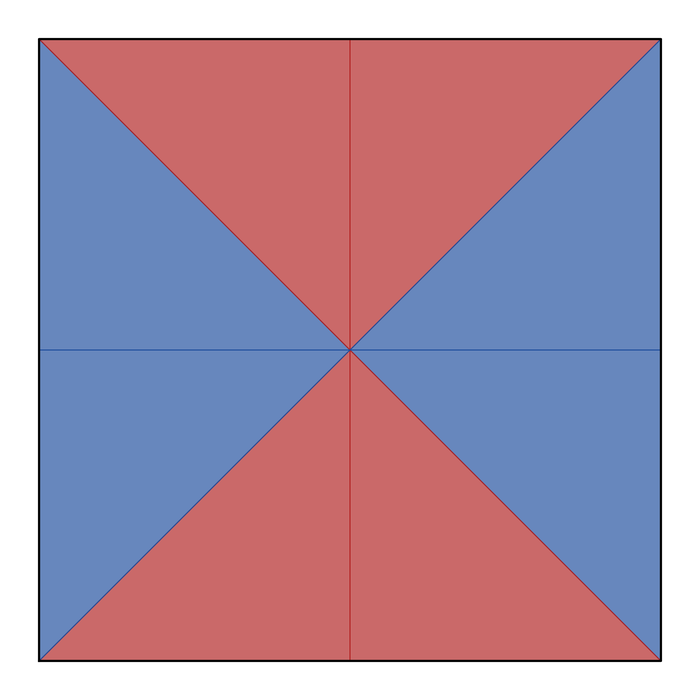}}} & \makebox[0.110\textwidth][l]{\hyperref[fig:2D-square-4mmm]{2D Case 04}} \\
\cmidrule{1-5}
\makebox[0.050\textwidth][c]{$\Gamma^{+}_{2}$} & \makebox[0.150\textwidth][l]{${}^{1}4/{}^{1}m{}^{2}m{}^{2}m$} & \makebox[0.115\textwidth][l]{$g$-wave} & \makebox[0.095\textwidth][c]{\raisebox{-0.5\height}{\includegraphics[width=2.0cm]{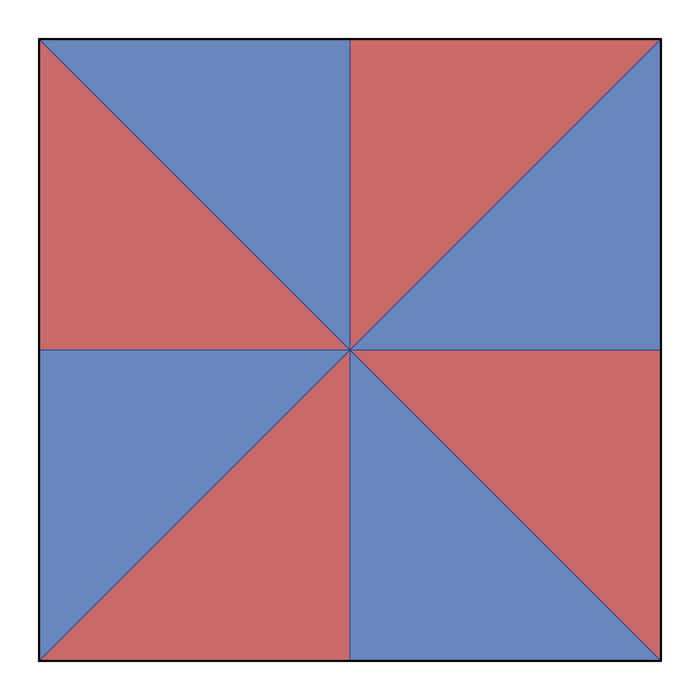}}} & \makebox[0.110\textwidth][l]{\hyperref[fig:2D-square-4mmm]{2D Case 05}} \\
\end{tabular} \\
\cmidrule{2-2}
\makebox[0.070\textwidth][l]{$4/m$} &
\begin{tabular}[c]{@{}l@{\hspace{8pt}}l@{\hspace{8pt}}l@{\hspace{8pt}}l@{\hspace{28pt}}l@{}}
\makebox[0.050\textwidth][c]{$\Gamma^{+}_{2}$} & \makebox[0.150\textwidth][l]{${}^{2}4/{}^{1}m$} & \makebox[0.115\textwidth][l]{$d$-wave} & \makebox[0.095\textwidth][c]{\raisebox{-0.5\height}{\includegraphics[width=2.0cm]{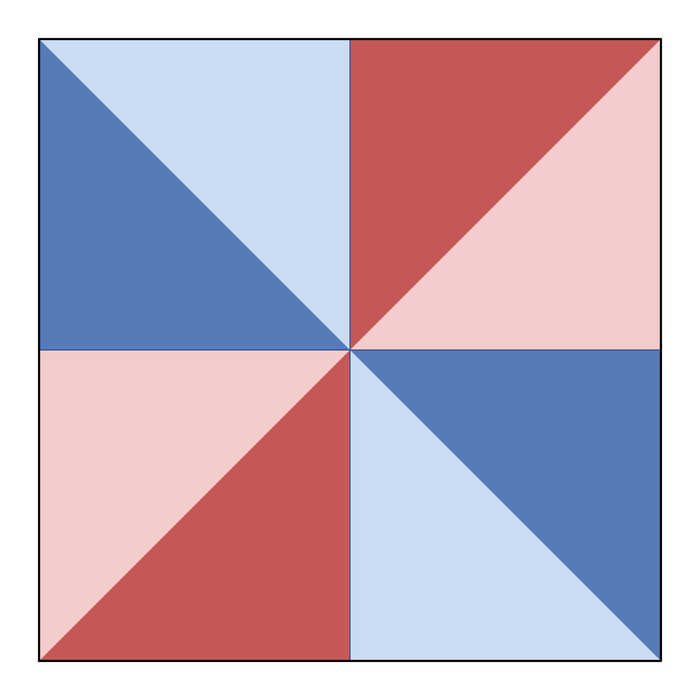}}} & \makebox[0.110\textwidth][l]{\hyperref[fig:2D-square-4m]{2D Case 06}} \\
\end{tabular} \\
\end{tabular} \\
\end{tabular} \\
\midrule
\begin{tabular}[c]{@{}l@{\hspace{8pt}}l@{}}
\makebox[0.120\textwidth][l]{Rectangular} &
\begin{tabular}[c]{@{}l@{\hspace{8pt}}l@{}}
\makebox[0.070\textwidth][l]{{\renewcommand{\arraystretch}{1.05}\begin{tabular}[c]{@{}l@{}}$mmm$\\ $2/m$\end{tabular}}} &
\begin{tabular}[c]{@{}l@{\hspace{8pt}}l@{\hspace{8pt}}l@{\hspace{8pt}}l@{\hspace{28pt}}l@{}}
\makebox[0.050\textwidth][c]{{\renewcommand{\arraystretch}{1.05}\begin{tabular}[c]{@{}c@{}}$\Gamma^{+}_{4}$\\ $\Gamma^{+}_{2}$\end{tabular}}} & \makebox[0.150\textwidth][l]{{\renewcommand{\arraystretch}{1.05}\begin{tabular}[c]{@{}l@{}}${}^{2}m{}^{2}m{}^{1}m$\\ ${}^{2}2/{}^{2}m_x$\end{tabular}}} & \makebox[0.115\textwidth][l]{$d$-wave} & \makebox[0.095\textwidth][c]{\raisebox{-0.5\height}{\includegraphics[width=2.0cm]{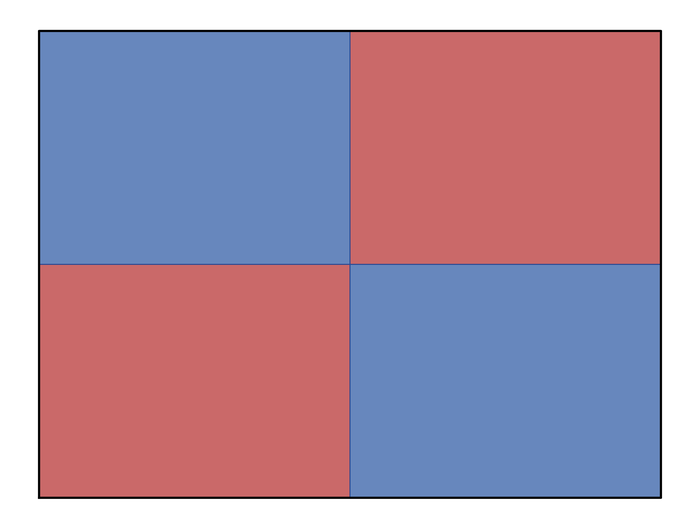}}} & \makebox[0.110\textwidth][l]{{\renewcommand{\arraystretch}{1.05}\begin{tabular}[c]{@{}l@{}}\hyperref[fig:2D-rectangular]{2D Case 07}\\ \hyperref[fig:2D-rectangular]{2D Case 08}\end{tabular}}} \\
\end{tabular} \\
\end{tabular} \\
\end{tabular} \\
\midrule
\begin{tabular}[c]{@{}l@{\hspace{8pt}}l@{}}
\makebox[0.120\textwidth][l]{\makecell[l]{Centered\\rectangular\\ ($a<b$)}} &
\begin{tabular}[c]{@{}l@{\hspace{8pt}}l@{}}
\makebox[0.070\textwidth][l]{{\renewcommand{\arraystretch}{1.05}\begin{tabular}[c]{@{}l@{}}$mmm$\\ $2/m$\end{tabular}}} &
\begin{tabular}[c]{@{}l@{\hspace{8pt}}l@{\hspace{8pt}}l@{\hspace{8pt}}l@{\hspace{28pt}}l@{}}
\makebox[0.050\textwidth][c]{{\renewcommand{\arraystretch}{1.05}\begin{tabular}[c]{@{}c@{}}$\Gamma^{+}_{4}$\\ $\Gamma^{+}_{2}$\end{tabular}}} & \makebox[0.150\textwidth][l]{{\renewcommand{\arraystretch}{1.05}\begin{tabular}[c]{@{}l@{}}${}^{2}m{}^{2}m{}^{1}m$\\ ${}^{2}2/{}^{2}m_x$\end{tabular}}} & \makebox[0.115\textwidth][l]{$d$-wave} & \makebox[0.095\textwidth][c]{\raisebox{-0.5\height}{\includegraphics[width=2.0cm]{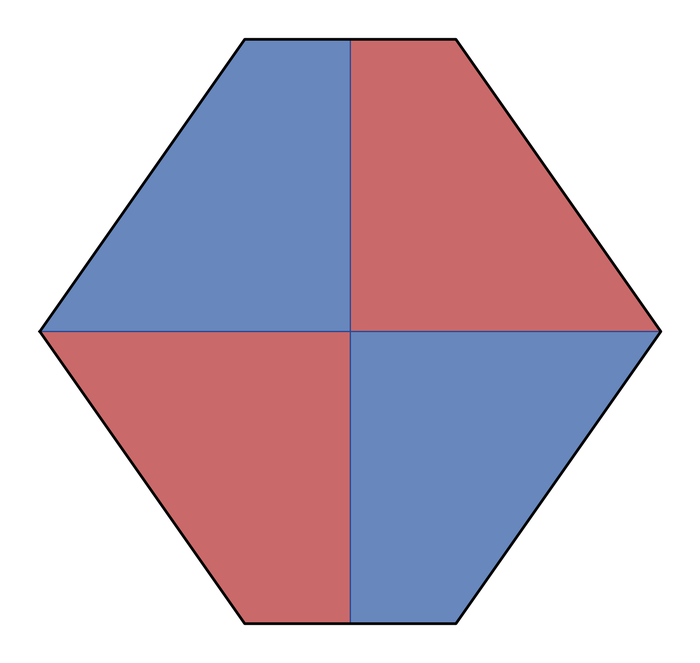}}} & \makebox[0.110\textwidth][l]{{\renewcommand{\arraystretch}{1.05}\begin{tabular}[c]{@{}l@{}}\hyperref[fig:2D-centered-rectangular-alt]{2D Case 09a}\\ \hyperref[fig:2D-centered-rectangular-alt]{2D Case 10a}\end{tabular}}} \\
\end{tabular} \\
\end{tabular} \\
\end{tabular} \\
\midrule
\begin{tabular}[c]{@{}l@{\hspace{8pt}}l@{}}
\makebox[0.120\textwidth][l]{\makecell[l]{Centered\\rectangular\\ ($a>b$)}} &
\begin{tabular}[c]{@{}l@{\hspace{8pt}}l@{}}
\makebox[0.070\textwidth][l]{{\renewcommand{\arraystretch}{1.05}\begin{tabular}[c]{@{}l@{}}$mmm$\\ $2/m$\end{tabular}}} &
\begin{tabular}[c]{@{}l@{\hspace{8pt}}l@{\hspace{8pt}}l@{\hspace{8pt}}l@{\hspace{28pt}}l@{}}
\makebox[0.050\textwidth][c]{{\renewcommand{\arraystretch}{1.05}\begin{tabular}[c]{@{}c@{}}$\Gamma^{+}_{4}$\\ $\Gamma^{+}_{2}$\end{tabular}}} & \makebox[0.150\textwidth][l]{{\renewcommand{\arraystretch}{1.05}\begin{tabular}[c]{@{}l@{}}${}^{2}m{}^{2}m{}^{1}m$\\ ${}^{2}2/{}^{2}m_x$\end{tabular}}} & \makebox[0.115\textwidth][l]{$d$-wave} & \makebox[0.095\textwidth][c]{\raisebox{-0.5\height}{\includegraphics[width=2.0cm]{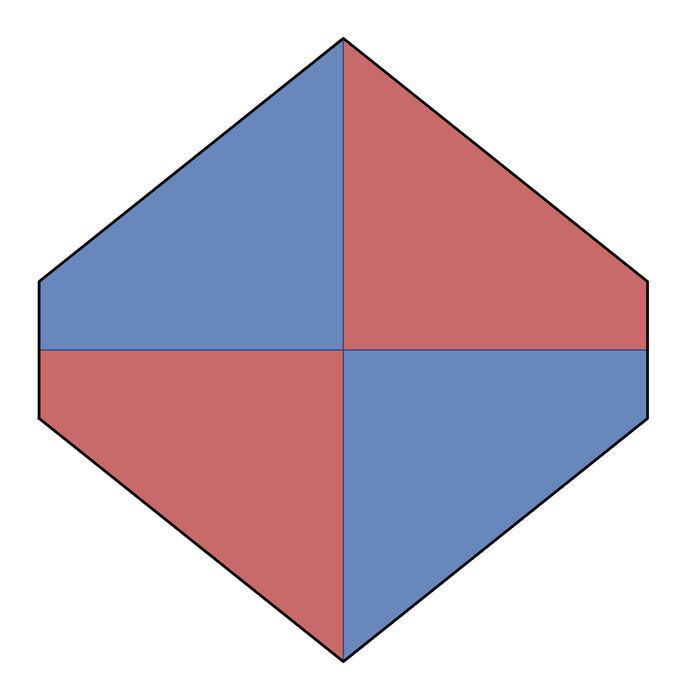}}} & \makebox[0.110\textwidth][l]{{\renewcommand{\arraystretch}{1.05}\begin{tabular}[c]{@{}l@{}}\hyperref[fig:2D-centered-rectangular-agt]{2D Case 09b}\\ \hyperref[fig:2D-centered-rectangular-agt]{2D Case 10b}\end{tabular}}} \\
\end{tabular} \\
\end{tabular} \\
\end{tabular} \\
\bottomrule
\end{tabular}
\renewcommand{\arraystretch}{1}
\end{table}

\clearpage
\subsection{Cubic}
\label{sec:case-library-cubic}

\subsubsection{cP2}

\noindent\textbf{Case 01, \hpkot{cP2}: ${}^{1}m{}^{1}\bar{3}{}^{2}m$, bulk $i$-wave}\par
\begin{figure}[H]
\centering
\includegraphics[width=0.9\textwidth,keepaspectratio]{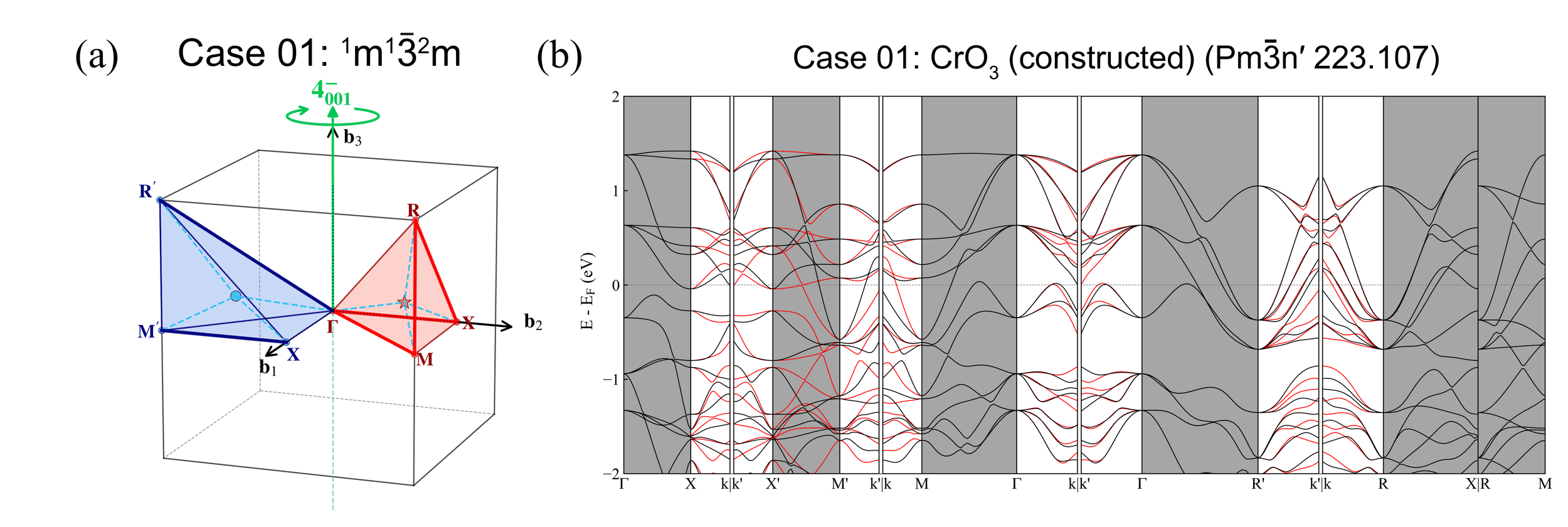}
\caption{Cubic cP2 BZ and band for the spin-Laue-group $^{1}m^{1}\bar{3}^{2}m$, using CrO$_3$ (constructed) (MSG without SOC: $Pm\bar{3}n'$, BNS 223.107, Type III) as a representative example.}
\label{fig:cP2}
\end{figure}

\begin{equation}
\Gamma\text{--}\underbrace{X\text{--}k\,|\,k'\text{--}X'}_{}\text{--}\underbrace{M'\text{--}k'\,|\,k\text{--}M}_{}\text{--}\underbrace{\Gamma\text{--}k\,|\,k'\text{--}\Gamma}_{}\text{--}\underbrace{R'\text{--}k'\,|\,k\text{--}R}_{}\text{--}X\,|\,R\text{--}M .
\end{equation}

\subsubsection{cF2}

\noindent\textbf{Case 02, \hpkot{cF2}: ${}^{1}m{}^{1}\bar{3}{}^{2}m$, bulk $i$-wave}\par
\begin{figure}[H]
\centering
\includegraphics[width=0.9\textwidth,keepaspectratio]{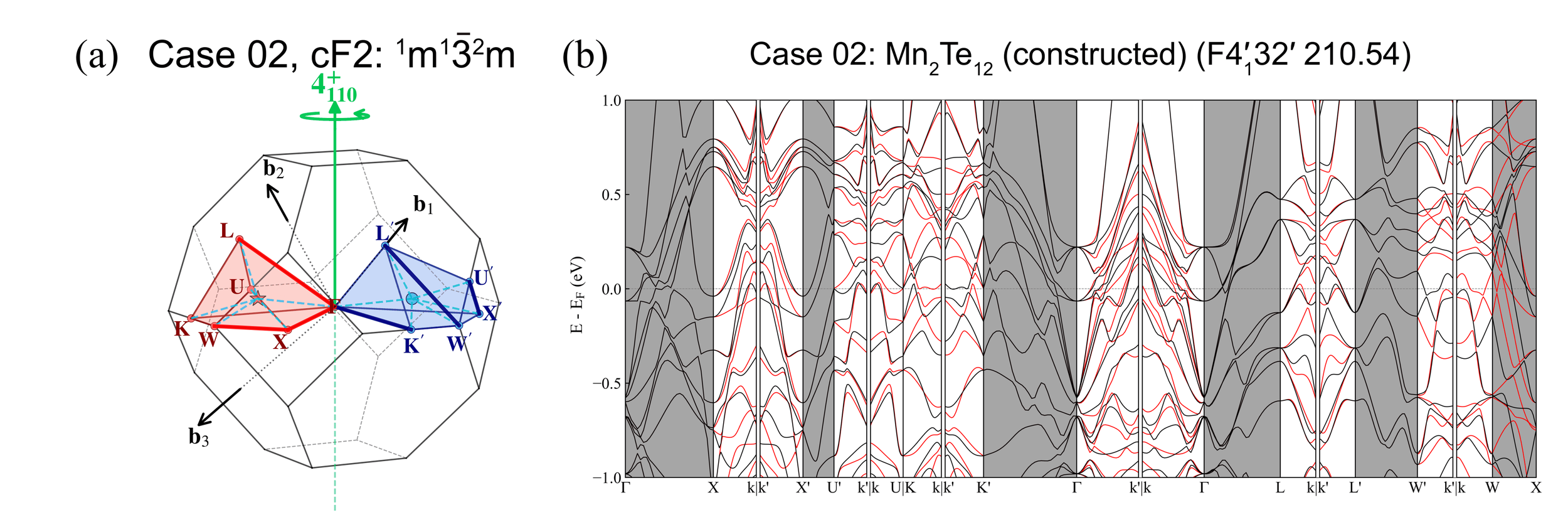}
\caption{Cubic cF2 BZ and band for the spin-Laue-group $^{1}m^{1}\bar{3}^{2}m$, using Mn$_2$Te$_{12}$ (constructed) (MSG without SOC: $F4_1'32'$, BNS 210.54, Type III) as a representative example.}
\label{fig:cF2}
\end{figure}

\begin{equation}
\Gamma\text{--}\underbrace{X\text{--}k\,|\,k'\text{--}X'}_{}\text{--}\underbrace{U'\text{--}k'\,|\,k\text{--}U}_{}\,|\,\underbrace{K\text{--}k\,|\,k'\text{--}K'}_{}\text{--}\underbrace{\Gamma\text{--}k'\,|\,k\text{--}\Gamma}_{}\text{--}\underbrace{L\text{--}k\,|\,k'\text{--}L'}_{}\text{--}\underbrace{W'\text{--}k'\,|\,k\text{--}W}_{}\text{--}X .
\end{equation}

\subsubsection{cI1}

\noindent\textbf{Case 03, \hpkot{cI1}: ${}^{1}m{}^{1}\bar{3}{}^{2}m$, bulk $i$-wave}\par
\begin{figure}[H]
\centering
\includegraphics[width=0.9\textwidth,keepaspectratio]{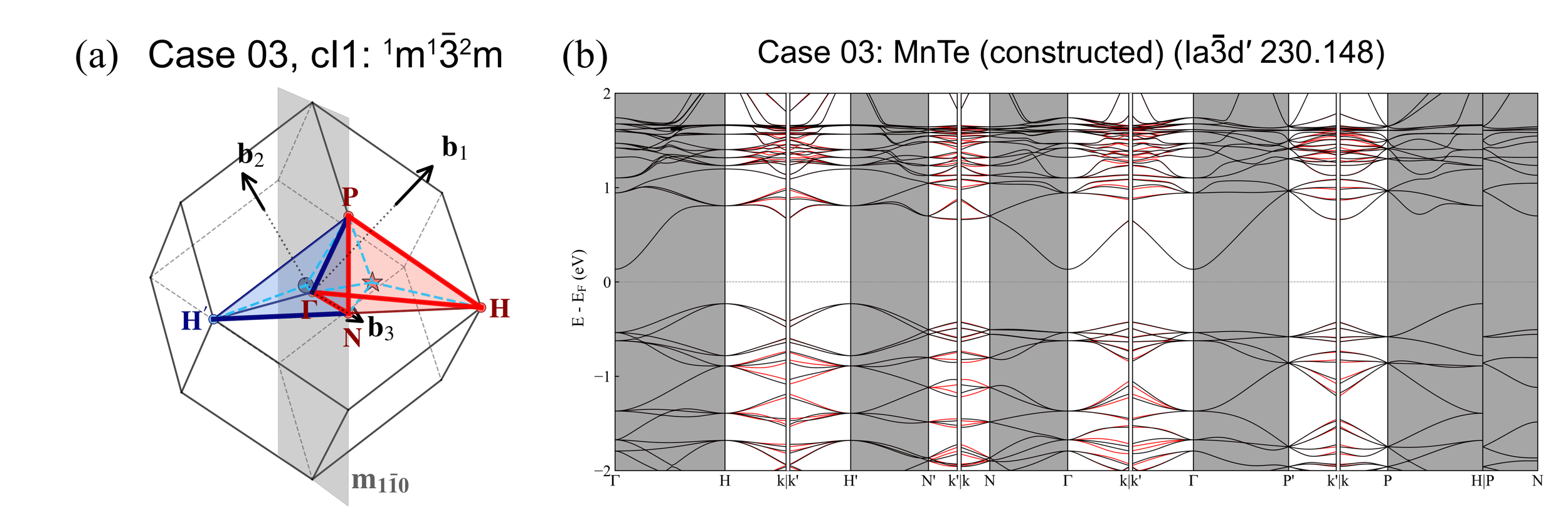}
\caption{Cubic cI1 BZ and band for the spin-Laue-group $^{1}m^{1}\bar{3}^{2}m$, using MnTe (constructed) (MSG without SOC: $Ia\bar{3}d'$, BNS 230.148, Type III) as a representative example.}
\label{fig:cI1}
\end{figure}

\begin{equation}
\Gamma\text{--}\underbrace{H\text{--}k\,|\,k'\text{--}H'}_{}\text{--}\underbrace{N'\text{--}k'\,|\,k\text{--}N}_{}\text{--}\underbrace{\Gamma\text{--}k\,|\,k'\text{--}\Gamma}_{}\text{--}\underbrace{P'\text{--}k'\,|\,k\text{--}P}_{}\text{--}H\,|\,P\text{--}N .
\end{equation}

\subsection{Hexagonal}

\subsubsection{hP2 (6/mmm)}
\noindent\textbf{Case 04: ${}^{2}6/{}^{2}m{}^{2}m{}^{1}m$, bulk $g$-wave}\par
\noindent\textbf{Case 05: ${}^{2}6/{}^{2}m{}^{1}m{}^{2}m$, bulk $g$-wave}\par
\noindent\textbf{Case 06: ${}^{1}6/{}^{1}m{}^{2}m{}^{2}m$, planar $i$-wave}\par
\begin{figure}[H]
\centering
\includegraphics[width=0.9\textwidth,keepaspectratio]{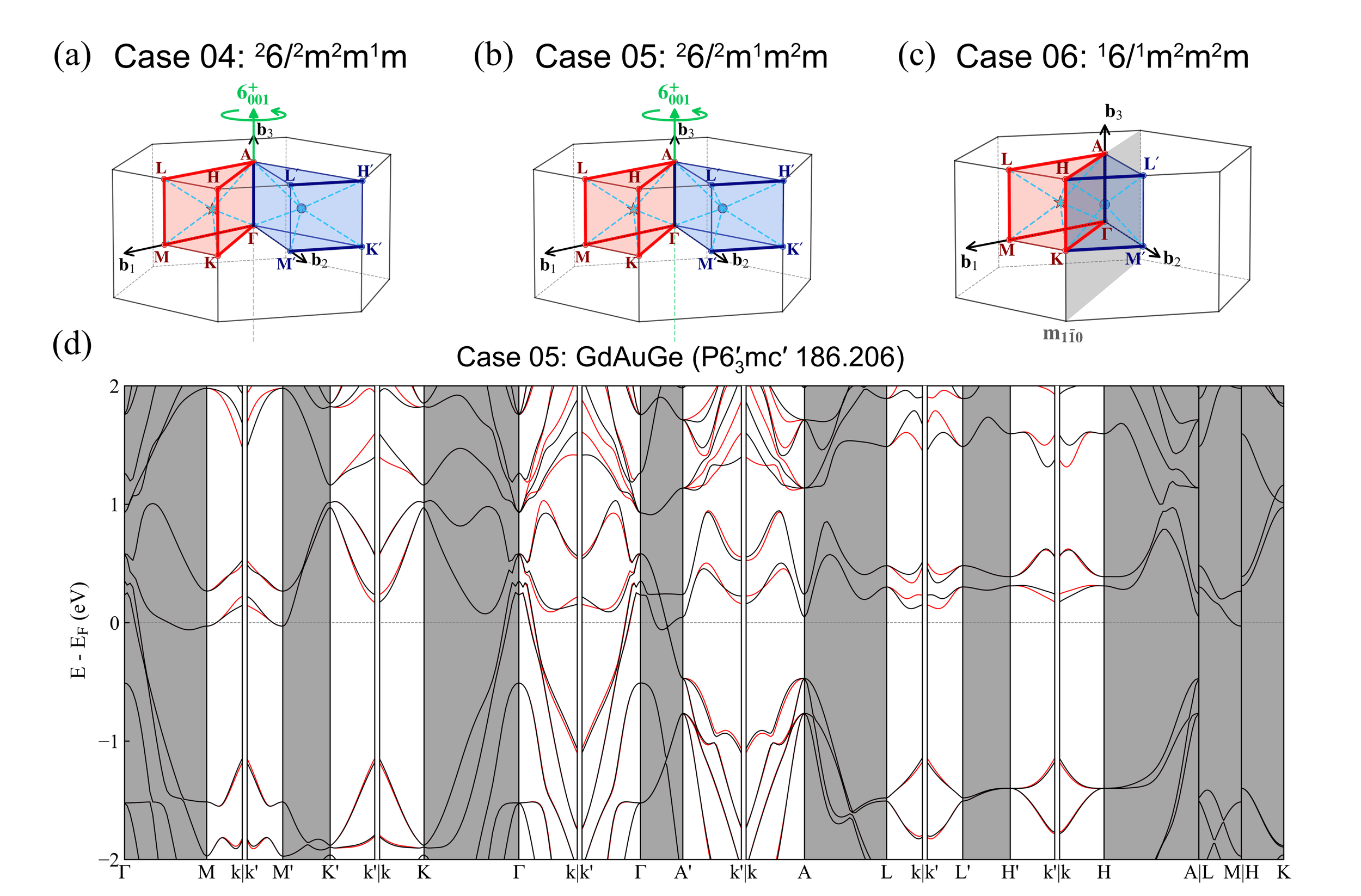}
\caption{Hexagonal hP2 BZ and band for the three spin Laue groups of the $6/mmm$ Laue group, using GdAuGe (MSG without SOC: $P6_3'mc'$, BNS 186.206, Type III) as a representative example. (a)--(c) show Cases 04--06; since all three share the same k-path, only one band structure is shown, corresponding to Case 05 ($^{2}6/^{2}m^{1}m^{2}m$).}
\label{fig:hP2-6/mmm}
\end{figure}

\begin{equation}
\Gamma\text{--}
\underbrace{M\text{--}k\,|\,k'\text{--}M'}_{}
\text{--}
\underbrace{K'\text{--}k'\,|\,k\text{--}K}_{}
\text{--}
\underbrace{\Gamma\text{--}k\,|\,k'\text{--}\Gamma}_{}
\text{--}
\underbrace{A'\text{--}k'\,|\,k\text{--}A}_{}
\text{--}
\underbrace{L\text{--}k\,|\,k'\text{--}L'}_{}
\text{--}
\underbrace{H'\text{--}k'\,|\,k\text{--}H}_{}
\text{--}A
\,|\,L\text{--}M
\,|\,H\text{--}K .
\end{equation}

\subsubsection{hP2 (6/m)}

\noindent\textbf{Case 07, \hpkot{hP2}: ${}^{2}6/{}^{2}m$, bulk $g$-wave}\par

\begin{figure}[H]
\centering
\includegraphics[width=0.9\textwidth,keepaspectratio]{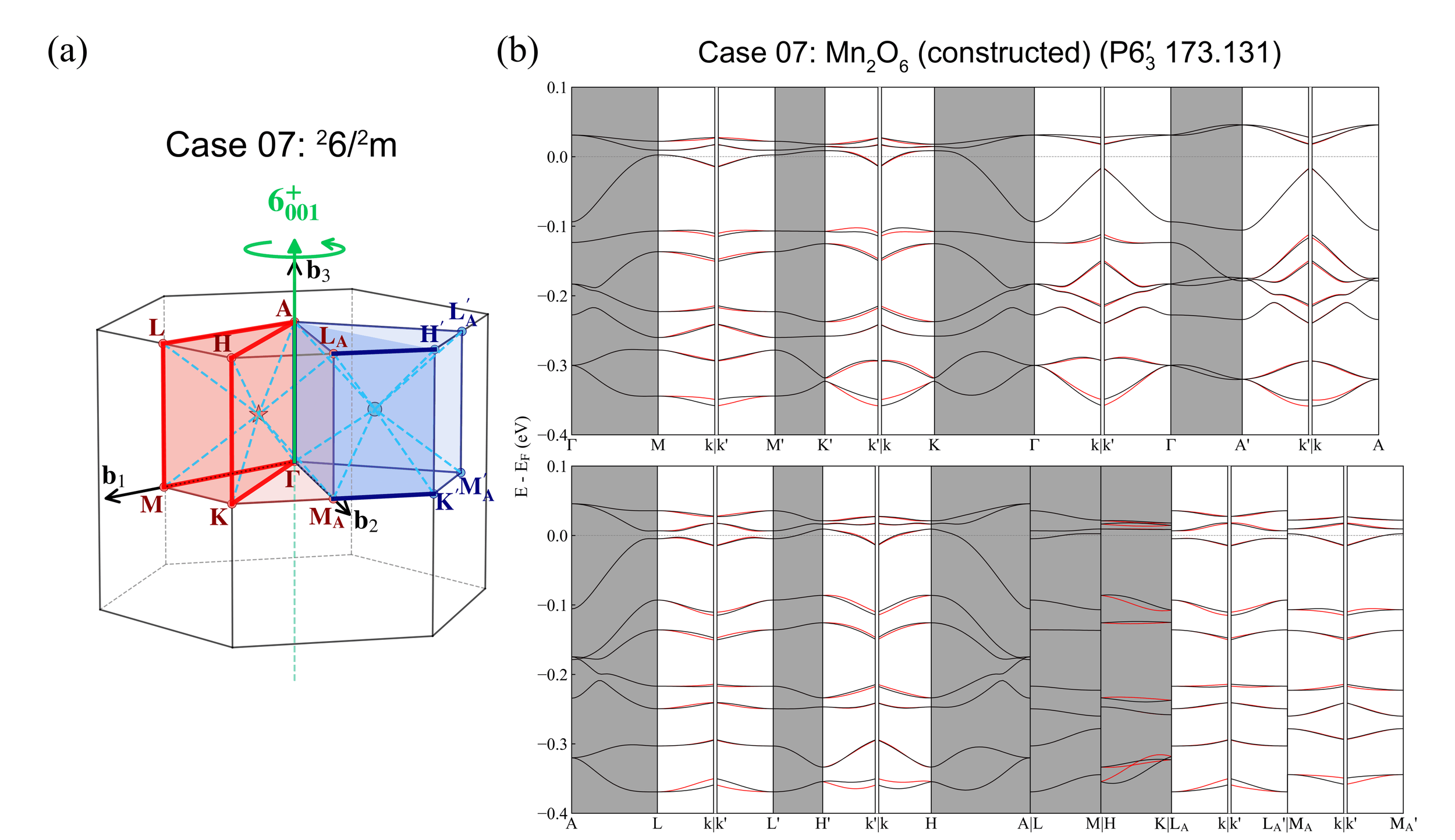}
\caption{Hexagonal hP2 BZ and band for the spin-Laue-group $^{2}6/^{2}m$, using Mn$_2$O$_6$ (constructed) (MSG without SOC: $P6_3'$, BNS 173.131, Type III) as a representative example.}
\label{fig:hP2-6/m}
\end{figure}

\begin{equation}
\begin{split}
\Gamma
&\text{--}\underbrace{M\text{--}k\,|\,k'\text{--}M'}_{}\text{--}\underbrace{K'\text{--}k'\,|\,k\text{--}K}_{}\text{--}\underbrace{\Gamma\text{--}k\,|\,k'\text{--}\Gamma}_{}\text{--}\underbrace{A'\text{--}k'\,|\,k\text{--}A}_{}\text{--}\underbrace{L\text{--}k\,|\,k'\text{--}L'}_{}
\\
&\text{--}\underbrace{H'\text{--}k'\,|\,k\text{--}H}_{}\text{--}A\,|\,L\text{--}M\,|\,H\text{--}K\,|\,\underbrace{L_A\text{--}k\,|\,k'\text{--}L_A'}_{}\,|\,\underbrace{M_A\text{--}k\,|\,k'\text{--}M_A'}_{} .
\end{split}
\end{equation}

\subsection{Trigonal}

\subsubsection{hP1}

\noindent\textbf{Case 08, \hpkot{hP1}: ${}^{1}\bar{3}{}^{2}m$, bulk $g$-wave}\par
\begin{figure}[H]
\centering
\includegraphics[width=0.9\textwidth,keepaspectratio]{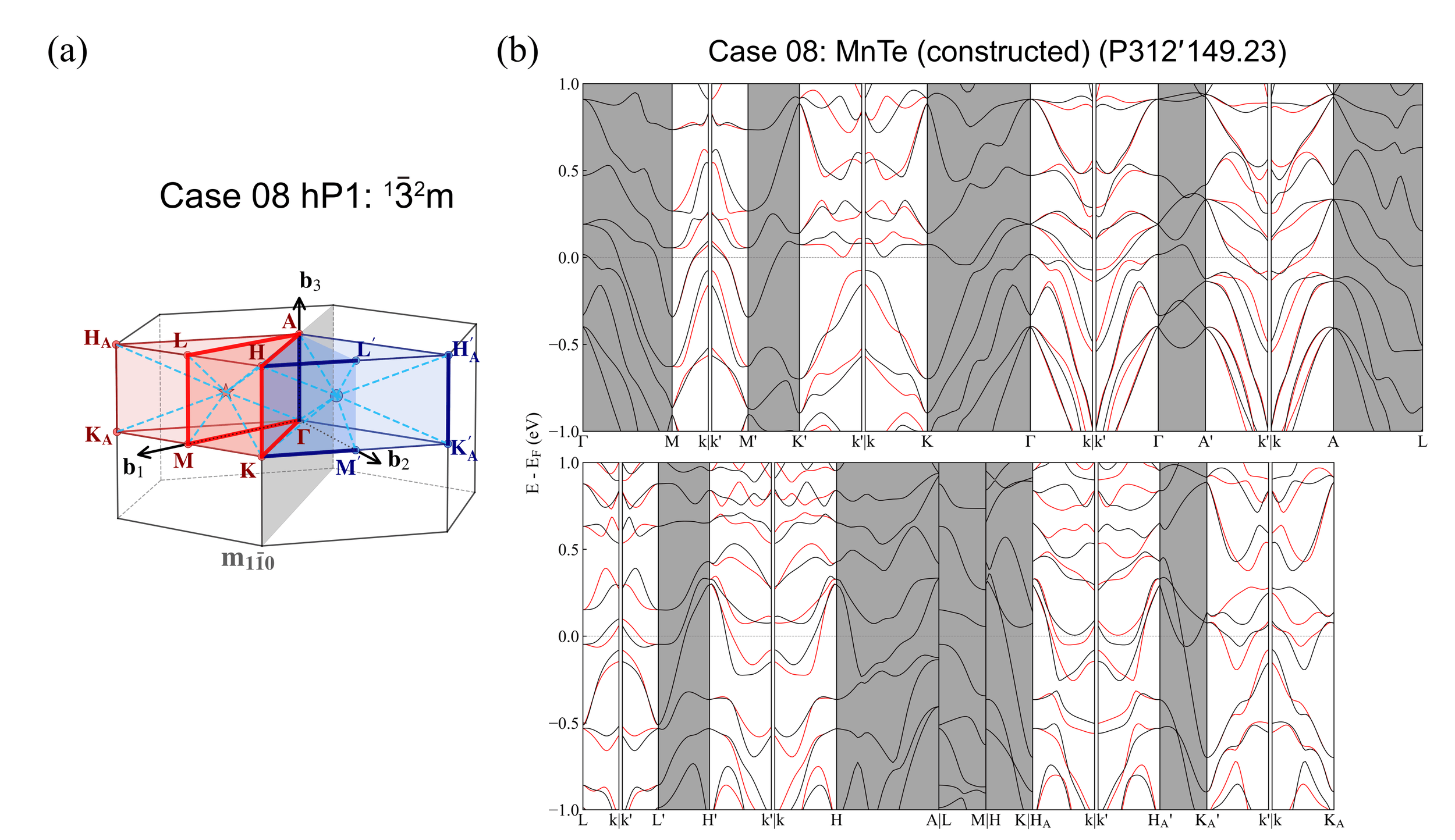}
\caption{Hexagonal hP1 BZ and band for the trigonal spin-Laue-group $^{1}\bar{3}^{2}m$, using MnTe (constructed) (MSG without SOC: $P312'$, BNS 149.23, Type III) as a representative example.}
\label{fig:hP1-3m}
\end{figure}

\begin{equation}
\begin{split}
\Gamma
&\text{--}\underbrace{M\text{--}k\,|\,k'\text{--}M'}_{}\text{--}\underbrace{K'\text{--}k'\,|\,k\text{--}K}_{}\text{--}\underbrace{\Gamma\text{--}k\,|\,k'\text{--}\Gamma}_{}\text{--}\underbrace{A'\text{--}k'\,|\,k\text{--}A}_{}\text{--}\underbrace{L\text{--}k\,|\,k'\text{--}L'}_{}\text{--}\underbrace{H'\text{--}k'\,|\,k\text{--}H}_{}\text{--}A
\\
&\,|\,L\text{--}M\,|\,H\text{--}K\,|\,\underbrace{H_A\text{--}k\,|\,k'\text{--}H_A'}_{}\text{--}\underbrace{K_A'\text{--}k'\,|\,k\text{--}K_A}_{} .
\end{split}
\end{equation}

\subsubsection{hP2}

\noindent\textbf{Case 09, \hpkot{hP2}: ${}^{1}\bar{3}{}^{2}m$, bulk $g$-wave}\par
\begin{figure}[H]
\centering
\includegraphics[width=0.9\textwidth,keepaspectratio]{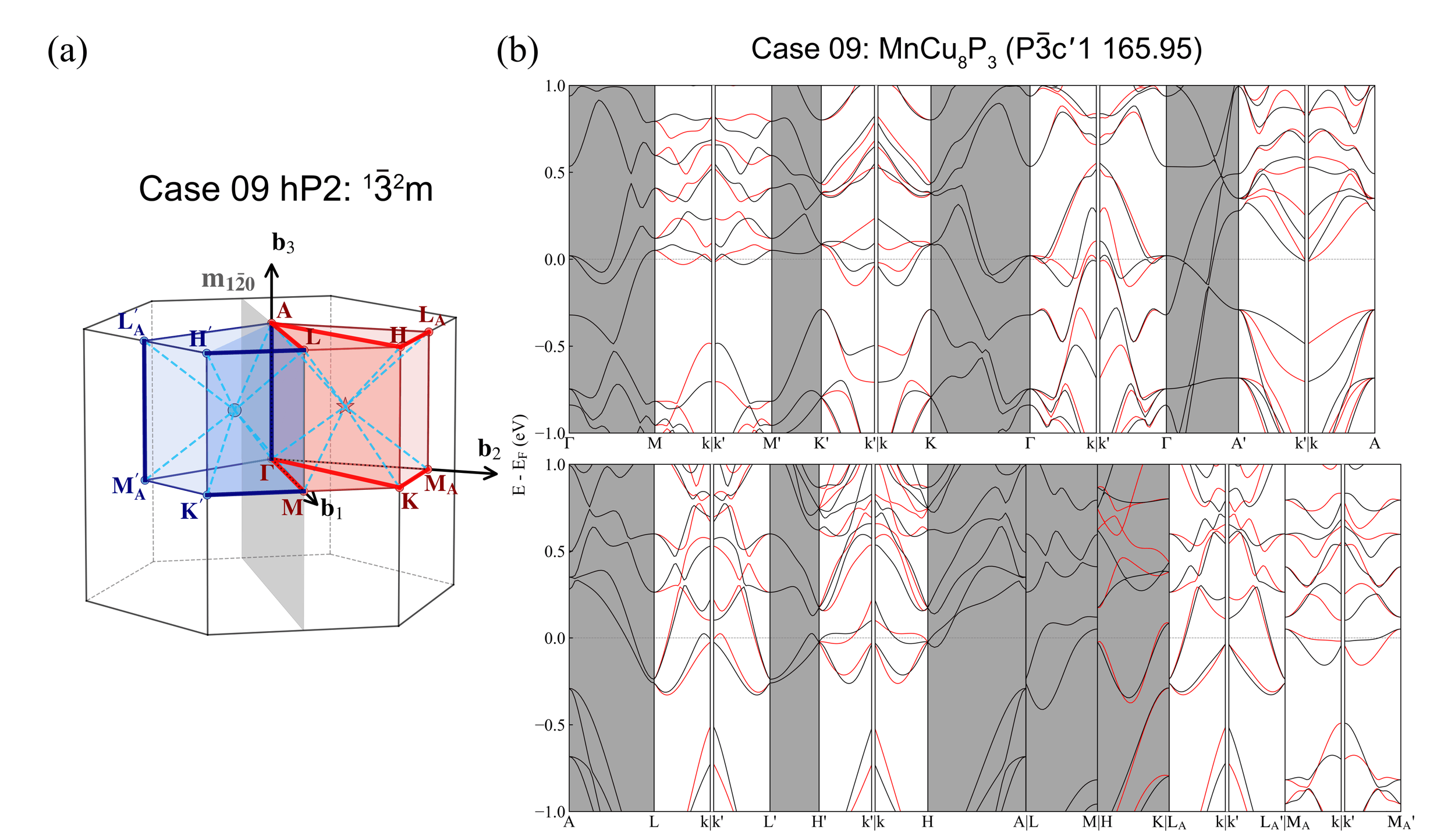}
\caption{Hexagonal hP2 BZ and band for the trigonal spin-Laue-group $^{1}\bar{3}^{2}m$, using MnCu$_8$P$_3$ (MSG without SOC: $P\bar{3}c'1$, BNS 165.95, Type III) as a representative example.}
\label{fig:hP2-3m}
\end{figure}

\begin{equation}
\begin{split}
\Gamma
&\text{--}\underbrace{M\text{--}k\,|\,k'\text{--}M'}_{}\text{--}\underbrace{K'\text{--}k'\,|\,k\text{--}K}_{}\text{--}\underbrace{\Gamma\text{--}k\,|\,k'\text{--}\Gamma}_{}\text{--}\underbrace{A'\text{--}k'\,|\,k\text{--}A}_{}\text{--}\underbrace{L\text{--}k\,|\,k'\text{--}L'}_{}\text{--}\underbrace{H'\text{--}k'\,|\,k\text{--}H}_{}\text{--}A
\\
&\,|\,L\text{--}M\,|\,H\text{--}K\,|\,\underbrace{L_A\text{--}k\,|\,k'\text{--}L_A'}_{}\,|\,\underbrace{M_A\text{--}k\,|\,k'\text{--}M_A'}_{} .
\end{split}
\end{equation}

\subsubsection{hR1}

\noindent\textbf{Case 10, \hpkot{hR1}: ${}^{1}\bar{3}{}^{2}m$, bulk $g$-wave}\par
\begin{figure}[H]
\centering
\includegraphics[width=0.9\textwidth,keepaspectratio]{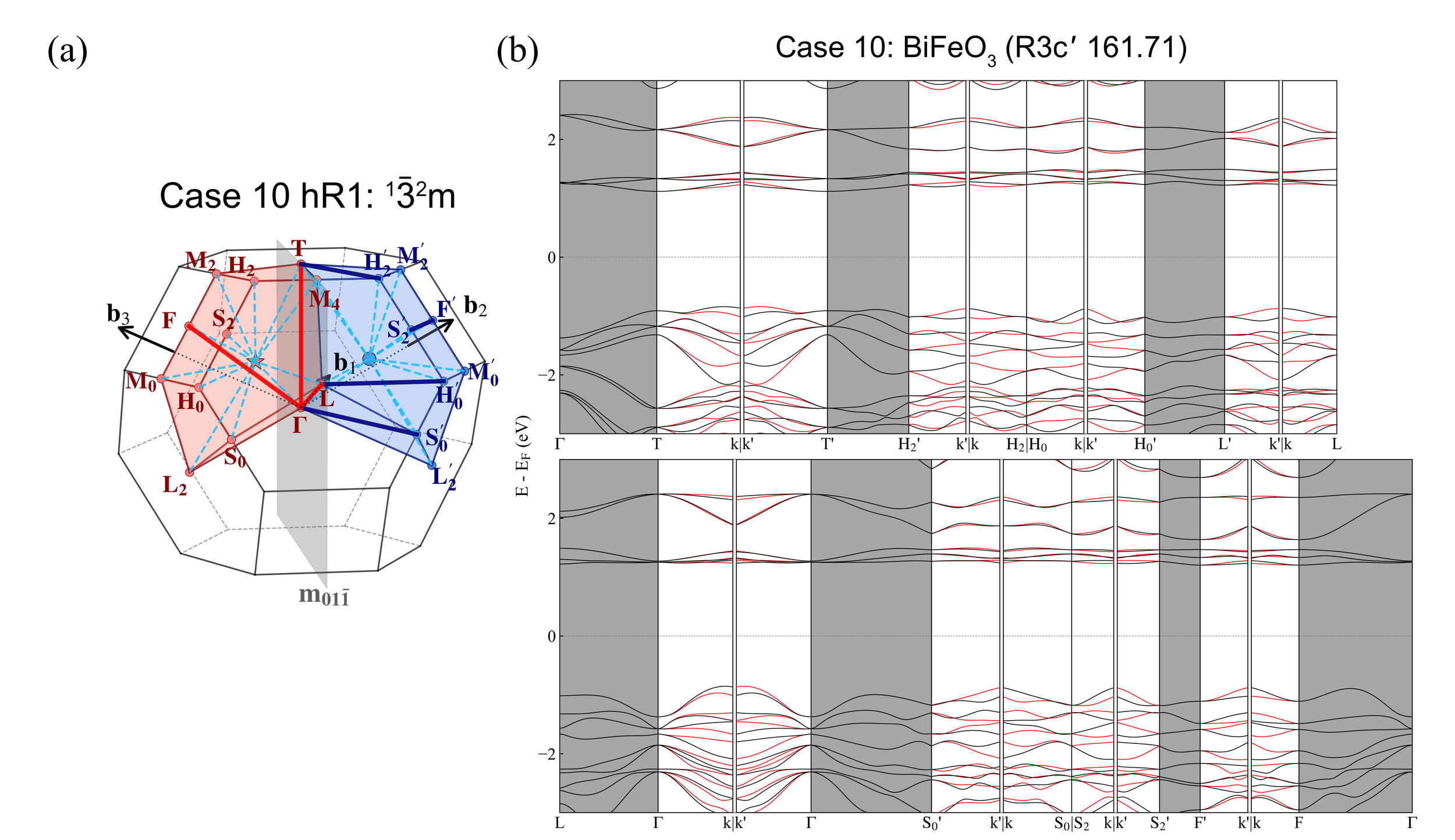}
\caption{Rhombohedral hR1 BZ and band for the trigonal spin-Laue-group $^{1}\bar{3}^{2}m$, using BiFeO$_3$ (MSG without SOC: $R3c'$, BNS 161.71, Type III) as a representative example.}
\label{fig:hR1-3m}
\end{figure}

\begin{equation}
\Gamma\text{--}\underbrace{T\text{--}k\,|\,k'\text{--}T'}_{}\text{--}\underbrace{H_2'\text{--}k'\,|\,k\text{--}H_2}_{}\,|\,\underbrace{H_0\text{--}k\,|\,k'\text{--}H_0'}_{}\text{--}\underbrace{L'\text{--}k'\,|\,k\text{--}L}_{}\text{--}\underbrace{\Gamma\text{--}k\,|\,k'\text{--}\Gamma}_{}\text{--}\underbrace{S_0'\text{--}k'\,|\,k\text{--}S_0}_{}\,|\,\underbrace{S_2\text{--}k\,|\,k'\text{--}S_2'}_{}\text{--}\underbrace{F'\text{--}k'\,|\,k\text{--}F}_{}\text{--}\Gamma .
\end{equation}

\subsubsection{hR2}

\noindent\textbf{Case 11, \hpkot{hR2}: ${}^{1}\bar{3}{}^{2}m$, bulk $g$-wave}\par
\begin{figure}[H]
\centering
\includegraphics[width=0.9\textwidth,keepaspectratio]{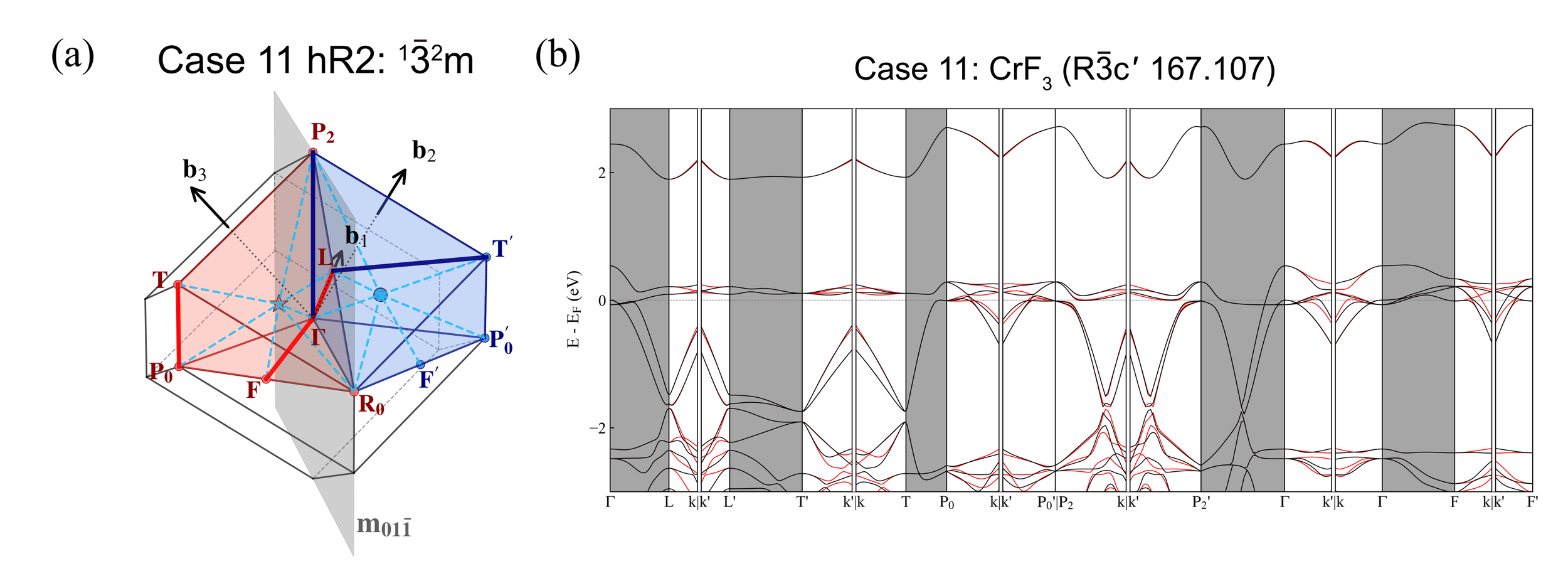}
\caption{Rhombohedral hR2 BZ and band for the trigonal spin-Laue-group $^{1}\bar{3}^{2}m$, using CrF$_3$ (mp-560338) (MSG without SOC: $R\bar{3}c'$, BNS 167.107, Type III) as a representative example.}
\label{fig:hR2-3m}
\end{figure}

\begin{equation}
\Gamma\text{--}\underbrace{L\text{--}k\,|\,k'\text{--}L'}_{}\text{--}\underbrace{T'\text{--}k'\,|\,k\text{--}T}_{}\text{--}\underbrace{P_0\text{--}k\,|\,k'\text{--}P_0'}_{}\,|\,\underbrace{P_2\text{--}k\,|\,k'\text{--}P_2'}_{}\text{--}\underbrace{\Gamma\text{--}k'\,|\,k\text{--}\Gamma}_{}\text{--}\underbrace{F\text{--}k\,|\,k'\text{--}F'}_{} .
\end{equation}

\subsection{Tetragonal}

\subsubsection{tP1 (4/mmm)}
\noindent\textbf{Case 12: ${}^{2}4/{}^{1}m{}^{2}m{}^{1}m$, planar $d$-wave}\par
\noindent\textbf{Case 13: ${}^{2}4/{}^{1}m{}^{1}m{}^{2}m$, planar $d$-wave}\par
\noindent\textbf{Case 14: ${}^{1}4/{}^{1}m{}^{2}m{}^{2}m$, planar $g$-wave}\par
\begin{figure}[H]
\centering
\includegraphics[width=0.9\textwidth,keepaspectratio]{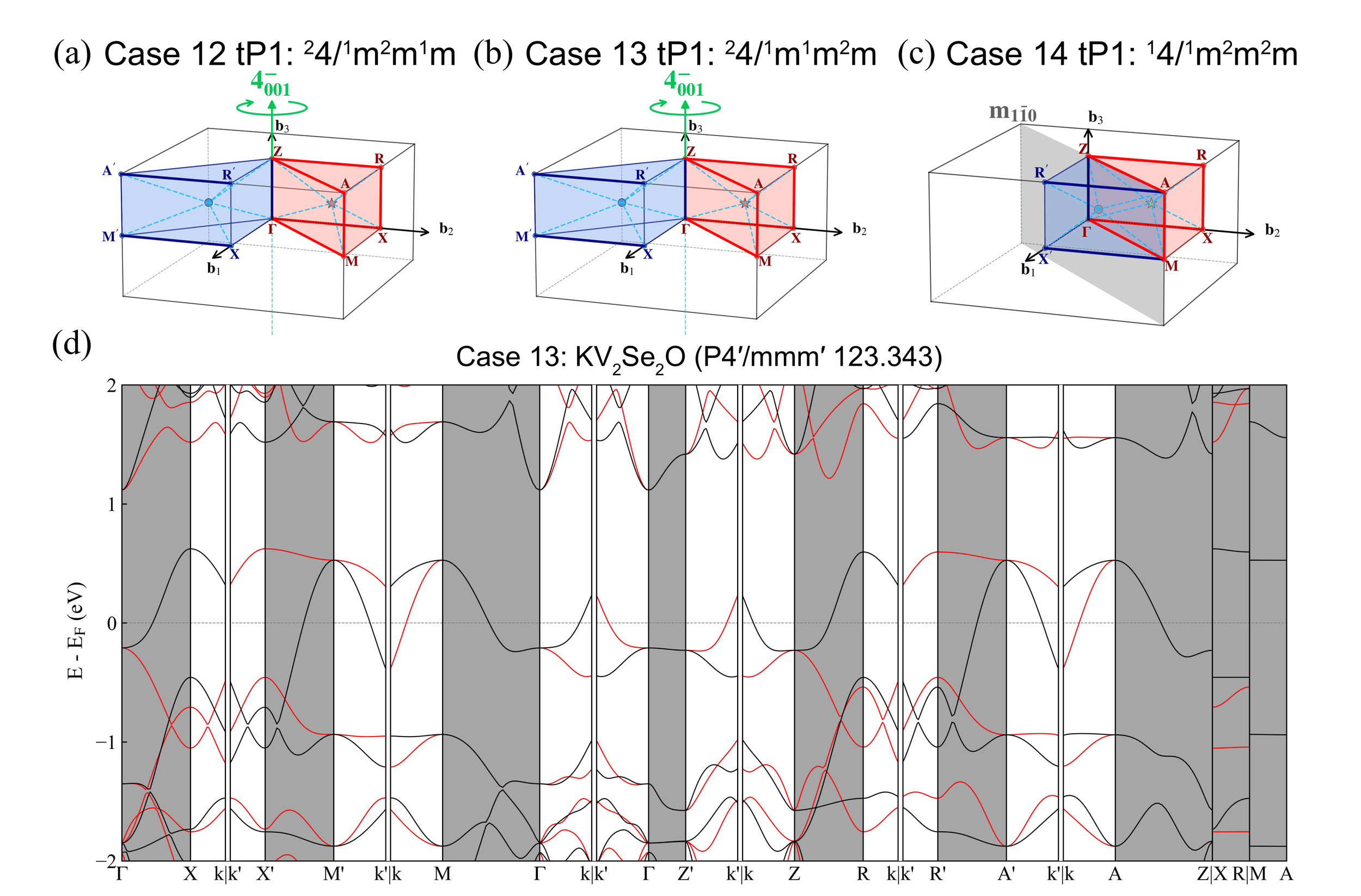}
\caption{Tetragonal tP1 BZ and band for the three spin Laue groups of the $4/mmm$ Laue group, using KV$_2$Se$_2$O (MSG without SOC: $P4'/mmm'$, BNS 123.343, Type III) as a representative example. (a)--(c) show Cases 12--14; since all three share the same k-path, only one band structure is shown, corresponding to Case 13 ($^{2}4/^{1}m^{1}m^{2}m$).}
\label{fig:tP1-4/mmm}
\end{figure}

\begin{equation}
\Gamma\text{--}\underbrace{X\text{--}k\,|\,k'\text{--}X'}_{}\text{--}\underbrace{M'\text{--}k'\,|\,k\text{--}M}_{}\text{--}\underbrace{\Gamma\text{--}k\,|\,k'\text{--}\Gamma}_{}\text{--}\underbrace{Z'\text{--}k'\,|\,k\text{--}Z}_{}\text{--}\underbrace{R\text{--}k\,|\,k'\text{--}R'}_{}\text{--}\underbrace{A'\text{--}k'\,|\,k\text{--}A}_{}\text{--}Z\,|\,X\text{--}R\,|\,M\text{--}A .
\end{equation}

\subsubsection{tI1 (4/mmm)}
\noindent\textbf{Case 15: ${}^{2}4/{}^{1}m{}^{2}m{}^{1}m$, planar $d$-wave}\par
\noindent\textbf{Case 16: ${}^{2}4/{}^{1}m{}^{1}m{}^{2}m$, planar $d$-wave}\par
\noindent\textbf{Case 17: ${}^{1}4/{}^{1}m{}^{2}m{}^{2}m$, planar $g$-wave}\par
\begin{figure}[H]
\centering
\includegraphics[width=0.9\textwidth,keepaspectratio]{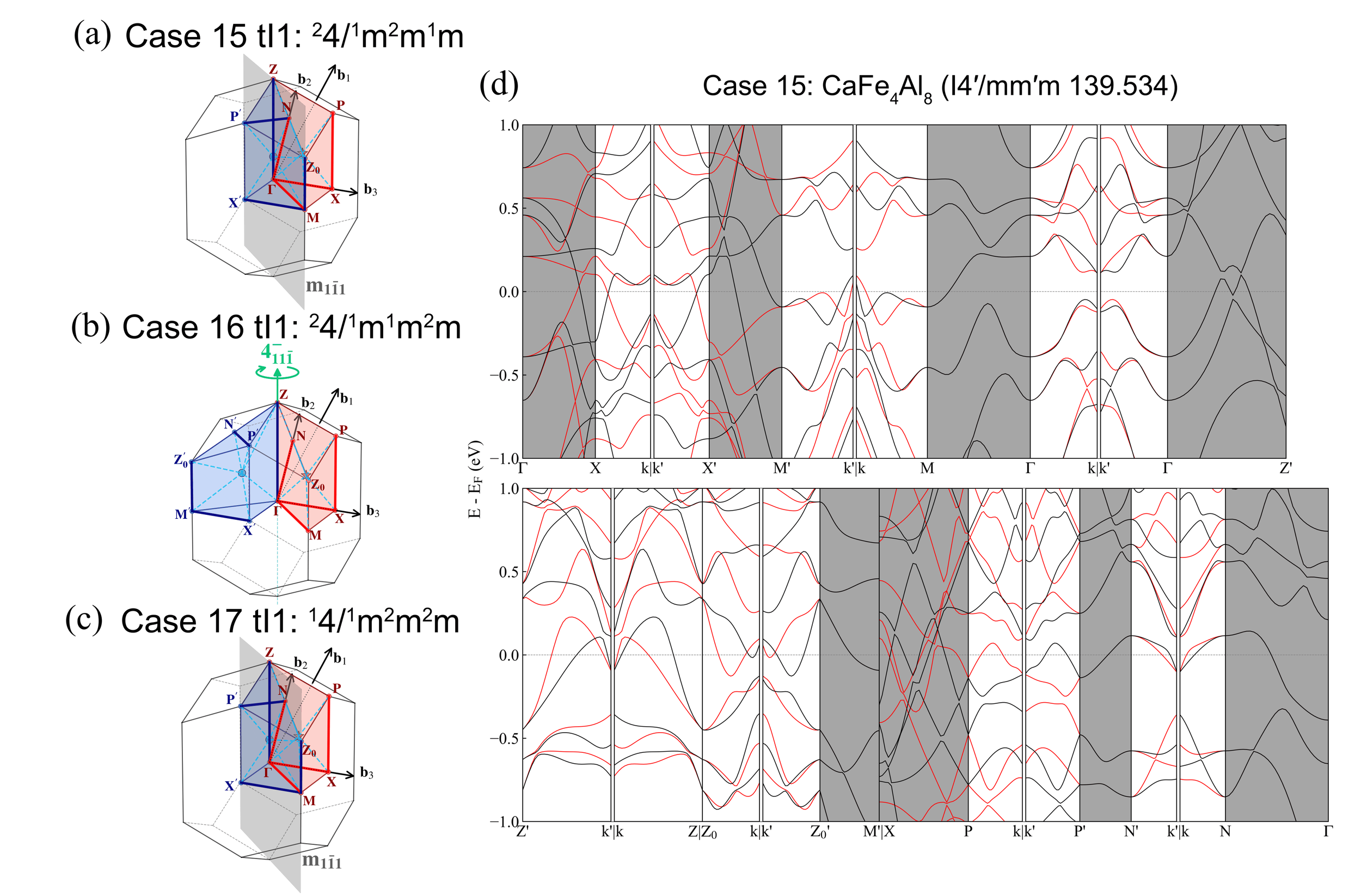}
\caption{Tetragonal tI1 BZ and band for the three spin Laue groups of the $4/mmm$ Laue group, using CaFe$_4$Al$_8$ (MAGNDATA 0.236) (MSG without SOC: $I4'/mm'm$, BNS 139.534, Type III) as a representative example. (a)--(c) show Cases 15--17; since all three share the same k-path, only one band structure is shown, corresponding to Case 15 ($^{2}4/^{1}m^{2}m^{1}m$).}
\label{fig:tI1-4/mmm}
\end{figure}

\begin{equation}
\Gamma\text{--}\underbrace{X\text{--}k\,|\,k'\text{--}X'}_{}\text{--}\underbrace{M'\text{--}k'\,|\,k\text{--}M}_{}\text{--}\underbrace{\Gamma\text{--}k\,|\,k'\text{--}\Gamma}_{}\text{--}\underbrace{Z'\text{--}k'\,|\,k\text{--}Z}_{}\,|\,\underbrace{Z_0\text{--}k\,|\,k'\text{--}Z_0'}_{}\text{--}M'\,|\,X\text{--}\underbrace{P\text{--}k\,|\,k'\text{--}P'}_{}\text{--}\underbrace{N'\text{--}k'\,|\,k\text{--}N}_{}\text{--}\Gamma .
\end{equation}

\subsubsection{tI2 (4/mmm)}
\noindent\textbf{Case 18: ${}^{2}4/{}^{1}m{}^{2}m{}^{1}m$, planar $d$-wave}\par
\noindent\textbf{Case 19: ${}^{2}4/{}^{1}m{}^{1}m{}^{2}m$, planar $d$-wave}\par
\noindent\textbf{Case 20: ${}^{1}4/{}^{1}m{}^{2}m{}^{2}m$, planar $g$-wave}\par
\begin{figure}[H]
\centering
\includegraphics[width=0.9\textwidth,keepaspectratio]{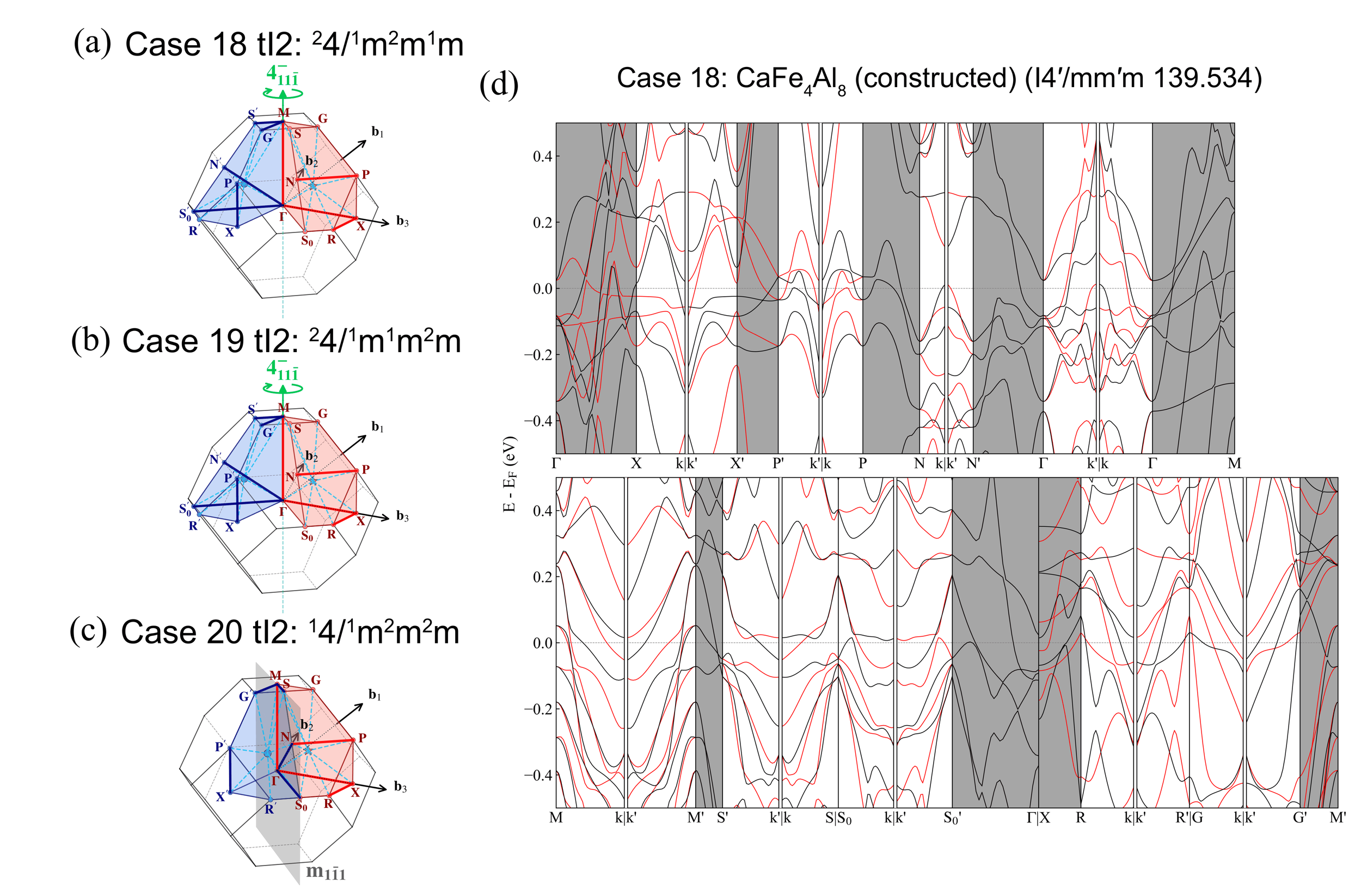}
\caption{Tetragonal tI2 BZ and band for the three spin Laue groups of the $4/mmm$ Laue group, using CaFe$_4$Al$_8$ (constructed, based on MAGNDATA 0.236, made by letting $c>a$) (MSG without SOC: $I4'/mm'm$, BNS 139.534, Type III) as a representative example. (a)--(c) show Cases 18--20; since all three share the same k-path, only one band structure is shown, corresponding to Case 18 ($^{2}4/^{1}m^{2}m^{1}m$).}
\label{fig:tI2-4/mmm}
\end{figure}

\begin{equation}
\begin{split}
\Gamma
&\text{--}\underbrace{X\text{--}k\,|\,k'\text{--}X'}_{}\text{--}\underbrace{P'\text{--}k'\,|\,k\text{--}P}_{}\text{--}\underbrace{N\text{--}k\,|\,k'\text{--}N'}_{}\text{--}\underbrace{\Gamma\text{--}k'\,|\,k\text{--}\Gamma}_{}\text{--}\underbrace{M\text{--}k\,|\,k'\text{--}M'}_{}\text{--}\underbrace{S'\text{--}k'\,|\,k\text{--}S}_{}
\\
&\,|\,\underbrace{S_0\text{--}k\,|\,k'\text{--}S_0'}_{}\text{--}\Gamma\,|\,X\text{--}\underbrace{R\text{--}k\,|\,k'\text{--}R'}_{}\,|\,\underbrace{G\text{--}k\,|\,k'\text{--}G'}_{}\text{--}M' .
\end{split}
\end{equation}

\subsubsection{tP1 (4/m)}

\noindent\textbf{Case 21, \hpkot{tP1}: ${}^{2}4/{}^{1}m$, planar $d$-wave}\par
\begin{figure}[H]
\centering
\includegraphics[width=0.9\textwidth,keepaspectratio]{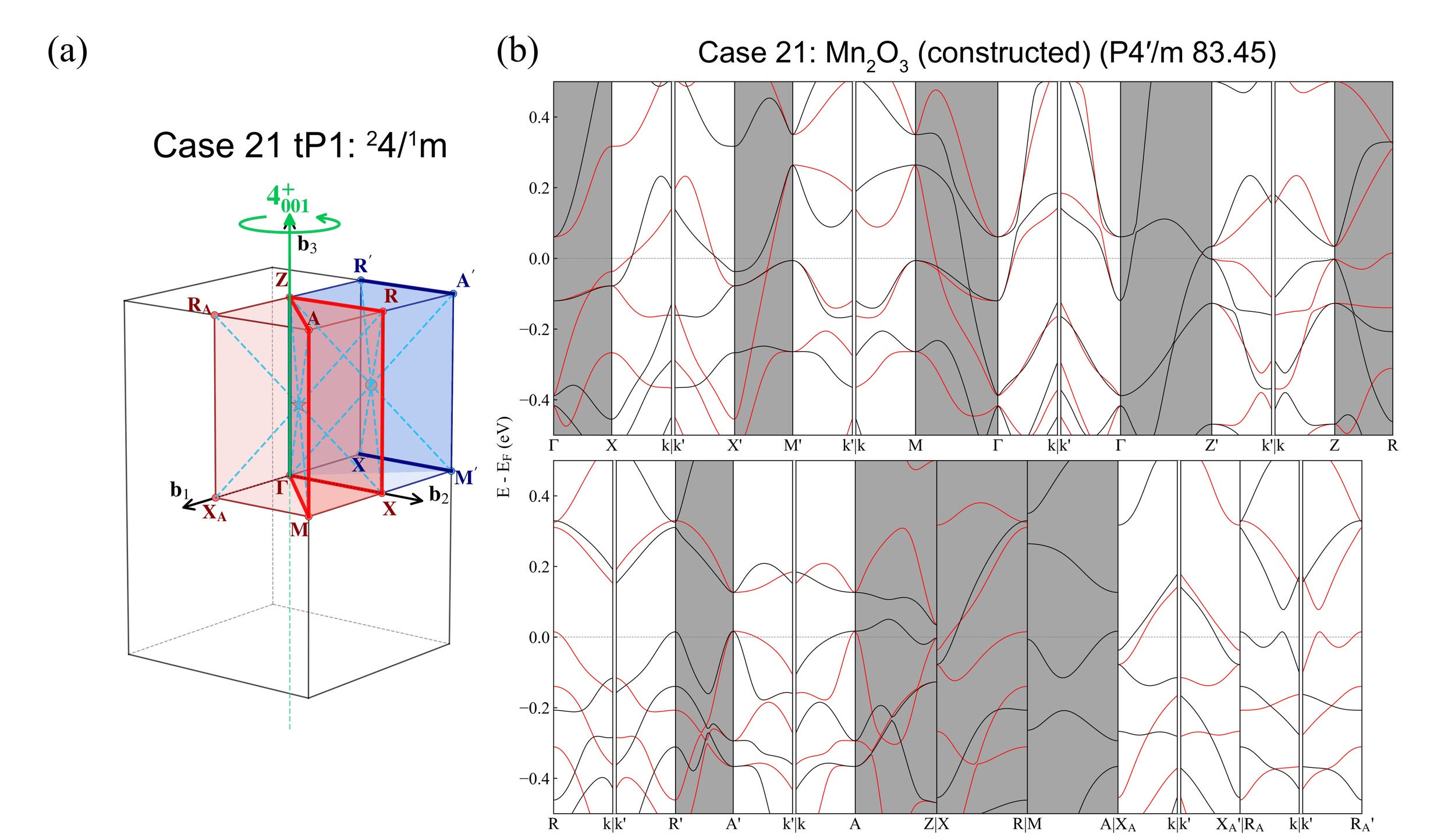}
\caption{Tetragonal tP1 BZ and band for the spin-Laue-group $^{2}4/^{1}m$, using Mn$_2$O$_3$ (constructed) (MSG without SOC: $P4'/m$, BNS 83.45, Type III) as a representative example.}
\label{fig:tP1-4/m}
\end{figure}

\begin{equation}
\begin{split}
\Gamma
&\text{--}\underbrace{X\text{--}k\,|\,k'\text{--}X'}_{}\text{--}\underbrace{M'\text{--}k'\,|\,k\text{--}M}_{}\text{--}\underbrace{\Gamma\text{--}k\,|\,k'\text{--}\Gamma}_{}\text{--}\underbrace{Z'\text{--}k'\,|\,k\text{--}Z}_{}\text{--}\underbrace{R\text{--}k\,|\,k'\text{--}R'}_{}\text{--}\underbrace{A'\text{--}k'\,|\,k\text{--}A}_{}\text{--}Z
\\
&\,|\,X\text{--}R\,|\,M\text{--}A\,|\,\underbrace{X_A\text{--}k\,|\,k'\text{--}X_A'}_{}\,|\,\underbrace{R_A\text{--}k\,|\,k'\text{--}R_A'}_{} .
\end{split}
\end{equation}

\subsubsection{tI1 (4/m)}

\noindent\textbf{Case 22, \hpkot{tI1}: ${}^{2}4/{}^{1}m$, planar $d$-wave}\par
\begin{figure}[H]
\centering
\includegraphics[width=0.9\textwidth,keepaspectratio]{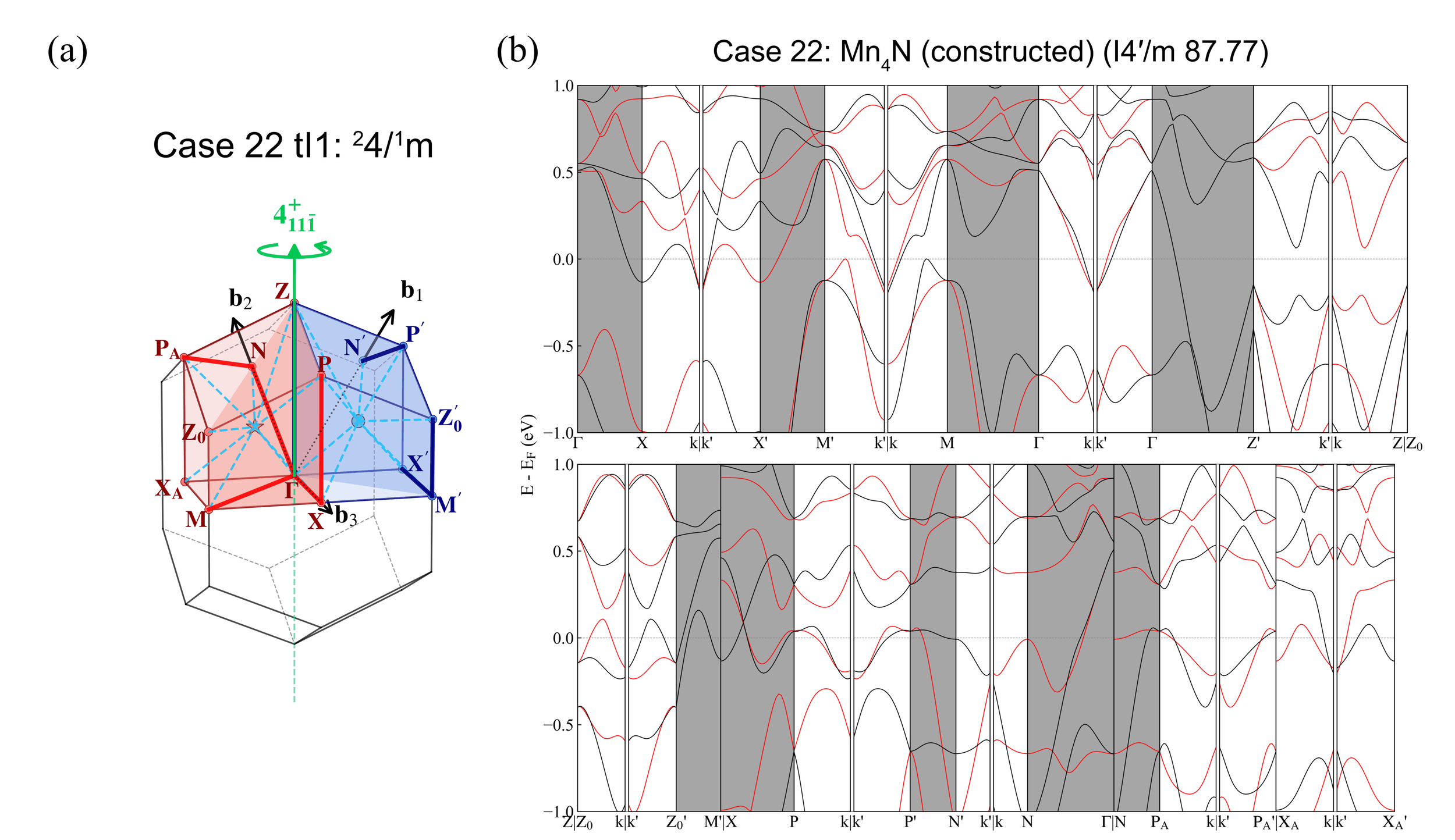}
\caption{Tetragonal tI1 BZ and band for the spin-Laue-group $^{2}4/^{1}m$, using Mn$_4$N (constructed) (MSG without SOC: $I4'/m$, BNS 87.77, Type III) as a representative example.}
\label{fig:tI1-4/m}
\end{figure}

\begin{equation}
\begin{split}
\Gamma
&\text{--}\underbrace{X\text{--}k\,|\,k'\text{--}X'}_{}\text{--}\underbrace{M'\text{--}k'\,|\,k\text{--}M}_{}\text{--}\underbrace{\Gamma\text{--}k\,|\,k'\text{--}\Gamma}_{}\text{--}\underbrace{Z'\text{--}k'\,|\,k\text{--}Z}_{}\,|\,\underbrace{Z_0\text{--}k\,|\,k'\text{--}Z_0'}_{}\text{--}M'\,|\,X
\\
&\text{--}\underbrace{P\text{--}k\,|\,k'\text{--}P'}_{}\text{--}\underbrace{N'\text{--}k'\,|\,k\text{--}N}_{}\text{--}\Gamma\,|\,N\text{--}\underbrace{P_A\text{--}k\,|\,k'\text{--}P_A'}_{}\,|\,\underbrace{X_A\text{--}k\,|\,k'\text{--}X_A'}_{} .
\end{split}
\end{equation}

\subsubsection{tI2 (4/m)}

\noindent\textbf{Case 23, \hpkot{tI2}: ${}^{2}4/{}^{1}m$, planar $d$-wave}\par
\begin{figure}[H]
\centering
\includegraphics[width=0.9\textwidth,keepaspectratio]{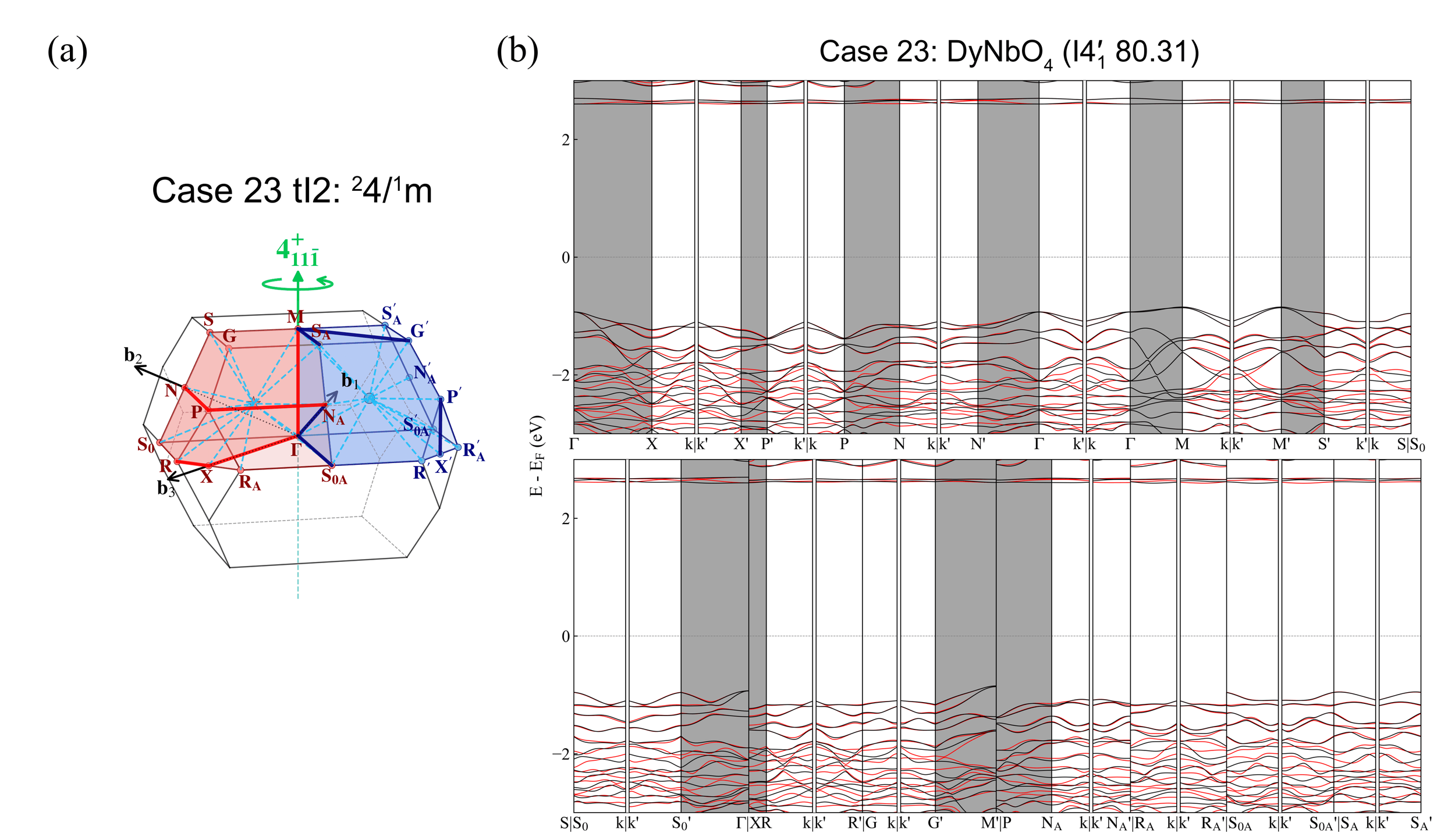}
\caption{Tetragonal tI2 BZ and band for the spin-Laue-group $^{2}4/^{1}m$, using DyNbO$_4$ (mp-1440576) (MSG without SOC: $I4_1'$, BNS 80.31, Type III) as a representative example.}
\label{fig:tI2-4/m}
\end{figure}

\begin{equation}
\begin{split}
\Gamma
&\text{--}\underbrace{X\text{--}k\,|\,k'\text{--}X'}_{}\text{--}\underbrace{P'\text{--}k'\,|\,k\text{--}P}_{}\text{--}\underbrace{N\text{--}k\,|\,k'\text{--}N'}_{}\text{--}\underbrace{\Gamma\text{--}k'\,|\,k\text{--}\Gamma}_{}\text{--}\underbrace{M\text{--}k\,|\,k'\text{--}M'}_{}\text{--}\underbrace{S'\text{--}k'\,|\,k\text{--}S}_{}\,|\,\underbrace{S_0\text{--}k\,|\,k'\text{--}S_0'}_{}\text{--}\Gamma\,|\,X
\\
&\text{--}\underbrace{R\text{--}k\,|\,k'\text{--}R'}_{}\,|\,\underbrace{G\text{--}k\,|\,k'\text{--}G'}_{}\text{--}M'\,|\,P\text{--}\underbrace{N_A\text{--}k\,|\,k'\text{--}N_A'}_{}\,|\,\underbrace{R_A\text{--}k\,|\,k'\text{--}R_A'}_{}\,|\,\underbrace{S_{0A}\text{--}k\,|\,k'\text{--}S_{0A}'}_{}\,|\,\underbrace{S_A\text{--}k\,|\,k'\text{--}S_A'}_{} .
\end{split}
\end{equation}

\subsection{Orthorhombic}

\subsubsection{oP1}
\noindent\textbf{Case 24: ${}^{2}m{}^{2}m{}^{1}m$, planar $d$-wave}\par
\noindent\textbf{Case 25: ${}^{2}m{}^{1}m{}^{2}m$, bulk $d$-wave}\par
\noindent\textbf{Case 26: ${}^{1}m{}^{2}m{}^{2}m$, bulk $d$-wave}\par
\begin{figure}[H]
\centering
\includegraphics[width=0.9\textwidth,keepaspectratio]{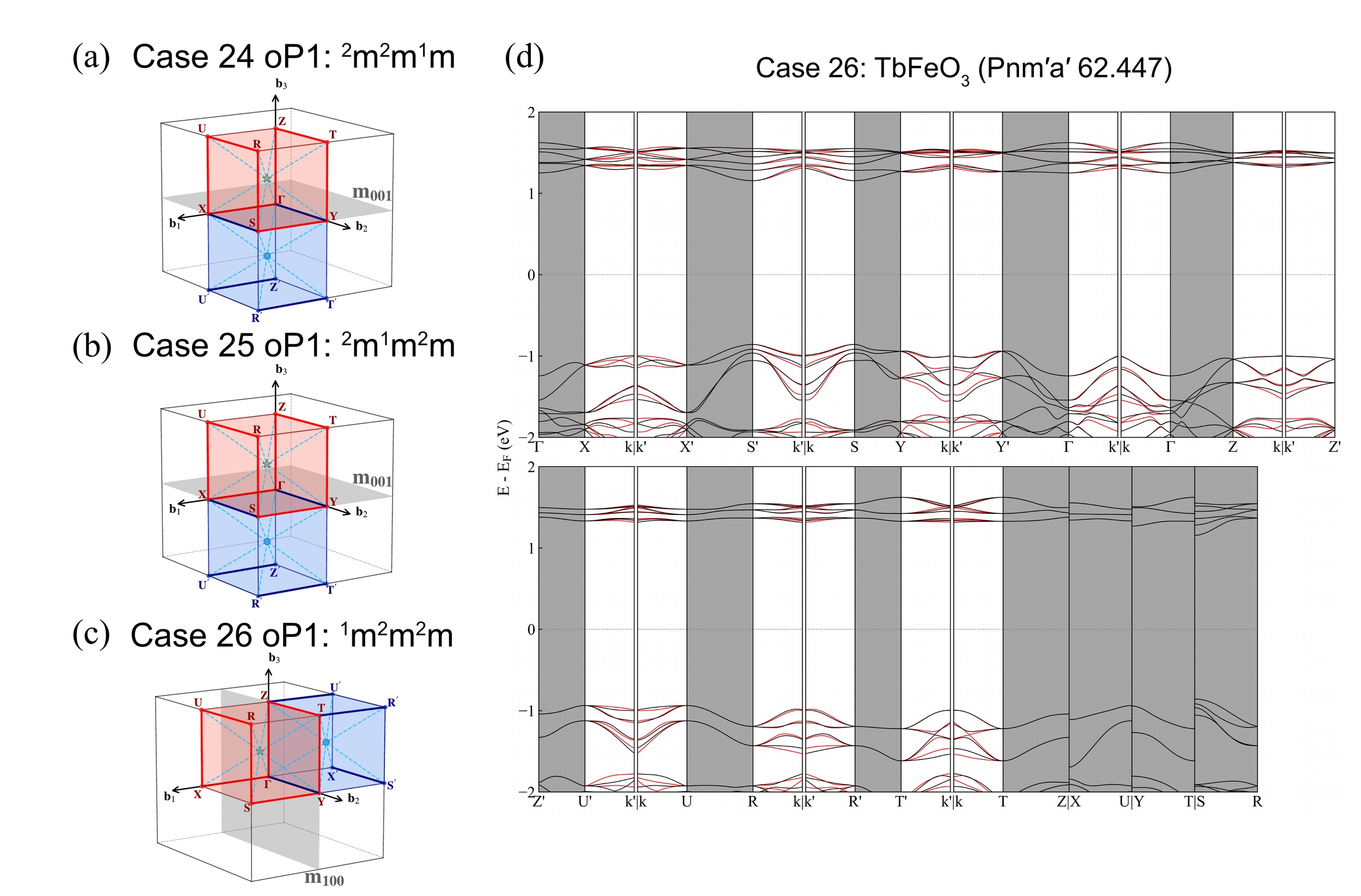}
\caption{Orthorhombic oP1 BZ and band for the three spin Laue groups of the $mmm$ Laue group, using TbFeO$_3$ (MAGNDATA 0.351) (MSG without SOC: $Pnm'a'$, BNS 62.447, Type III) as a representative example. (a)--(c) show Cases 24--26; since all three share the same k-path, only one band structure is shown, corresponding to Case 26 ($^{1}m^{2}m^{2}m$).}
\label{fig:oP1}
\end{figure}

\begin{equation}
\begin{split}
\Gamma
&\text{--}\underbrace{X\text{--}k\,|\,k'\text{--}X'}_{}\text{--}\underbrace{S'\text{--}k'\,|\,k\text{--}S}_{}\text{--}\underbrace{Y\text{--}k\,|\,k'\text{--}Y'}_{}\text{--}\underbrace{\Gamma\text{--}k'\,|\,k\text{--}\Gamma}_{}\text{--}\underbrace{Z\text{--}k\,|\,k'\text{--}Z'}_{}
\\
&\text{--}\underbrace{U'\text{--}k'\,|\,k\text{--}U}_{}\text{--}\underbrace{R\text{--}k\,|\,k'\text{--}R'}_{}
\text{--}\underbrace{T'\text{--}k'\,|\,k\text{--}T}_{}\text{--}Z\,|\,X\text{--}U\,|\,Y\text{--}T\,|\,S\text{--}R .
\end{split}
\end{equation}

\subsubsection{oI1}
\noindent\textbf{Case 27: ${}^{2}m{}^{2}m{}^{1}m$, planar $d$-wave}\par
\noindent\textbf{Case 28: ${}^{2}m{}^{1}m{}^{2}m$, bulk $d$-wave}\par
\noindent\textbf{Case 29: ${}^{1}m{}^{2}m{}^{2}m$, bulk $d$-wave}\par
\begin{figure}[H]
\centering
\includegraphics[width=0.9\textwidth,keepaspectratio]{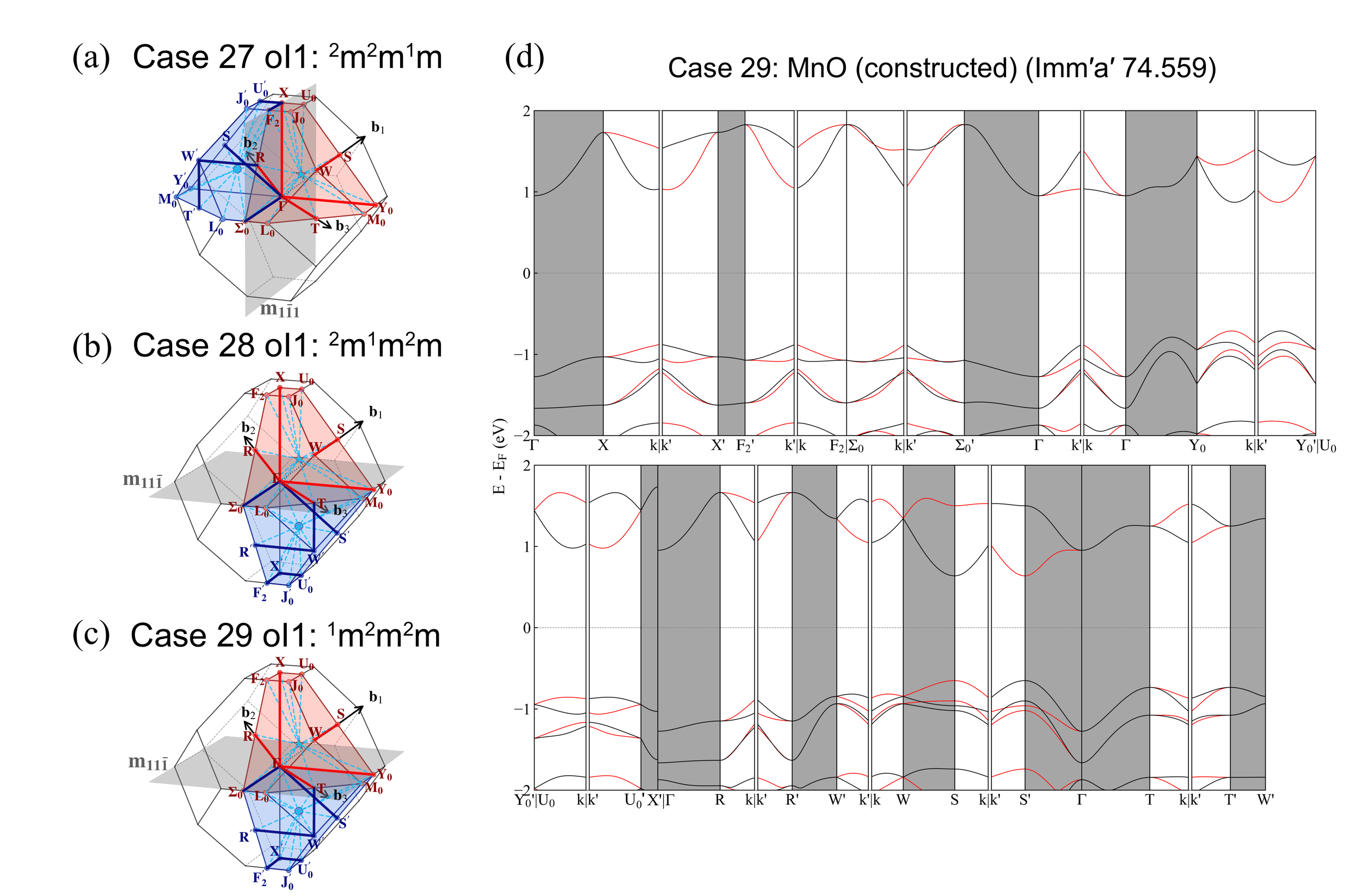}
\caption{Orthorhombic oI1 BZ and band for the three spin Laue groups of the $mmm$ Laue group, using MnO (constructed) (MSG without SOC: $Imm'a'$, BNS 74.559, Type III) as a representative example. (a)--(c) show Cases 27--29; since all three share the same k-path, only one band structure is shown, corresponding to Case 29 ($^{1}m^{2}m^{2}m$).}
\label{fig:oI1}
\end{figure}

\begin{equation}
\begin{split}
\Gamma
&\text{--}\underbrace{X\text{--}k\,|\,k'\text{--}X'}_{}\text{--}\underbrace{F_2'\text{--}k'\,|\,k\text{--}F_2}_{}\,|\,\underbrace{\Sigma_0\text{--}k\,|\,k'\text{--}\Sigma_0'}_{}\text{--}\underbrace{\Gamma\text{--}k'\,|\,k\text{--}\Gamma}_{}\text{--}\underbrace{Y_0\text{--}k\,|\,k'\text{--}Y_0'}_{}\,|\,\underbrace{U_0\text{--}k\,|\,k'\text{--}U_0'}_{}\text{--}X'
\\
&\,|\,\Gamma\text{--}\underbrace{R\text{--}k\,|\,k'\text{--}R'}_{}\text{--}\underbrace{W'\text{--}k'\,|\,k\text{--}W}_{}\text{--}\underbrace{S\text{--}k\,|\,k'\text{--}S'}_{}\text{--}\Gamma\text{--}\underbrace{T\text{--}k\,|\,k'\text{--}T'}_{}\text{--}W' .
\end{split}
\end{equation}

\subsubsection{oI2}
\noindent\textbf{Case 30: ${}^{2}m{}^{2}m{}^{1}m$, planar $d$-wave}\par
\noindent\textbf{Case 31: ${}^{2}m{}^{1}m{}^{2}m$, bulk $d$-wave}\par
\noindent\textbf{Case 32: ${}^{1}m{}^{2}m{}^{2}m$, bulk $d$-wave}\par
\begin{figure}[H]
\centering
\includegraphics[width=0.9\textwidth,keepaspectratio]{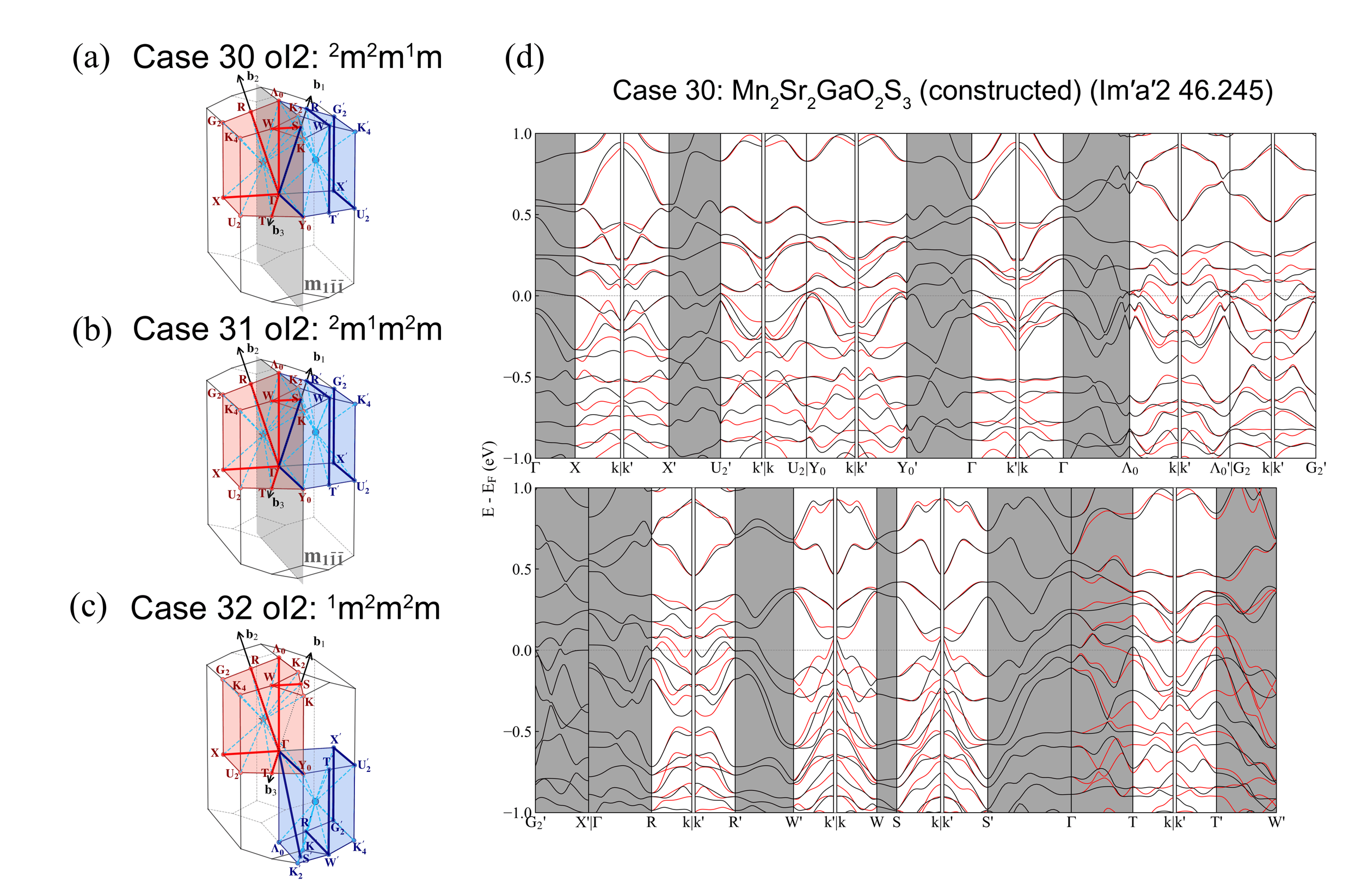}
\caption{Orthorhombic oI2 BZ and band for the three spin Laue groups of the $mmm$ Laue group, using Mn$_2$Sr$_2$GaO$_2$S$_3$ (constructed, based on MAGNDATA 0.823) (MSG without SOC: $Im'a'2$, BNS 46.245, Type III) as a representative example. (a)--(c) show Cases 30--32; since all three share the same k-path, only one band structure is shown, corresponding to Case 30 ($^{2}m^{2}m^{1}m$). The spin-flip operation chosen in (c) is $2_{1\bar{1}1}$ with rotation axis along $\Gamma$--$Y_0$.}
\label{fig:oI2}
\end{figure}

\begin{equation}
\begin{split}
\Gamma
&\text{--}\underbrace{X\text{--}k\,|\,k'\text{--}X'}_{}\text{--}\underbrace{U_2'\text{--}k'\,|\,k\text{--}U_2}_{}\,|\,\underbrace{Y_0\text{--}k\,|\,k'\text{--}Y_0'}_{}\text{--}\underbrace{\Gamma\text{--}k'\,|\,k\text{--}\Gamma}_{}\text{--}\underbrace{\Lambda_0\text{--}k\,|\,k'\text{--}\Lambda_0'}_{}\,|\,\underbrace{G_2\text{--}k\,|\,k'\text{--}G_2'}_{}\text{--}X'
\\
&\,|\,\Gamma\text{--}\underbrace{R\text{--}k\,|\,k'\text{--}R'}_{}\text{--}\underbrace{W'\text{--}k'\,|\,k\text{--}W}_{}\text{--}\underbrace{S\text{--}k\,|\,k'\text{--}S'}_{}\text{--}\Gamma\text{--}\underbrace{T\text{--}k\,|\,k'\text{--}T'}_{}\text{--}W' .
\end{split}
\end{equation}

\subsubsection{oI3}
\noindent\textbf{Case 33: ${}^{2}m{}^{2}m{}^{1}m$, planar $d$-wave}\par
\noindent\textbf{Case 34: ${}^{2}m{}^{1}m{}^{2}m$, bulk $d$-wave}\par
\noindent\textbf{Case 35: ${}^{1}m{}^{2}m{}^{2}m$, bulk $d$-wave}\par
\begin{figure}[H]
\centering
\includegraphics[width=0.9\textwidth,keepaspectratio]{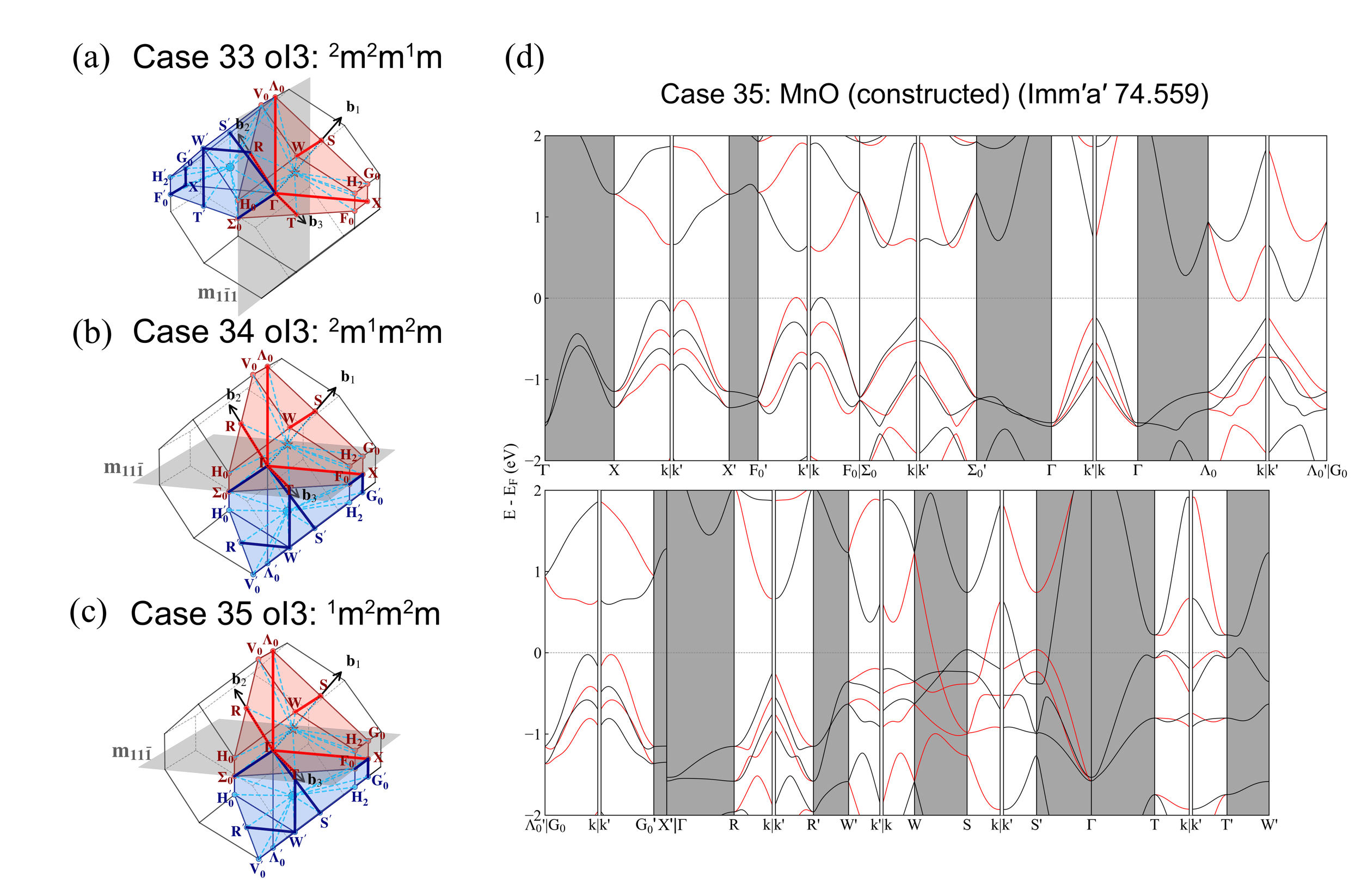}
\caption{Orthorhombic oI3 BZ and band for the three spin Laue groups of the $mmm$ Laue group, using MnO (constructed) (MSG without SOC: $Imm'a'$, BNS 74.559, Type III) as a representative example. (a)--(c) show Cases 33--35; since all three share the same k-path, only one band structure is shown, corresponding to Case 35 ($^{1}m^{2}m^{2}m$).}
\label{fig:oI3}
\end{figure}

\begin{equation}
\begin{split}
\Gamma
&\text{--}\underbrace{X\text{--}k\,|\,k'\text{--}X'}_{}\text{--}\underbrace{F_0'\text{--}k'\,|\,k\text{--}F_0}_{}\,|\,\underbrace{\Sigma_0\text{--}k\,|\,k'\text{--}\Sigma_0'}_{}\text{--}\underbrace{\Gamma\text{--}k'\,|\,k\text{--}\Gamma}_{}\text{--}\underbrace{\Lambda_0\text{--}k\,|\,k'\text{--}\Lambda_0'}_{}\,|\,\underbrace{G_0\text{--}k\,|\,k'\text{--}G_0'}_{}\text{--}X'
\\
&\,|\,\Gamma\text{--}\underbrace{R\text{--}k\,|\,k'\text{--}R'}_{}\text{--}\underbrace{W'\text{--}k'\,|\,k\text{--}W}_{}\text{--}\underbrace{S\text{--}k\,|\,k'\text{--}S'}_{}\text{--}\Gamma\text{--}\underbrace{T\text{--}k\,|\,k'\text{--}T'}_{}\text{--}W' .
\end{split}
\end{equation}

\subsubsection{oC1 (oA1)}
\noindent\textbf{Case 36: ${}^{2}m{}^{2}m{}^{1}m$, planar $d$-wave}\par
\noindent\textbf{Case 37: ${}^{2}m{}^{1}m{}^{2}m$, bulk $d$-wave}\par
\noindent\textbf{Case 38: ${}^{1}m{}^{2}m{}^{2}m$, bulk $d$-wave}\par
\begin{figure}[H]
\centering
\includegraphics[width=0.9\textwidth,keepaspectratio]{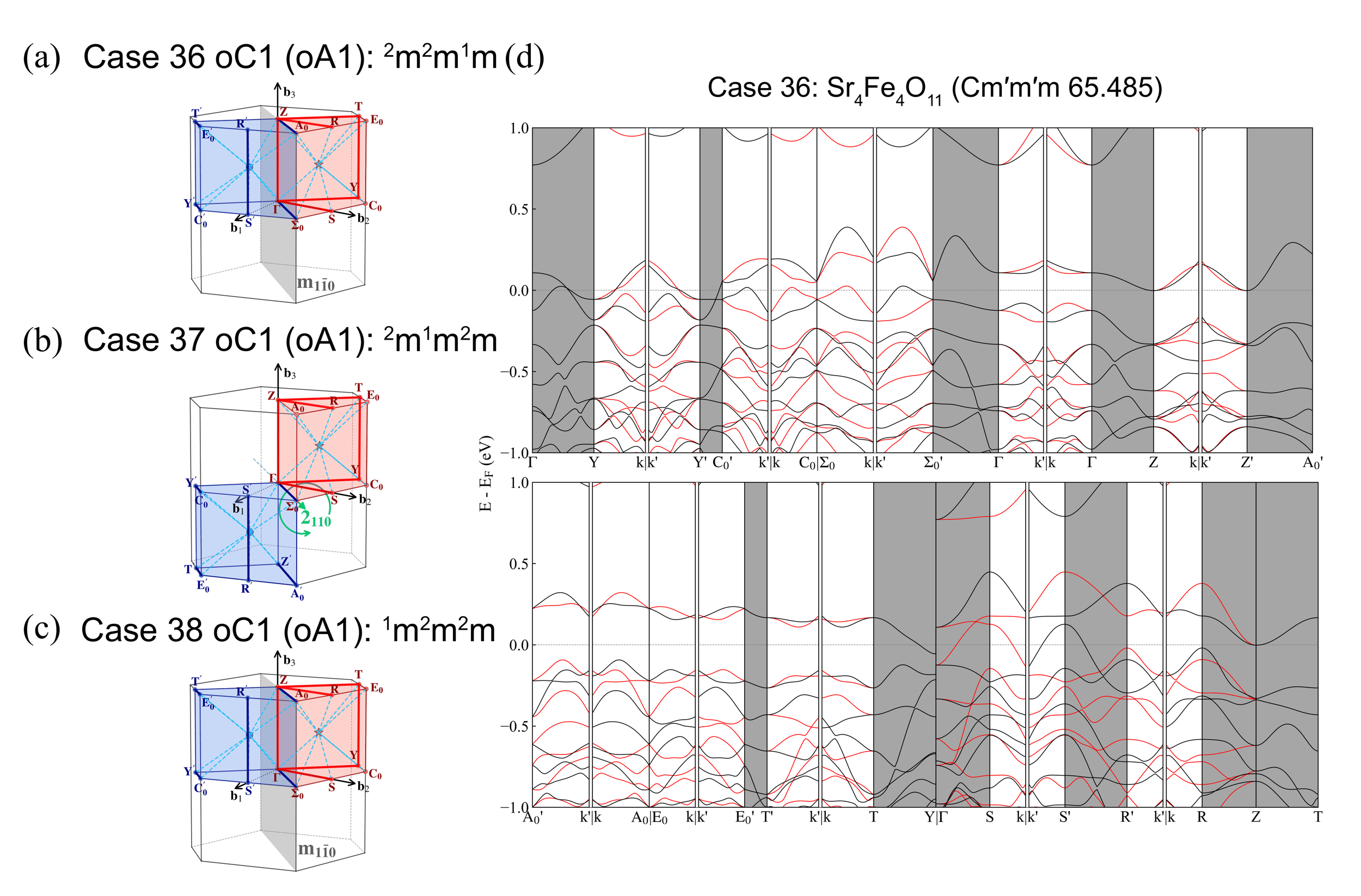}
\caption{Orthorhombic oC1 (oA1) BZ and band for the three spin Laue groups of the $mmm$ Laue group, using Sr$_4$Fe$_4$O$_{11}$ (MAGNDATA 0.402) (MSG without SOC: $Cm'm'm$, BNS 65.485, Type III) as a representative example. (a)--(c) show Cases 36--38; since all three share the same k-path, only one band structure is shown, corresponding to Case 36 ($^{2}m^{2}m^{1}m$).}
\label{fig:oC1}
\end{figure}

\begin{equation}
\begin{split}
\Gamma
&\text{--}\underbrace{Y\text{--}k\,|\,k'\text{--}Y'}_{}\text{--}\underbrace{C_0'\text{--}k'\,|\,k\text{--}C_0}_{}\,|\,\underbrace{\Sigma_0\text{--}k\,|\,k'\text{--}\Sigma_0'}_{}\text{--}\underbrace{\Gamma\text{--}k'\,|\,k\text{--}\Gamma}_{}\text{--}\underbrace{Z\text{--}k\,|\,k'\text{--}Z'}_{}\text{--}\underbrace{A_0'\text{--}k'\,|\,k\text{--}A_0}_{}\,
\\
&|\,\underbrace{E_0\text{--}k\,|\,k'\text{--}E_0'}_{}\text{--}\underbrace{T'\text{--}k'\,|\,k\text{--}T}_{}\text{--}Y\,|\,\Gamma\text{--}\underbrace{S\text{--}k\,|\,k'\text{--}S'}_{}\text{--}\underbrace{R'\text{--}k'\,|\,k\text{--}R}_{}\text{--}Z\text{--}T .
\end{split}
\end{equation}

\subsubsection{oC2 (oA2)}
\noindent\textbf{Case 39: ${}^{2}m{}^{2}m{}^{1}m$, planar $d$-wave}\par
\noindent\textbf{Case 40: ${}^{2}m{}^{1}m{}^{2}m$, bulk $d$-wave}\par
\noindent\textbf{Case 41: ${}^{1}m{}^{2}m{}^{2}m$, bulk $d$-wave}\par
\begin{figure}[H]
\centering
\includegraphics[width=0.9\textwidth,keepaspectratio]{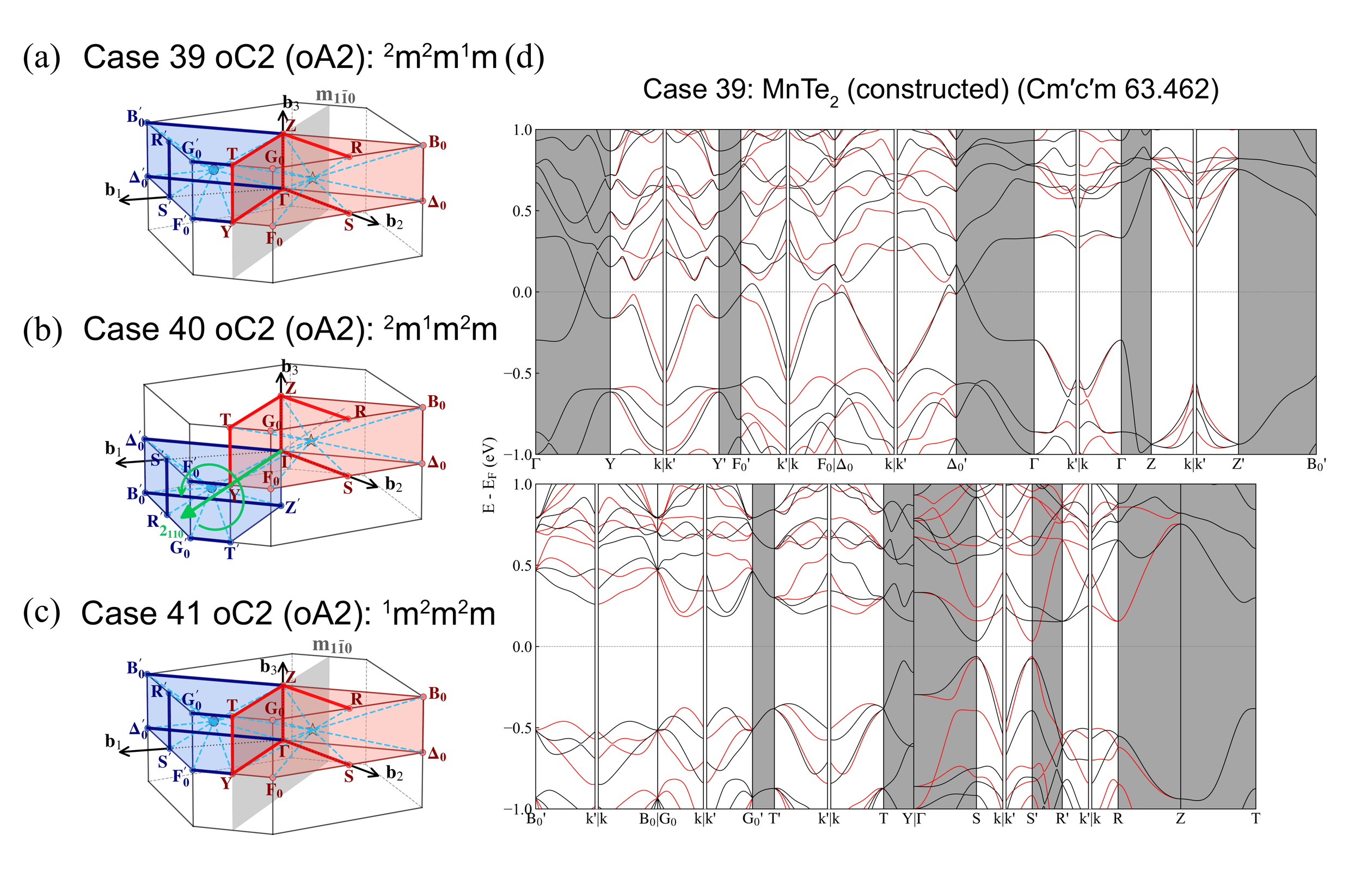}
\caption{Orthorhombic oC2 (oA2) BZ and band for the three spin Laue groups of the $mmm$ Laue group, using MnTe$_2$ (constructed) (MSG without SOC: $Cm'c'm$, BNS 63.462, Type III) as a representative example. (a)--(c) show Cases 39--41; since all three share the same k-path, only one band structure is shown, corresponding to Case 39 ($^{2}m^{2}m^{1}m$).}
\label{fig:oC2}
\end{figure}

\begin{equation}
\begin{split}
\Gamma
&\text{--}\underbrace{Y\text{--}k\,|\,k'\text{--}Y'}_{}\text{--}\underbrace{F_0'\text{--}k'\,|\,k\text{--}F_0}_{}\,|\,\underbrace{\Delta_0\text{--}k\,|\,k'\text{--}\Delta_0'}_{}\text{--}\underbrace{\Gamma\text{--}k'\,|\,k\text{--}\Gamma}_{}\text{--}\underbrace{Z\text{--}k\,|\,k'\text{--}Z'}_{}
\\
&\text{--}\underbrace{B_0'\text{--}k'\,|\,k\text{--}B_0}_{}\,|\,\underbrace{G_0\text{--}k\,|\,k'\text{--}G_0'}_{}\text{--}\underbrace{T'\text{--}k'\,|\,k\text{--}T}_{}\text{--}Y\,|\,\Gamma\text{--}\underbrace{S\text{--}k\,|\,k'\text{--}S'}_{}\text{--}\underbrace{R'\text{--}k'\,|\,k\text{--}R}_{}\text{--}Z\text{--}T .
\end{split}
\end{equation}

\subsubsection{oF1}
\noindent\textbf{Case 42: ${}^{2}m{}^{2}m{}^{1}m$, planar $d$-wave}\par
\noindent\textbf{Case 43: ${}^{2}m{}^{1}m{}^{2}m$, bulk $d$-wave}\par
\noindent\textbf{Case 44: ${}^{1}m{}^{2}m{}^{2}m$, bulk $d$-wave}\par
\begin{figure}[H]
\centering
\includegraphics[width=0.9\textwidth,keepaspectratio]{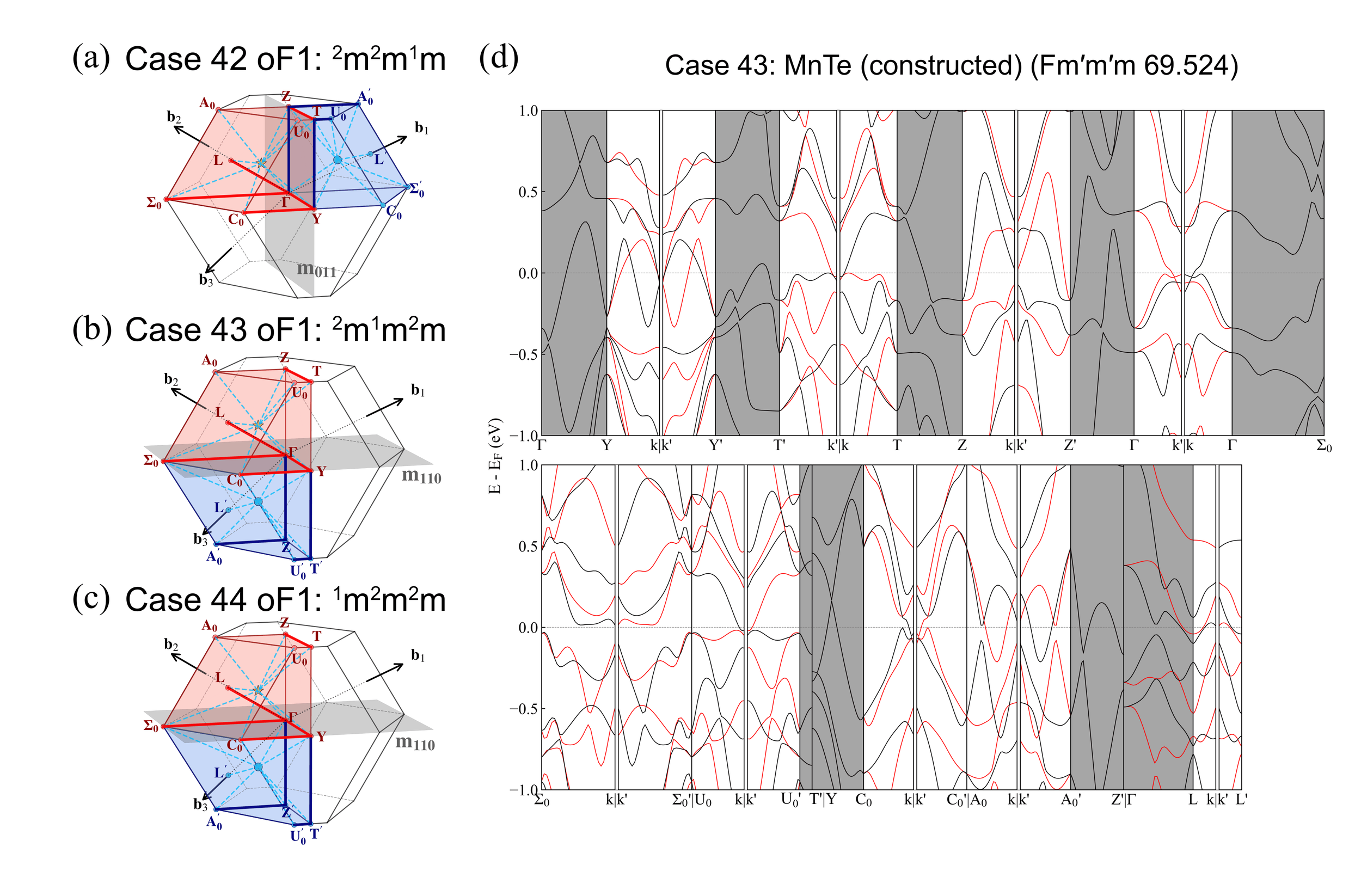}
\caption{Orthorhombic oF1 BZ and band for the three spin Laue groups of the $mmm$ Laue group, using MnTe (constructed) (MSG without SOC: $Fm'm'm$, BNS 69.524, Type III) as a representative example. (a)--(c) show Cases 42--44; since all three share the same k-path, only one band structure is shown, corresponding to Case 43 ($^{2}m^{1}m^{2}m$).}
\label{fig:oF1}
\end{figure}

\begin{equation}
\begin{split}
\Gamma
&\text{--}\underbrace{Y\text{--}k\,|\,k'\text{--}Y'}_{}\text{--}\underbrace{T'\text{--}k'\,|\,k\text{--}T}_{}\text{--}\underbrace{Z\text{--}k\,|\,k'\text{--}Z'}_{}\text{--}\underbrace{\Gamma\text{--}k'\,|\,k\text{--}\Gamma}_{}\text{--}\underbrace{\Sigma_0\text{--}k\,|\,k'\text{--}\Sigma_0'}_{}\,|\,\underbrace{U_0\text{--}k\,|\,k'\text{--}U_0'}_{}
\\
&\text{--}T'\,|\,Y\text{--}\underbrace{C_0\text{--}k\,|\,k'\text{--}C_0'}_{}\,|\,\underbrace{A_0\text{--}k\,|\,k'\text{--}A_0'}_{}\text{--}Z'\,|\,\Gamma\text{--}\underbrace{L\text{--}k\,|\,k'\text{--}L'}_{} .
\end{split}
\end{equation}

\subsubsection{oF2}
\noindent\textbf{Case 45: ${}^{2}m{}^{2}m{}^{1}m$, planar $d$-wave}\par
\noindent\textbf{Case 46: ${}^{2}m{}^{1}m{}^{2}m$, bulk $d$-wave}\par
\noindent\textbf{Case 47: ${}^{1}m{}^{2}m{}^{2}m$, bulk $d$-wave}\par
\begin{figure}[H]
\centering
\includegraphics[width=0.9\textwidth,keepaspectratio]{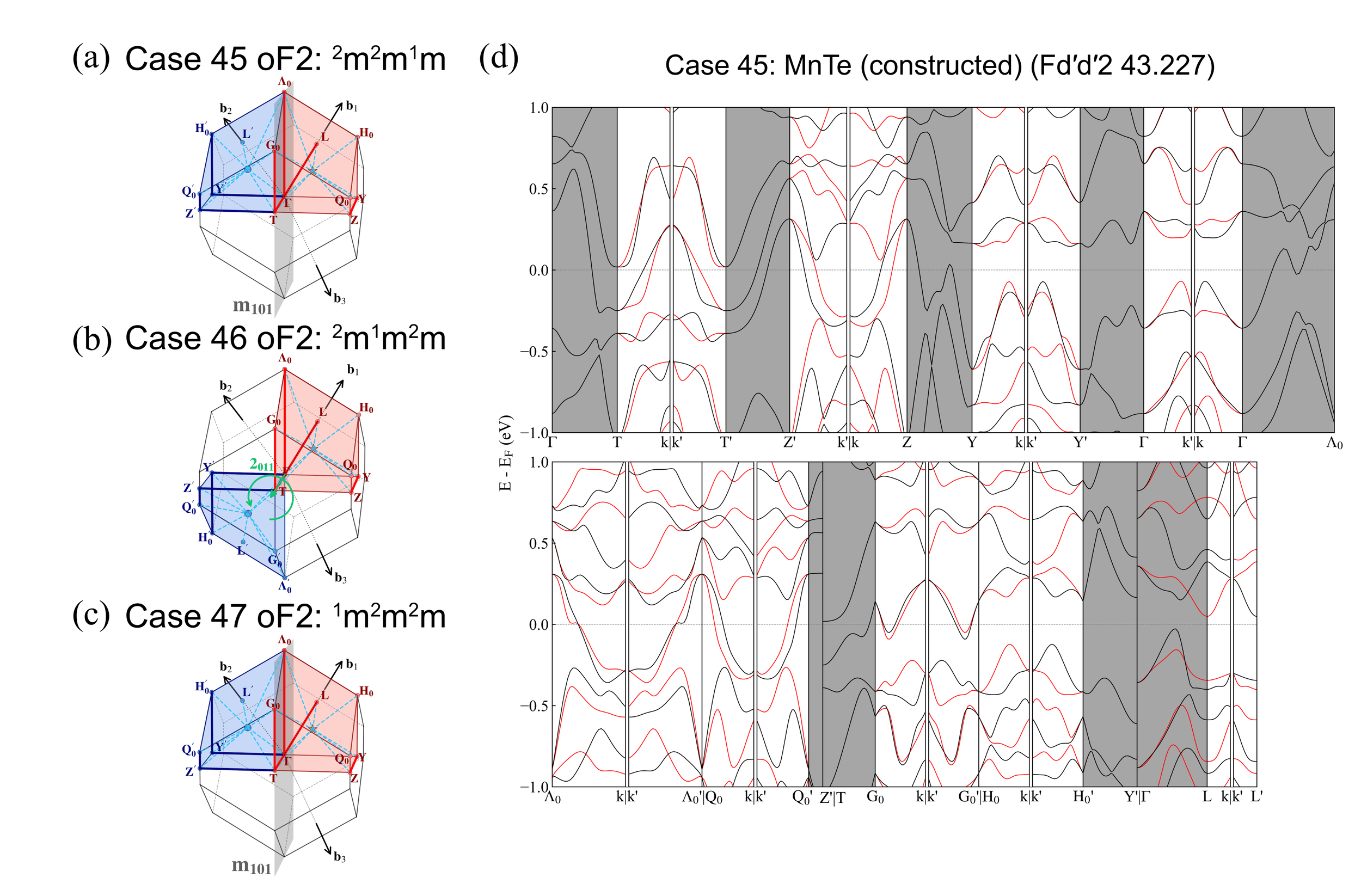}
\caption{Orthorhombic oF2 BZ and band for the three spin Laue groups of the $mmm$ Laue group, using MnTe (constructed) (MSG without SOC: $Fd'd'2$, BNS 43.227, Type III) as a representative example. (a)--(c) show Cases 45--47; since all three share the same k-path, only one band structure is shown, corresponding to Case 45 ($^{2}m^{2}m^{1}m$).}
\label{fig:oF2}
\end{figure}

\begin{equation}
\begin{split}
\Gamma
&\text{--}\underbrace{T\text{--}k\,|\,k'\text{--}T'}_{}\text{--}\underbrace{Z'\text{--}k'\,|\,k\text{--}Z}_{}\text{--}\underbrace{Y\text{--}k\,|\,k'\text{--}Y'}_{}\text{--}\underbrace{\Gamma\text{--}k'\,|\,k\text{--}\Gamma}_{}\text{--}\underbrace{\Lambda_0\text{--}k\,|\,k'\text{--}\Lambda_0'}_{}\,|\,\underbrace{Q_0\text{--}k\,|\,k'\text{--}Q_0'}_{}
\\
&\text{--}Z'\,|\,T\text{--}\underbrace{G_0\text{--}k\,|\,k'\text{--}G_0'}_{}\,|\,\underbrace{H_0\text{--}k\,|\,k'\text{--}H_0'}_{}\text{--}Y'\,|\,\Gamma\text{--}\underbrace{L\text{--}k\,|\,k'\text{--}L'}_{} .
\end{split}
\end{equation}

\subsubsection{oF3}
\noindent\textbf{Case 48: ${}^{2}m{}^{2}m{}^{1}m$, planar $d$-wave}\par
\noindent\textbf{Case 49: ${}^{2}m{}^{1}m{}^{2}m$, bulk $d$-wave}\par
\noindent\textbf{Case 50: ${}^{1}m{}^{2}m{}^{2}m$, bulk $d$-wave}\par
\begin{figure}[H]
\centering
\includegraphics[width=0.9\textwidth,keepaspectratio]{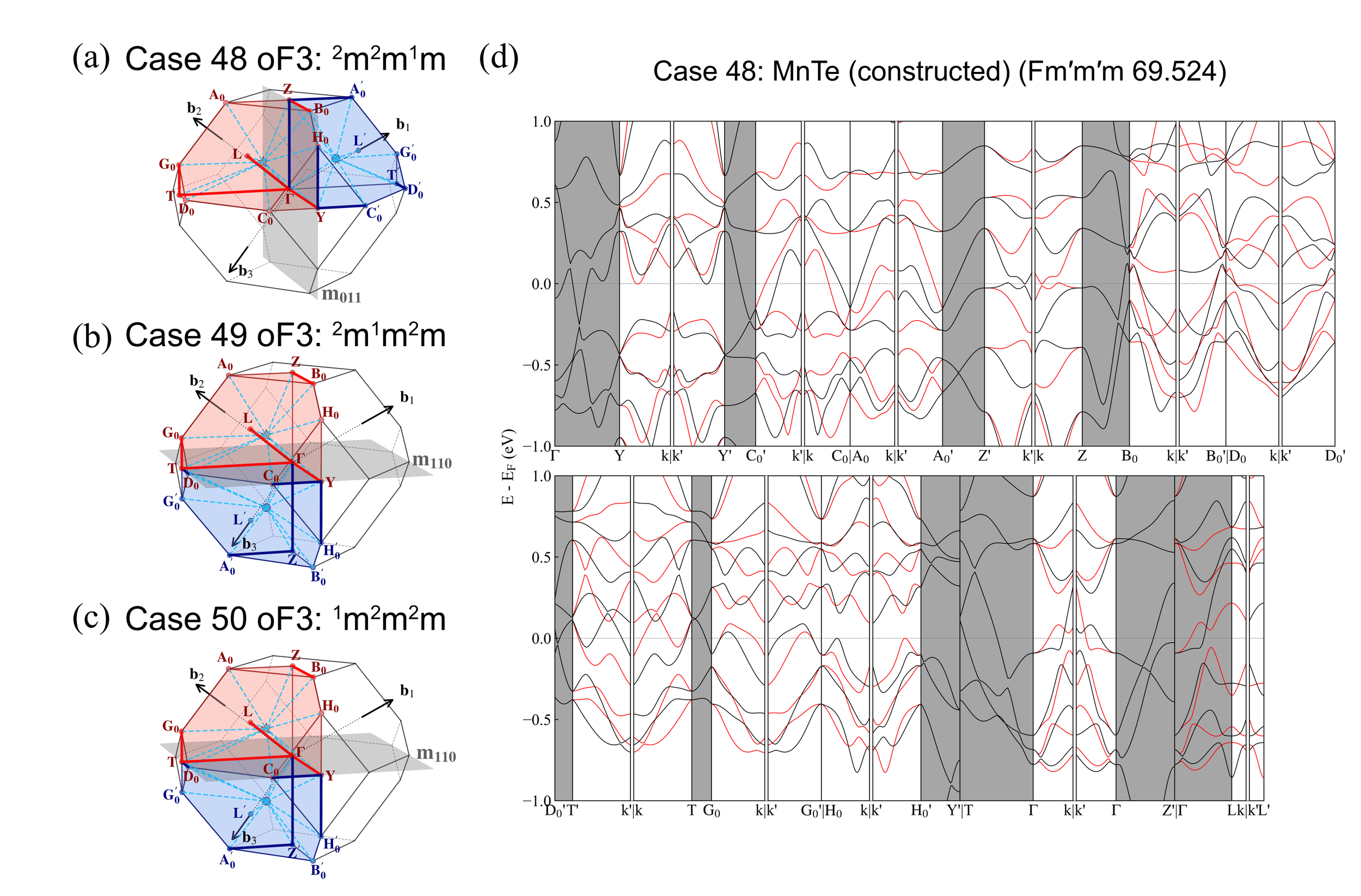}
\caption{Orthorhombic oF3 BZ and band for the three spin Laue groups of the $mmm$ Laue group, using MnTe (constructed) (MSG without SOC: $Fm'm'm$, BNS 69.524, Type III) as a representative example. (a)--(c) show Cases 48--50; since all three share the same k-path, only one band structure is shown, corresponding to Case 48 ($^{2}m^{2}m^{1}m$).}
\label{fig:oF3}
\end{figure}

\begin{equation}
\begin{split}
\Gamma
&\text{--}\underbrace{Y\text{--}k\,|\,k'\text{--}Y'}_{}\text{--}\underbrace{C_0'\text{--}k'\,|\,k\text{--}C_0}_{}\,|\,\underbrace{A_0\text{--}k\,|\,k'\text{--}A_0'}_{}\text{--}\underbrace{Z'\text{--}k'\,|\,k\text{--}Z}_{}\text{--}\underbrace{B_0\text{--}k\,|\,k'\text{--}B_0'}_{}\,|\,\underbrace{D_0\text{--}k\,|\,k'\text{--}D_0'}_{}
\\
&\text{--}\underbrace{T'\text{--}k'\,|\,k\text{--}T}_{}\text{--}\underbrace{G_0\text{--}k\,|\,k'\text{--}G_0'}_{}\,|\,\underbrace{H_0\text{--}k\,|\,k'\text{--}H_0'}_{}\text{--}Y'\,|\,T\text{--}\underbrace{\Gamma\text{--}k\,|\,k'\text{--}\Gamma}_{}\text{--}Z'\,|\,\Gamma\text{--}\underbrace{L\text{--}k\,|\,k'\text{--}L'}_{} .
\end{split}
\end{equation}

\subsection{Monoclinic}
\label{sec:case-library-monoclinic}

\subsubsection{mP1}

\noindent\textbf{Case 51, \hpkot{mP1}: ${}^{2}2/{}^{2}m$, bulk $d$-wave}\par
\begin{figure}[H]
\centering
\includegraphics[width=0.9\textwidth,keepaspectratio]{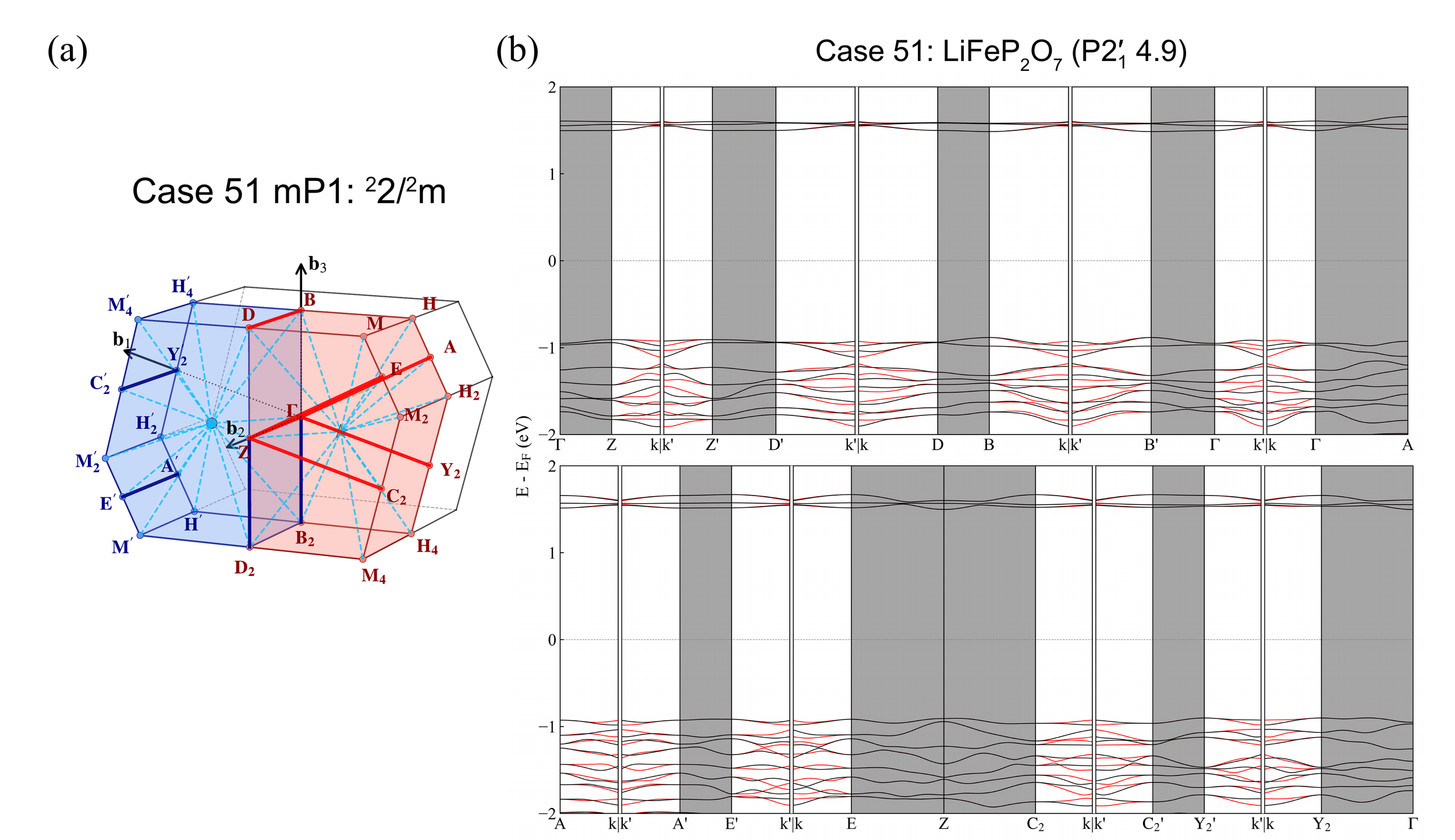}
\caption{Monoclinic mP1 BZ and band for the spin-Laue-group $^{2}2/^{2}m$, using LiFeP$_2$O$_7$ (MAGNDATA 0.83) (MSG without SOC: $P2_1'$, BNS 4.9, Type III) as a representative example. The spin-flip operation chosen in (a) is $2_{010}$ with rotation axis along the $y$ axis ($\Gamma$--$Z$ direction).}
\label{fig:mP1}
\end{figure}

\begin{equation}
\Gamma\text{--}\underbrace{Z\text{--}k\,|\,k'\text{--}Z'}_{}\text{--}\underbrace{D'\text{--}k'\,|\,k\text{--}D}_{}\text{--}\underbrace{B\text{--}k\,|\,k'\text{--}B'}_{}\text{--}\underbrace{\Gamma\text{--}k'\,|\,k\text{--}\Gamma}_{}\text{--}\underbrace{A\text{--}k\,|\,k'\text{--}A'}_{}\text{--}\underbrace{E'\text{--}k'\,|\,k\text{--}E}_{}\text{--}Z\text{--}\underbrace{C_2\text{--}k\,|\,k'\text{--}C_2'}_{}\text{--}\underbrace{Y_2'\text{--}k'\,|\,k\text{--}Y_2}_{}\text{--}\Gamma .
\end{equation}

\subsubsection{mC1}

\noindent\textbf{Case 52, \hpkot{mC1}: ${}^{2}2/{}^{2}m$, bulk $d$-wave}\par
\begin{figure}[H]
\centering
\includegraphics[width=0.9\textwidth,keepaspectratio]{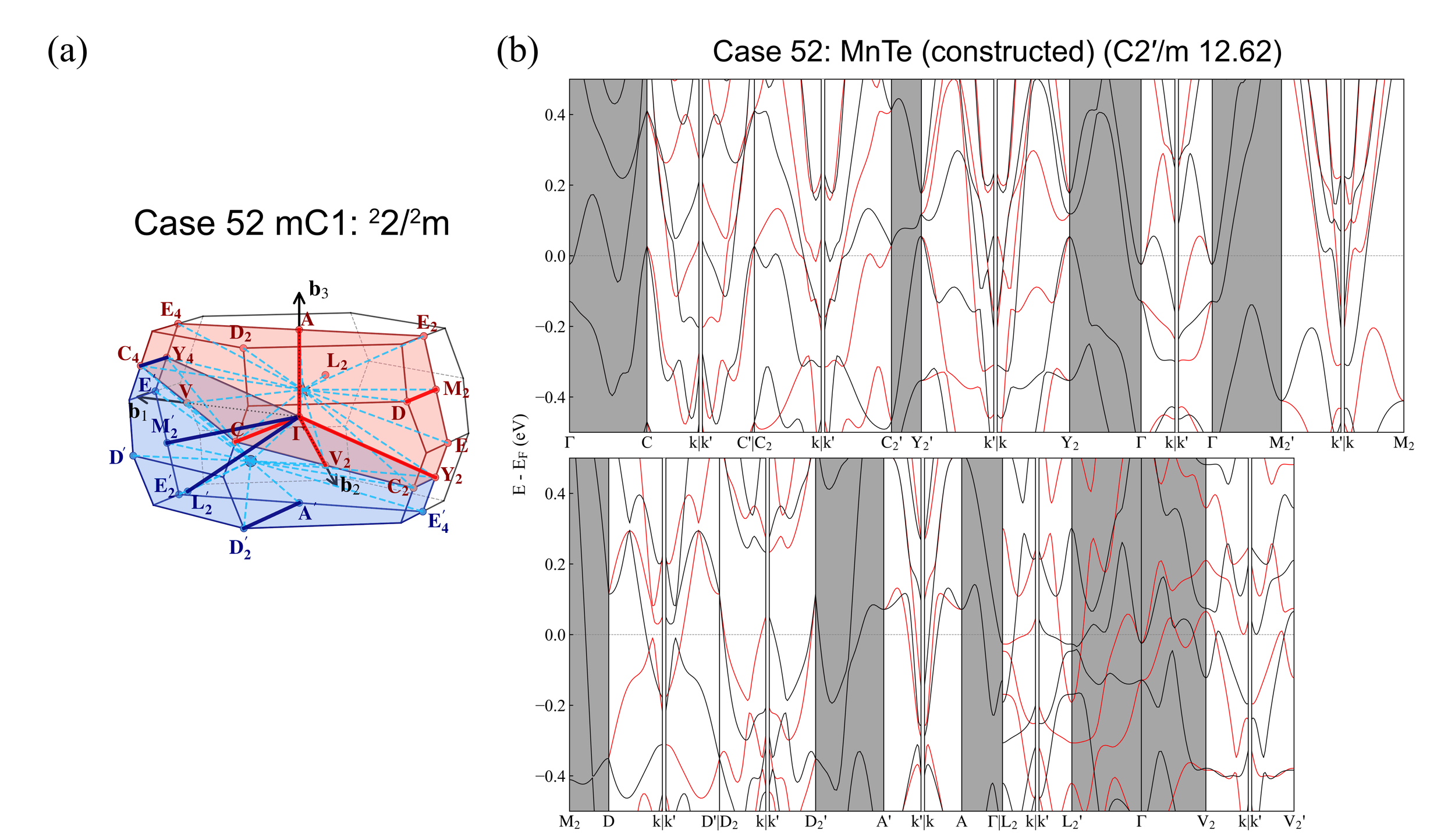}
\caption{Monoclinic mC1 BZ and band for the spin-Laue-group $^{2}2/^{2}m$, using MnTe (constructed) (MSG without SOC: $C2'/m'$, BNS 12.62, Type III) as a representative example. The spin-flip operation chosen in (a) is $2_{110}$ with rotation axis along the $y$ axis ($\Gamma$--$C$ direction).}
\label{fig:mC1}
\end{figure}

\begin{equation}
\begin{split}
\Gamma
&\text{--}\underbrace{C\text{--}k\,|\,k'\text{--}C'}_{}\,|\,\underbrace{C_2\text{--}k\,|\,k'\text{--}C_2'}_{}\text{--}\underbrace{Y_2'\text{--}k'\,|\,k\text{--}Y_2}_{}\text{--}\underbrace{\Gamma\text{--}k\,|\,k'\text{--}\Gamma}_{}\text{--}\underbrace{M_2'\text{--}k'\,|\,k\text{--}M_2}_{}\text{--}\underbrace{D\text{--}k\,|\,k'\text{--}D'}_{}
\\
&\,|\,\underbrace{D_2\text{--}k\,|\,k'\text{--}D_2'}_{}\text{--}\underbrace{A'\text{--}k'\,|\,k\text{--}A}_{}\text{--}\Gamma\,|\,\underbrace{L_2\text{--}k\,|\,k'\text{--}L_2'}_{}\text{--}\Gamma\text{--}\underbrace{V_2\text{--}k\,|\,k'\text{--}V_2'}_{} .
\end{split}
\end{equation}

\subsubsection{mC2}

\noindent\textbf{Case 53, \hpkot{mC2}: ${}^{2}2/{}^{2}m$, bulk $d$-wave}\par
\begin{figure}[H]
\centering
\includegraphics[width=0.9\textwidth,keepaspectratio]{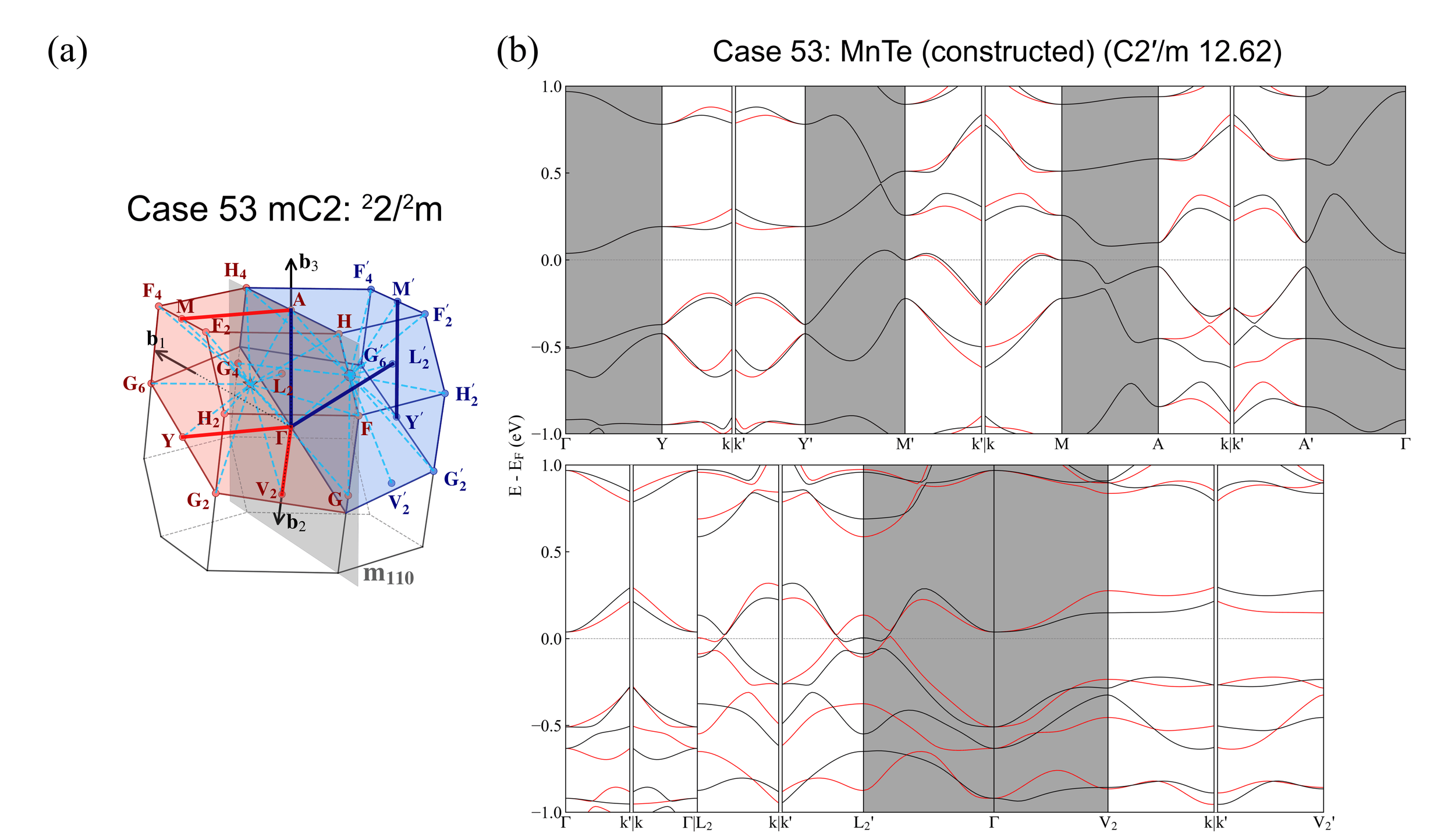}
\caption{Monoclinic mC2 BZ and band for the spin-Laue-group $^{2}2/^{2}m$, using MnTe (constructed) (MSG without SOC: $C2'/m'$, BNS 12.62, Type III) as a representative example.}
\label{fig:mC2}
\end{figure}

\begin{equation}
\Gamma\text{--}\underbrace{Y\text{--}k\,|\,k'\text{--}Y'}_{}\text{--}\underbrace{M'\text{--}k'\,|\,k\text{--}M}_{}\text{--}\underbrace{A\text{--}k\,|\,k'\text{--}A'}_{}\text{--}\underbrace{\Gamma\text{--}k'\,|\,k\text{--}\Gamma}_{}\,|\,\underbrace{L_2\text{--}k\,|\,k'\text{--}L_2'}_{}\text{--}\Gamma\text{--}\underbrace{V_2\text{--}k\,|\,k'\text{--}V_2'}_{} .
\end{equation}

\subsubsection{mC3}

\noindent\textbf{Case 54, \hpkot{mC3}: ${}^{2}2/{}^{2}m$, bulk $d$-wave}\par
\begin{figure}[H]
\centering
\includegraphics[width=0.9\textwidth,keepaspectratio]{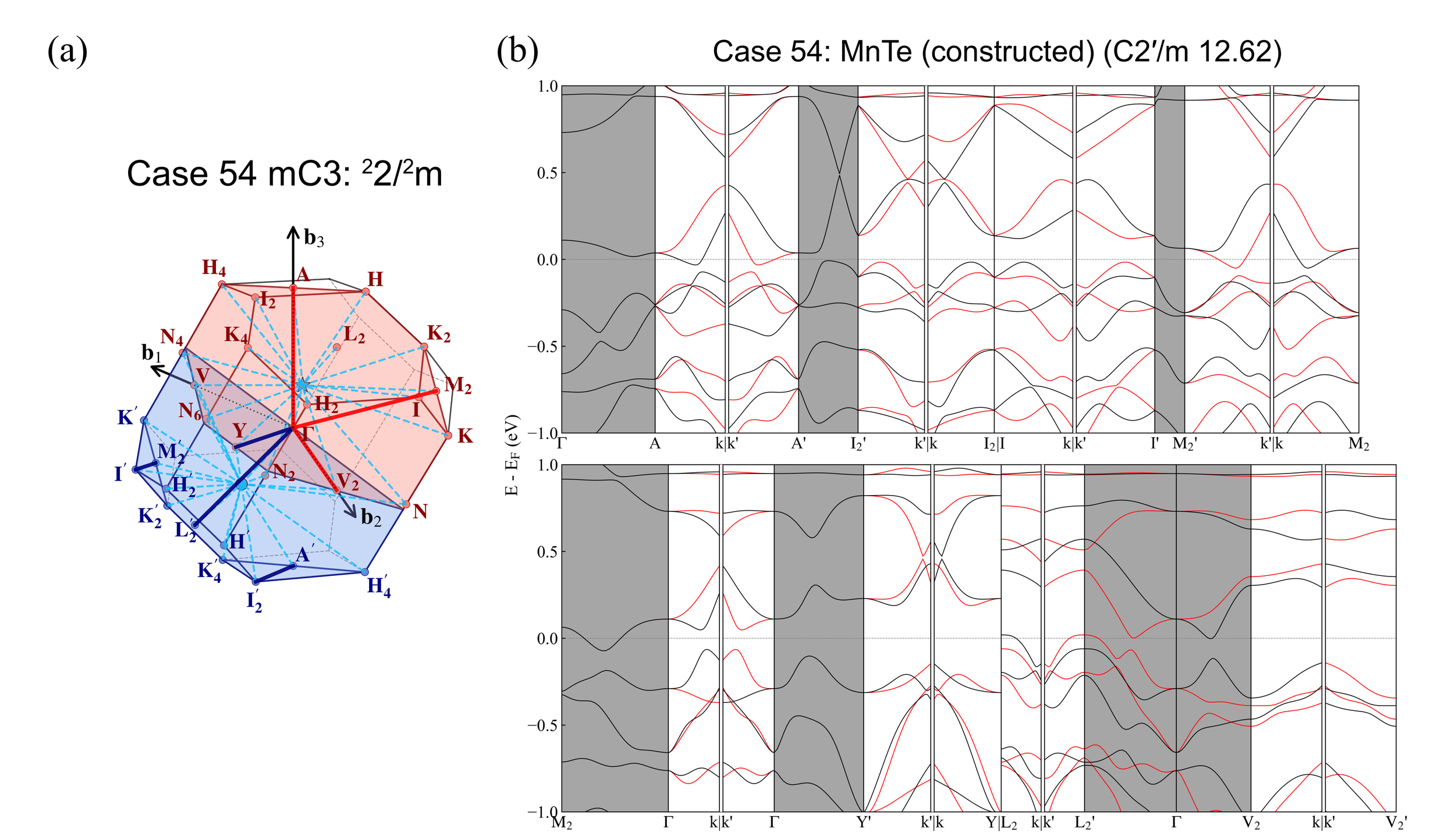}
\caption{Monoclinic mC3 BZ and band for the spin-Laue-group $^{2}2/^{2}m$, using MnTe (constructed) (MSG without SOC: $C2'/m'$, BNS 12.62, Type III) as a representative example. The spin-flip operation chosen in (a) is $2_{110}$ with rotation axis along the $y$ axis ($\Gamma$--$Y$ direction).}
\label{fig:mC3}
\end{figure}

\begin{equation}
\Gamma\text{--}\underbrace{A\text{--}k\,|\,k'\text{--}A'}_{}\text{--}\underbrace{I_2'\text{--}k'\,|\,k\text{--}I_2}_{}\,|\,\underbrace{I\text{--}k\,|\,k'\text{--}I'}_{}\text{--}\underbrace{M_2'\text{--}k'\,|\,k\text{--}M_2}_{}\text{--}\underbrace{\Gamma\text{--}k\,|\,k'\text{--}\Gamma}_{}\text{--}\underbrace{Y'\text{--}k'\,|\,k\text{--}Y}_{}\,|\,\underbrace{L_2\text{--}k\,|\,k'\text{--}L_2'}_{}\text{--}\Gamma\text{--}\underbrace{V_2\text{--}k\,|\,k'\text{--}V_2'}_{} .
\end{equation}

\subsection{Two-dimensional cases}
\label{sec:case-library-2d}

\subsubsection{Hexagonal}

\noindent\textbf{2D Case 01: ${}^{1}6/{}^{1}m{}^{2}m{}^{2}m$, $i$-wave}\par
\noindent\textbf{2D Case 02: ${}^{1}\bar{3}{}^{2}m$, $i$-wave}\par
\begin{figure}[H]
\centering
\includegraphics[width=0.9\textwidth,keepaspectratio]{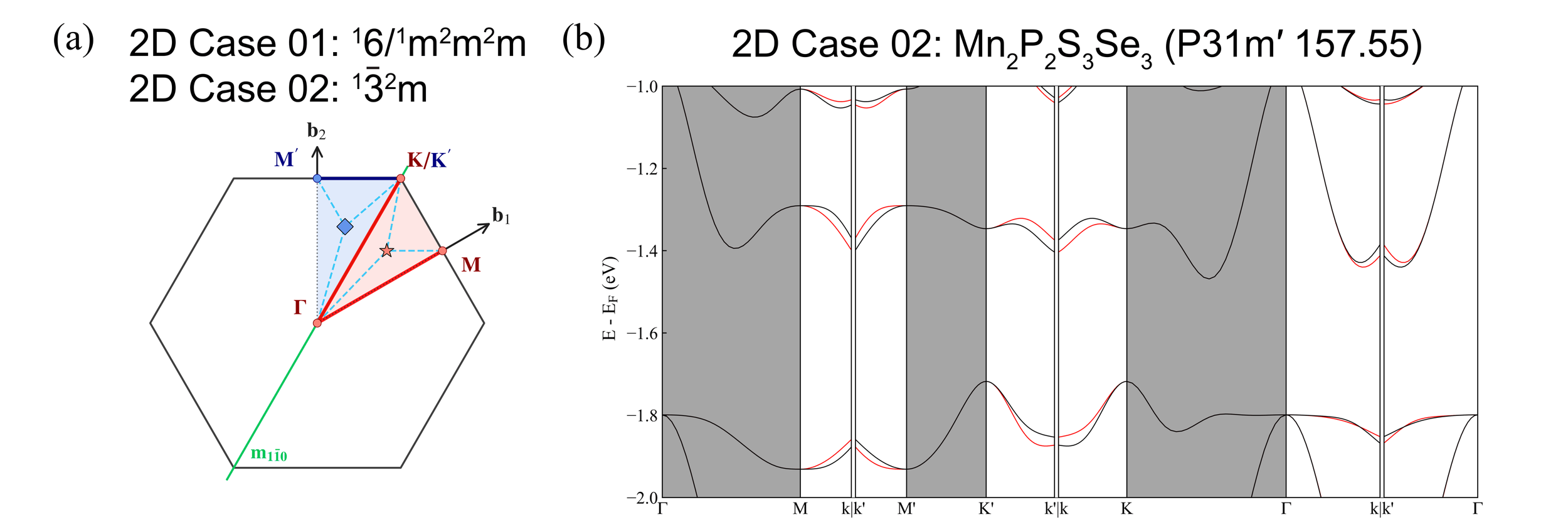}
\caption{2D Hexagonal BZ and band for the Laue groups $6/mmm$ and $\bar{3}m$, using Mn$_2$P$_2$S$_3$Se$_3$ (MSG without SOC: $P31m'$, BNS 157.55, Type III) as a representative example. (a) shows both 2D Cases 01 and 02; since both share the same in-plane k-path, only one band structure is shown, corresponding to 2D Case 02 (${}^{1}\bar{3}{}^{2}m$).}
\label{fig:2D-hexagonal}
\end{figure}

\begin{equation}
\Gamma\text{--}\underbrace{M\text{--}k\,|\,k'\text{--}M'}_{}\text{--}\underbrace{K'\text{--}k'\,|\,k\text{--}K}_{}\text{--}\underbrace{\Gamma\text{--}k\,|\,k'\text{--}\Gamma}_{} .
\end{equation}

\subsubsection{Square (4/mmm)}

\noindent\textbf{2D Case 03: ${}^{2}4/{}^{1}m{}^{2}m{}^{1}m$, $d$-wave}\par
\noindent\textbf{2D Case 04: ${}^{2}4/{}^{1}m{}^{1}m{}^{2}m$, $d$-wave}\par
\noindent\textbf{2D Case 05: ${}^{1}4/{}^{1}m{}^{2}m{}^{2}m$, $g$-wave}\par
\begin{figure}[H]
\centering
\includegraphics[width=0.9\textwidth,keepaspectratio]{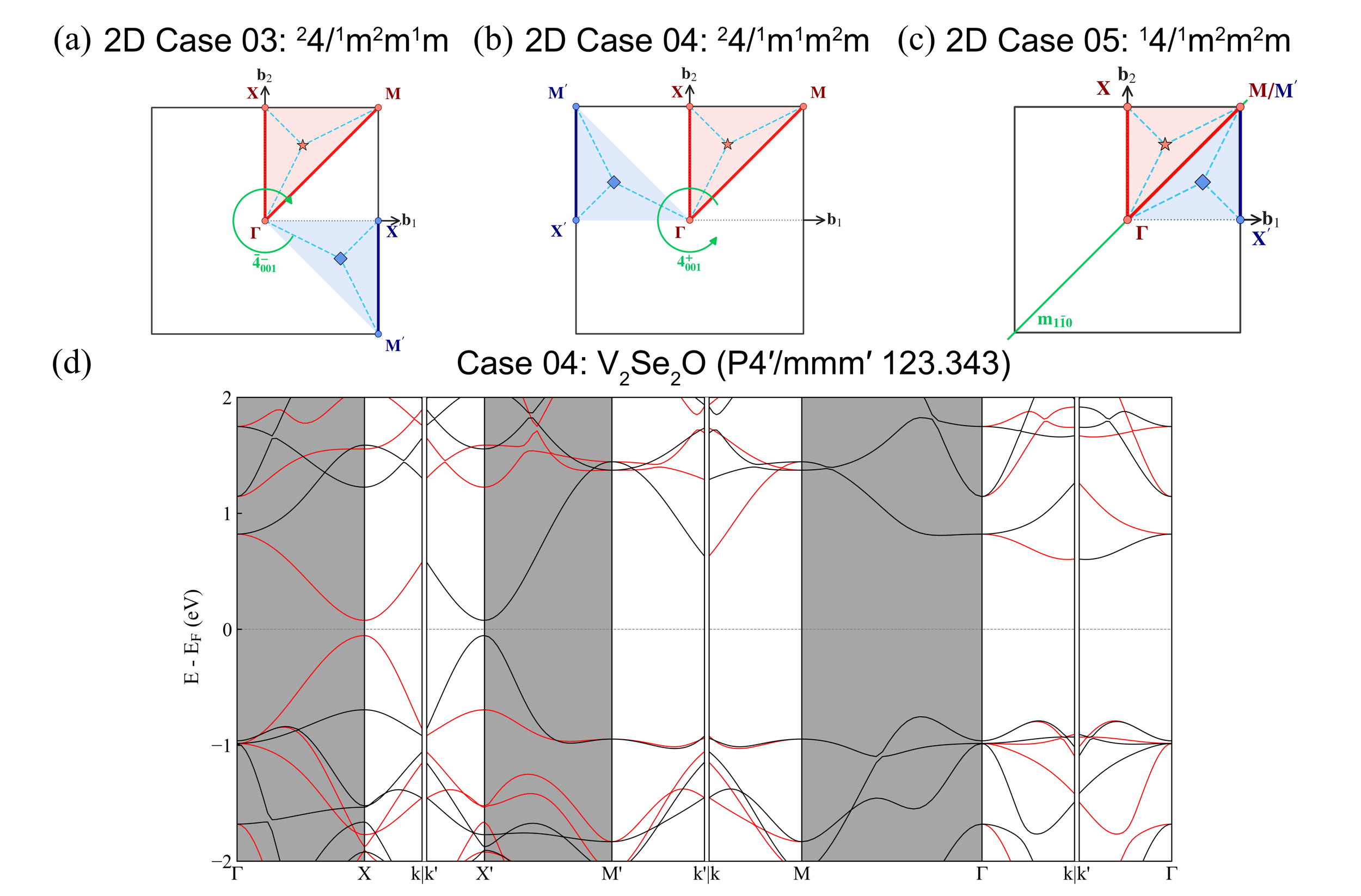}
\caption{2D Square BZ and band for the three spin Laue groups of the $4/mmm$ Laue group, using V$_2$Se$_2$O (MSG without SOC: $P4'/mmm'$, BNS 123.343, Type III) as a representative example. (a)--(c) show 2D Cases 03--05; since all three share the same in-plane k-path, only one band structure is shown, corresponding to 2D Case 04 (${}^{2}4/{}^{1}m{}^{1}m{}^{2}m$).}
\label{fig:2D-square-4mmm}
\end{figure}

\begin{equation}
\Gamma\text{--}\underbrace{X\text{--}k\,|\,k'\text{--}X'}_{}\text{--}\underbrace{M'\text{--}k'\,|\,k\text{--}M}_{}\text{--}\underbrace{\Gamma\text{--}k\,|\,k'\text{--}\Gamma}_{} .
\end{equation}

\subsubsection{Square (4/m)}

\noindent\textbf{2D Case 06: ${}^{2}4/{}^{1}m$, $d$-wave}\par
\begin{figure}[H]
\centering
\includegraphics[width=0.9\textwidth,keepaspectratio]{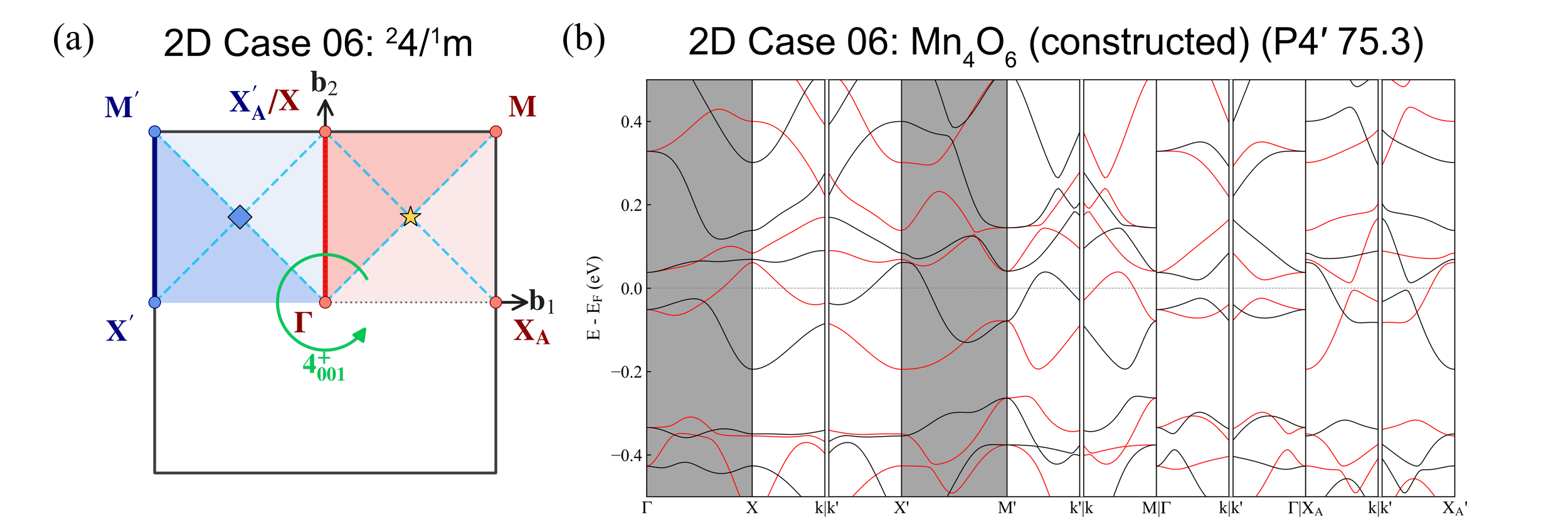}
\caption{2D Square BZ and band for the spin-Laue-group $^{2}4/^{1}m$, using Mn$_4$O$_6$ (constructed) (MSG without SOC: $P4'$, BNS 75.3, Type III) as a representative example. The $\Gamma$--$M$ line is not a high-symmetry line, and it coincides with the general-$k$ path $\Gamma$--$k$ and $k$--$M$. The conventional $\Gamma$--$X$--$M$--$\Gamma$ path is therefore reduced to $\Gamma$--$X$--$M$ to avoid redundancy.}
\label{fig:2D-square-4m}
\end{figure}

\begin{equation}
\Gamma\text{--}\underbrace{X\text{--}k\,|\,k'\text{--}X'}_{}\text{--}\underbrace{M'\text{--}k'\,|\,k\text{--}M}_{}\,|\,\underbrace{\Gamma\text{--}k\,|\,k'\text{--}\Gamma}_{}\,|\,\underbrace{X_A\text{--}k\,|\,k'\text{--}X_A'}_{} .
\end{equation}

\let\AlterSeekOrigNeedspace\Needspace
\renewcommand{\Needspace}[1]{}
\subsubsection{Rectangular}

\noindent\textbf{2D Case 07: ${}^{2}m{}^{2}m{}^{1}m$, $d$-wave}\par
\noindent\textbf{2D Case 08: ${}^{2}2/{}^{2}m_x$, $d$-wave}\par
\begin{figure}[H]
\centering
\includegraphics[width=0.9\textwidth,keepaspectratio]{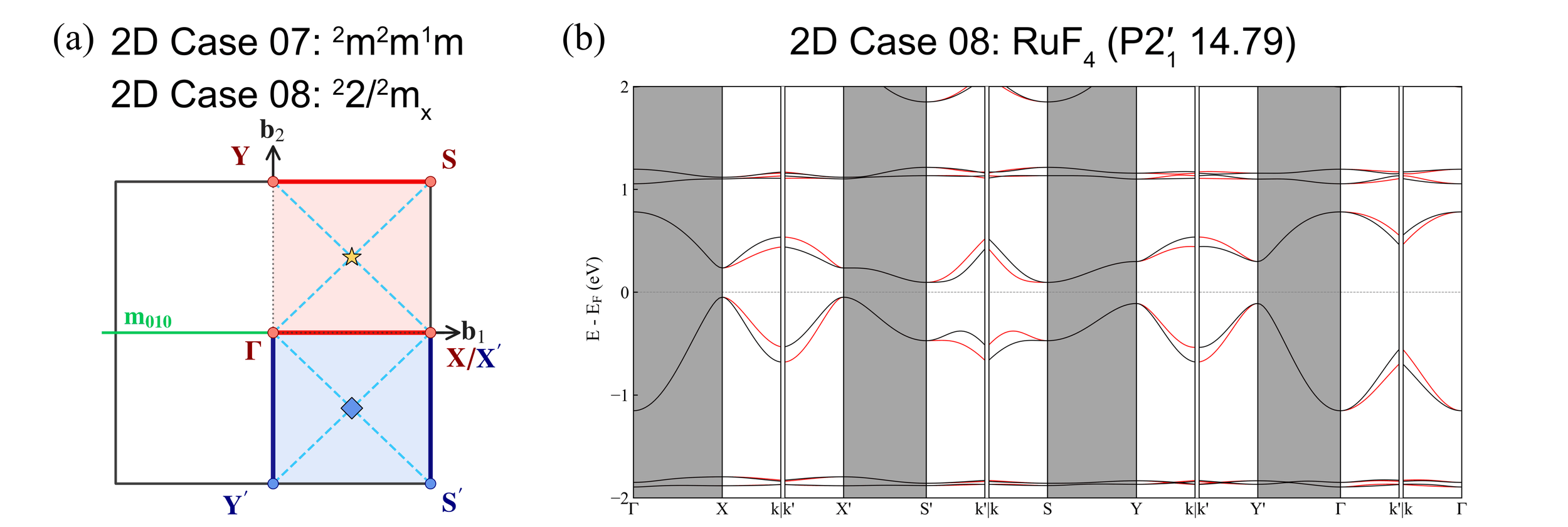}
\caption{2D Rectangular BZ and band for the Laue groups $mmm$ and $2/m$, using RuF$_4$ (MSG without SOC: $P2_1'/c'$, BNS 14.79, Type III) as a representative example. (a)--(b) show 2D Cases 07--08; both share the same rectangular 2D BZ, k-path, and high-symmetry point labels ($\Gamma$/$X$/$S$/$Y$). Only one band structure is shown, corresponding to 2D Case 08 (${}^{2}2/{}^{2}m_x$).}
\label{fig:2D-rectangular}
\end{figure}

\begin{equation}
\Gamma\text{--}\underbrace{X\text{--}k\,|\,k'\text{--}X'}_{}\text{--}\underbrace{S'\text{--}k'\,|\,k\text{--}S}_{}\text{--}\underbrace{Y\text{--}k\,|\,k'\text{--}Y'}_{}\text{--}\underbrace{\Gamma\text{--}k'\,|\,k\text{--}\Gamma}_{} .
\end{equation}

\let\Needspace\AlterSeekOrigNeedspace
\subsubsection{Centered rectangular ($a<b$)}

\noindent\textbf{2D Case 09a: ${}^{2}m{}^{2}m{}^{1}m$, $d$-wave}\par
\noindent\textbf{2D Case 10a: ${}^{2}2/{}^{2}m_x$, $d$-wave}\par
\begin{figure}[H]
\centering
\includegraphics[width=0.9\textwidth,keepaspectratio]{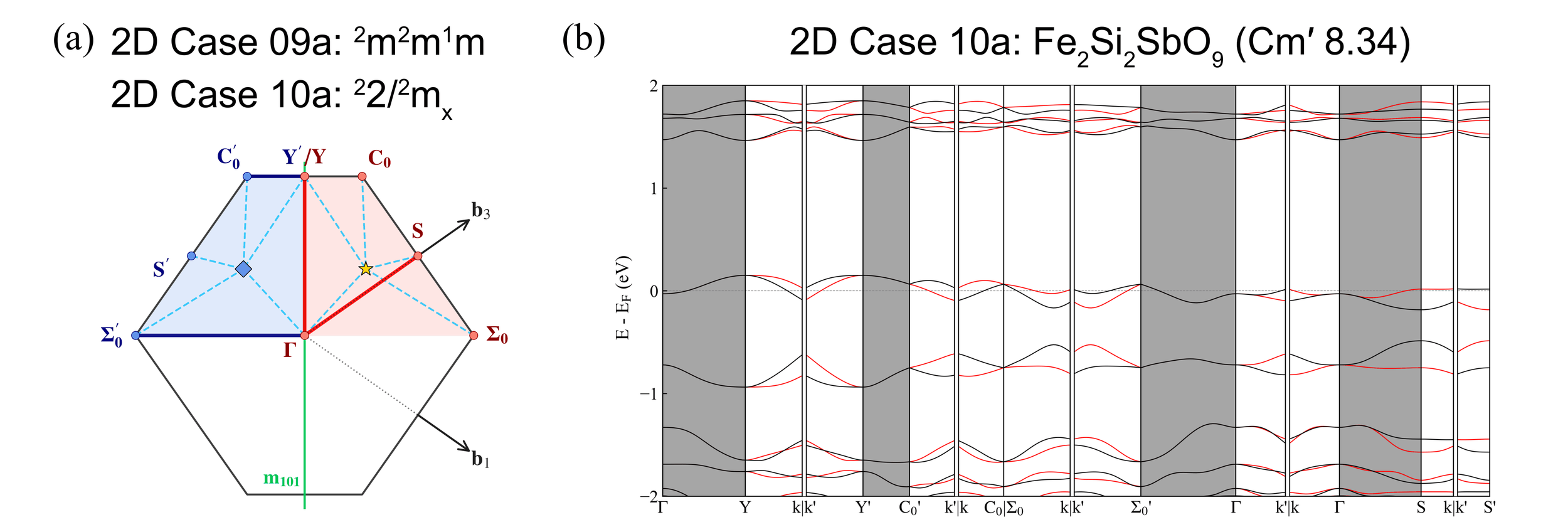}
\caption{2D Centered-rectangular ($a<b$) BZ and band for the Laue groups $mmm$ and $2/m$, using Fe$_2$Si$_2$SbO$_9$ (MSG without SOC: $Cm'$, BNS 8.34, Type III) as a representative example. (a)--(b) show 2D Cases 09a--10a, which share the same 2D BZ, k-path, and high-symmetry point labels ($\Gamma$/$Y$/$C_0$/$\Sigma_0$/$S$). Only one band structure is shown, corresponding to 2D Case 10a (${}^{2}2/{}^{2}m_x$).}
\label{fig:2D-centered-rectangular-alt}
\end{figure}

\begin{equation}
\Gamma\text{--}\underbrace{Y\text{--}k\,|\,k'\text{--}Y'}_{}\text{--}\underbrace{C_0'\text{--}k'\,|\,k\text{--}C_0}_{}\,|\,\underbrace{\Sigma_0\text{--}k\,|\,k'\text{--}\Sigma_0'}_{}\text{--}\underbrace{\Gamma\text{--}k'\,|\,k\text{--}\Gamma}_{}\text{--}\underbrace{S\text{--}k\,|\,k'\text{--}S'}_{} .
\end{equation}

\let\AlterSeekOrigNeedspace\Needspace
\renewcommand{\Needspace}[1]{}
\subsubsection{Centered rectangular ($a>b$)}

\noindent\textbf{2D Case 09b: ${}^{2}m{}^{2}m{}^{1}m$, $d$-wave}\par
\noindent\textbf{2D Case 10b: ${}^{2}2/{}^{2}m_x$, $d$-wave}\par
\begin{figure}[H]
\centering
\includegraphics[width=0.9\textwidth,keepaspectratio]{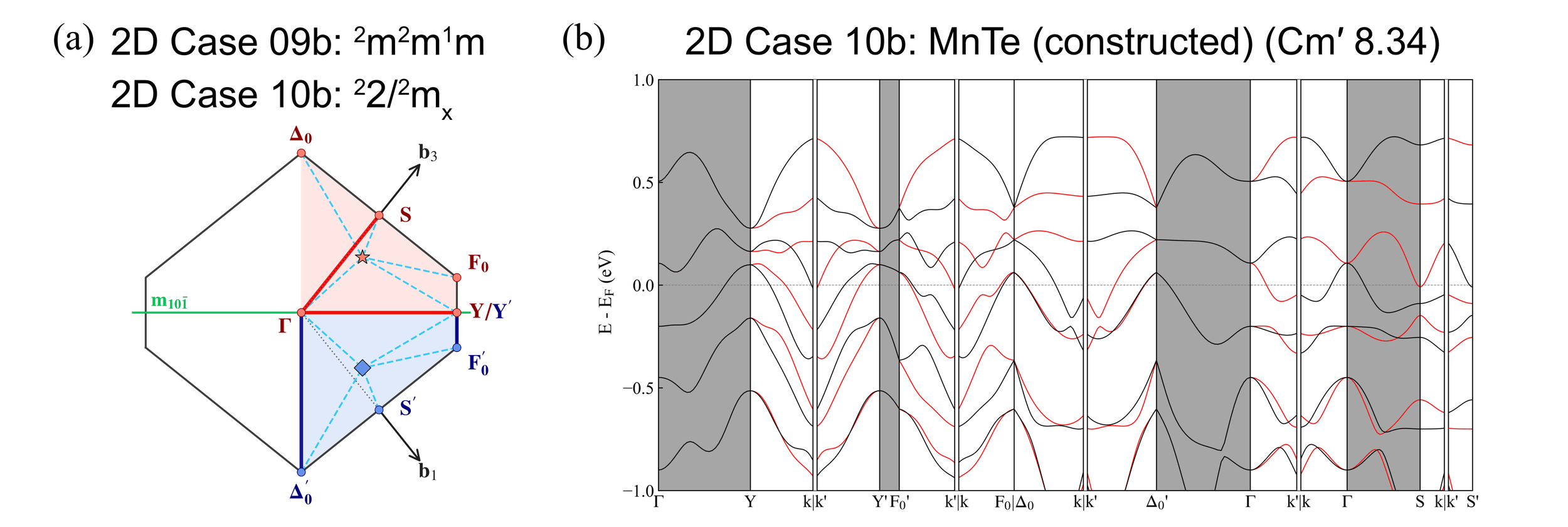}
\caption{2D Centered-rectangular ($a>b$) BZ and band for the Laue groups $mmm$ and $2/m$, using Mn$_2$Te$_5$ (MSG without SOC: $Cm'$, BNS 8.34, Type III) as a representative example. (a)--(b) show 2D Cases 09b--10b, which share the same 2D BZ, k-path, and high-symmetry point labels ($\Gamma$/$Y$/$F_0$/$\Delta_0$/$S$). Only one band structure is shown, corresponding to 2D Case 10b (${}^{2}2/{}^{2}m_x$).}
\label{fig:2D-centered-rectangular-agt}
\end{figure}

\begin{equation}
\Gamma\text{--}\underbrace{Y\text{--}k\,|\,k'\text{--}Y'}_{}\text{--}\underbrace{F_0'\text{--}k'\,|\,k\text{--}F_0}_{}\,|\,\underbrace{\Delta_0\text{--}k\,|\,k'\text{--}\Delta_0'}_{}\text{--}\underbrace{\Gamma\text{--}k'\,|\,k\text{--}\Gamma}_{}\text{--}\underbrace{S\text{--}k\,|\,k'\text{--}S'}_{} .
\end{equation}
\end{document}